\documentstyle[a4,11pt,axodraw]{article}

\def\ortho{{\bot}}
\def\today{\number\day\space
     \ifcase\month\or
       January\or February\or March\or April\or May\or June\or
       July\or August\or September\or October\or November\or December\fi
     \space\number\year}
\def\fanti(#1,#2){\BCirc(#1,#2){2}}
\def\freg(#1,#2){\Vertex(#1,#2){2}}
\def\dReg(#1,#2){\SetOffset(#1,#2)\BCirc(0,0){4}
    \Line(2.8,2.8)(-2.8,-2.8)
    \Line(2.8,-2.8)(-2.8,2.8)\SetOffset(0,0)}
\def\dAdj(#1,#2){\SetOffset(#1,#2)\BCirc(0,0){4}
    \Line(2.8,0)(-2.8,0)
    \Line(0,-2.8)(0,2.8)\SetOffset(0,0)}
\def\dEmp(#1,#2){\SetOffset(#1,#2)\BCirc(0,0){4}
	\Vertex(0,0){1.5} \SetOffset(0,0)}
\def\dotpr{\!\cdot\!}
\def\nn{\nonumber \\ &&}
\def\Tr{{\rm Tr~}}
\def\Str{{\rm Str~}}
\def\Ch{{\rm Ch}}
\def\ie{{\it i.e.}}
\def\eg{{\it e.g.}}
\def\half{\frac{1}{2}}
\def\intab(#1){&\hfill$\!\!#1\!\!$}

\begin{document}
\hfill
UM-TH-98-01 

\hfill
NIKHEF-98-004

\begin{center}
{\huge \bf Group theory factors for Feynman diagrams } \\[8mm] 
 T. van Ritbergen$^a$, A.N. Schellekens$^b$, J.A.M. Vermaseren$^b$ \\ [3mm]
\begin{itemize}
\item[$^a$]
 Randall Laboratory of Physics, University of Michigan,\\
 Ann Arbor, MI 48109, USA
\item[$^b$]
 NIKHEF, P.O. Box 41882, \\ 1009 DB, Amsterdam, The Netherlands \\
\end{itemize}
\end{center}
\begin{center}
\vspace{8mm}
\today
\vspace{8mm}
\end{center}

\begin{abstract}

We present algorithms for the group independent reduction of group theory
factors of Feynman diagrams. We also give formulas and values for a large
number of group invariants in which the group theory factors are expressed.
This includes formulas for various contractions of symmetric invariant
tensors, formulas and algorithms for the computation of characters
and generalized Dynkin indices and trace identities. 
Tables of all Dynkin indices for all exceptional algebras 
are presented, as well as all trace identities
to order equal to the dual Coxeter number. Further results
are available through efficient computer algorithms.
 
\end{abstract}
\newpage

\section{Introduction}

As the number of loops to which perturbative field theories are evaluated 
increases, the group\footnote{Since we are dealing with perturbation
theory we only encounter Lie algebras, and we are insensitive to the
global properties of the Lie group. Nevertheless, following standard
practice, we will often use the word ``group" rather than ``algebra".}
structure of the individual diagrams becomes more and 
more complicated. This problem has been recognized many years ago and on a 
group-by-group basis some very compact algorithms were 
proposed~\cite{cvitanovic} for their computation. Especially for the 
defining and the adjoint representations of the classical groups $SU(N)$, 
$SO(N)$ and $Sp(N)$ these algorithms can be implemented rather easily in a 
symbolic program that will then give the color trace of a diagram as a 
function of the parameter $N$~\cite{schladming}. The disadvantage of these 
algorithms is however that these results give no information about group 
invariants and hence it is only possible for very simple diagrams to 
generalize the results such that they are valid for arbitrary groups and 
arbitrary representations. Hence a different type of algorithm is needed, 
if one would like a more general answer. That such information is useful 
can be seen from some recent calculations in QCD \cite{QCD4} in which the 
representation in terms of invariants could show immediately why 
extrapolations of lower orders in perturbation theory could not be 
successful. In addition the presentation in terms of group invariants is 
more general and needs hardly any new work when one needs to apply it for 
different groups or representations. The need for this kind
of generality is clear, for example, from
grand unification and string theory, where all semi-simple Lie groups 
may occur.   

We consider non-abelian gauge theories based on simple compact Lie
groups. The extension to semi-simple algebras and additional $U(1)$
factors is then straightforward. The gauge bosons are assumed to couple
to matter in some irreducible representation $R$ of the gauge group. The
generalization to reducible representations is also straightforward. 
The group-theoretical quantities that appear in the initial expressions
are the structure constants $f^{abc}$ 
(appearing in gauge self-couplings and ghost couplings)
and the Lie-algebra generators 
$T_R^a$ in the representation $R$, appearing in the coupling of the
gauge bosons to matter. 
In this paper we consider only ``vacuum bubbles", \ie\ diagrams without
external lines. As far as the group theoretical factor is concerned,
our results are relevant for any diagram whose external lines carry
no gauge quantum number, or for the absolute 
value squared of any amplitude if one sums over the gauge quantum
numbers of all external lines. The group theoretical factor of other 
diagrams can be obtained by multiplying the diagram by projection 
operators.

The group theory factor of a vacuum bubble diagram consists of traces of
a certain number of matrices $T^a_R$, whose indices are fully 
contracted among each other and
with some combination of structure constants. Our goal is to
obtain an expression for this factor that is minimally
representation- or group-dependent.   
In principle,
this goal can be achieved as follows.

\begin{enumerate}
\item Express the traces in terms of symmetrized traces. This can always
be done at the expense of some additional factors $f^{abc}$

Now one may simplify the resulting expression further by observing
that the structure constants can be viewed as representation
matrices in the adjoint representation. This allows us to

\item Eliminate all closed loops of structure constants $f^{abc}$. 

This amounts to performing step 1.  on traces of adjoint
matrices $T^a_A$. Step 2. can also be performed in an algorithmic way to 
arbitrary order. However, the algorithm is not identical
to that of step 1
because of the special properties
of the adjoint representation.  
 
\item Express the symmetrized traces in terms of a standard basis of
symmetric invariant tensors. A Lie algebra of rank $r$ has precisely
$r$ such tensors \cite{Racah,Gelfand}.

At this point we have succeeded in expressing every group theory
factor
in terms of $r+1$ representation-independent quantities, namely
the symmetric tensors and $f^{abc}$. 
The representation dependence is encapsulated in terms of
(generalized)
indices \cite{patera}. We show how these indices can be computed for
any representation of any Lie-algebra to any desired order. 
This algorithm requires a convenient choice for the basis of tensors,
which is {\it not} the  
mathematically more elegant ``orthogonal" basis 
advocated in \cite{patera}. The result is also
to a reasonable extent group-independent. 
The only way group-dependence enters is trough the (non)-existence of
certain invariant tensors, but one may simply take all possible
tensors into account, and only eliminate them at the end.  
The only problem is that the group $SO(4N)$ has {\it two} distinct
tensors of rank $2N$. This case can rather easily be dealt with
explicitly.

Although our main goal has now been achieved, 
the result is expressed in terms of
many combinations of symmetric tensors and structure constants
that are not all independent. 
Unfortunately there do not seem to exist many mathematical
results regarding these invariants. In particular, we are not aware
of any theorem regarding the minimal number of invariant
combinations. For this reason the rest of our program is
limited to finite orders, and is not guaranteed to yield the
optimal answer in all cases. As a first step we

\item Eliminate as much as possible the 
remaining structure constants $f^{abc}$.

We do not know of a proof that this is always possible, and
in fact we have only been able to do this explicitly
up to a certain order. The first
object where we were unable to perform step 4 is built out of
two structure constants and three rank 4 symmetric tensors. 

\item If step 4 is completed, one is left with a fully contracted
combination of symmetric invariant tensors. We derive formulas expressing
many such contractions in terms of a few basic ones. 

The last step is 
essentially group dependent (and therefore somewhat outside
our main interest): 

\item Compute a formula for the basic invariants for each group.

Here  ``basic invariant" is any of the contracted combinations of 
symmetric tensors that could not be expressed in terms of others,
and any combination involving additional structure constants that
could not be eliminated.  
\end{enumerate}
 
None of these steps is new in itself, but we believe that in all
cases we are going considerably beyond previous results
(see {\it e.g} [7-25]).
Since 
the application we have in mind is to Feynman diagrams, it is
essential not just to develop an algorithm, but also to make
sure it can be carried out efficiently. Complicated group theory
factors appear only at higher orders in perturbation theory, which
implies that one must be able to deal with
a very large number of diagrams. 

The organization of this paper is as follows. In the next section
we give some definitions and conventions, and present some 
well-known general results on invariant tensors.
In section 3 we present the algorithm to perform step 1. Although this
is in principle straightforward, without proper 
care such an algorithm may quickly
get out of control. The same is true for step 2, which is 
presented in section 4. In section 5 we present the character method
for computing indices and symmetrized traces. This section is based on
results presented in \cite{ScWd} and \cite{patera} , the main novelty being
the extension to  
all higher indices of exceptional
algebras. In section 6 we discuss steps 4 and 5. Section 7 contains
some remarks regarding the advantages and disadvantages of the two
basis choices for the symmetric tensors. 

In appendix A we present explicit results for indices and
trace identities of the exceptional algebras; appendix B contains a
description of the computer methods used, and in appendix C we give
a few examples to demonstrate the efficiency of the algorithm.
Appendix D contains many explicit formulas for invariants (step 6).  
In appendix E we discuss chiral traces in $SO(2N)$

\section{Generalities}
\subsection{Definitions}
We consider simple Lie-algebras whose generators satisfy
the commutation relation 
\begin{equation}
 [ T^a, T^b]=if^{abc} T^c 
\end{equation}
Our conventions is to use hermitean generators $T^a$ and to choose the 
Killing form proportional to $\delta^{ab}$:
\begin{equation}
\label{eq:NormOne}
  \Tr T^a T^b \propto \delta^{ab}
\end{equation}
with a positive and representation dependent proportionality constant that
will be fixed later. With this convention the structure constants $f_{abc}$ 
are real and completely anti-symmetric.

\begin{figure}[htb]
\centering
\begin{picture}(168,304)(0,0)
\SetScale{0.8}
\SetPFont{Helvetica}{12}
\PText(6.5,372)(0)[r]{A}
\PText(7,332)(0)[r]{B}
\PText(7,292)(0)[r]{C}
\PText(7,242)(0)[r]{D}
\PText(7,207)(0)[r]{G}
\PText(7,172)(0)[r]{F}
\PText(7,122)(0)[r]{E}
\PText(7,72)(0)[r]{E}
\PText(7,22)(0)[r]{E}
\SetPFont{Helvetica}{9}
\PText(11,368)(0)[r]{r}
\PText(11,328)(0)[r]{r}
\PText(11,288)(0)[r]{r}
\PText(11,238)(0)[r]{r}
\PText(12,203)(0)[r]{2}
\PText(12,168)(0)[r]{4}
\PText(12.5,118)(0)[r]{6}
\PText(12.5,68)(0)[r]{7}
\PText(12.5,18)(0)[r]{8}
\SetPFont{Helvetica}{8}
	\Line(20,370)(120,370) \Line(160,370)(200,370)
	\DashLine(120,370)(160,370){4}
	\BCirc(20,370){2} \PText(20,360)(0)[b]{1}   
	\BCirc(50,370){2} \PText(50,360)(0)[b]{2}   
	\BCirc(80,370){2} \PText(80,360)(0)[b]{3}   
	\BCirc(110,370){2}\PText(110,360)(0)[b]{4}  
	\BCirc(170,370){2}\PText(171,360)(0)[b]{r-1}
	\BCirc(200,370){2}\PText(201,360)(0)[b]{r}  
%
	\Line(20,330)(120,330) \Line(160,330)(170,330)
	\DashLine(120,330)(160,330){4}
	\Line(170,331.5)(200,331.5) \Line(170,328.5)(200,328.5)
	\Line(187,330)(183,334) \Line(187,330)(183,326)
	\BCirc(20,330){2}  \PText(20,320)(0)[b]{1}   
	\BCirc(50,330){2}  \PText(50,320)(0)[b]{2}   
	\BCirc(80,330){2}  \PText(80,320)(0)[b]{3}   
	\BCirc(110,330){2} \PText(110,320)(0)[b]{4}  
	\BCirc(170,330){2} \PText(171,320)(0)[b]{r-1}
	\Vertex(200,330){2}\PText(201,320)(0)[b]{r}  
%
	\Line(20,290)(120,290) \Line(160,290)(170,290)
	\DashLine(120,290)(160,290){4}
	\Line(170,291.5)(200,291.5) \Line(170,288.5)(200,288.5)
	\Line(183,290)(187,294) \Line(183,290)(187,286)
	\Vertex(20,290){2} \PText(20,280)(0)[b]{1}   
	\Vertex(50,290){2} \PText(50,280)(0)[b]{2}   
	\Vertex(80,290){2} \PText(80,280)(0)[b]{3}   
	\Vertex(110,290){2}\PText(110,280)(0)[b]{4}  
	\Vertex(170,290){2}\PText(171,280)(0)[b]{r-1}
	\BCirc(200,290){2} \PText(201,280)(0)[b]{r}  
	\Line(20,240)(120,240) \Line(160,240)(170,240)
	\Line(170,240)(200,265) \Line(170,240)(200,215)
	\DashLine(120,240)(160,240){4}
	\BCirc(20,240){2} \PText(20,230)(0)[b]{1}   
	\BCirc(50,240){2} \PText(50,230)(0)[b]{2}   
	\BCirc(80,240){2} \PText(80,230)(0)[b]{3}   
	\BCirc(110,240){2}\PText(110,230)(0)[b]{4}  
	\BCirc(170,240){2}\PText(168,230)(0)[b]{r-2}
	\BCirc(200,215){2}\PText(205,216)(0)[l]{r-1}
	\BCirc(200,265){2}\PText(205,266)(0)[l]{r}  
	\Line(20,205)(55,205) \Line(20,207)(55,207) \Line(20,203)(55,203)
	\Line(39.5,205)(35.5,209) \Line(39.5,205)(35.5,201)
	\BCirc(20,205){2} \PText(20,195)(0)[b]{1} 
	\Vertex(55,205){2}\PText(55,195)(0)[b]{2} 
	\Line(20,170)(50,170) \Line(80,170)(110,170)
	\Line(50,171.5)(80,171.5) \Line(50,168.5)(80,168.5)
	\Line(67,170)(63,174)\Line(67,170)(63,166)
	\BCirc(20,170){2}  \PText(20,160)(0)[b]{1}  
	\BCirc(50,170){2}  \PText(50,160)(0)[b]{2}  
	\Vertex(80,170){2} \PText(80,160)(0)[b]{3}  
	\Vertex(110,170){2}\PText(110,160)(0)[b]{4} 
	\Line(20,120)(140,120) \Line(80,120)(80,150)
	\BCirc(20,120){2} \PText(20,110)(0)[b]{1}  
	\BCirc(50,120){2} \PText(50,110)(0)[b]{2}  
	\BCirc(80,120){2} \PText(80,110)(0)[b]{3}  
	\BCirc(110,120){2}\PText(110,110)(0)[b]{4} 
	\BCirc(140,120){2}\PText(141,110)(0)[b]{5} 
	\BCirc(80,150){2} \PText(85,151)(0)[l]{6}  
	\Line(20,70)(170,70) \Line(80,70)(80,100)
	\BCirc(20,70){2} \PText(20,60)(0)[b]{1}  
	\BCirc(50,70){2} \PText(50,60)(0)[b]{2}  
	\BCirc(80,70){2} \PText(80,60)(0)[b]{3}  
	\BCirc(110,70){2}\PText(110,60)(0)[b]{4} 
	\BCirc(140,70){2}\PText(141,60)(0)[b]{5} 
	\BCirc(170,70){2}\PText(171,60)(0)[b]{6} 
	\BCirc(80,100){2}\PText(85,101)(0)[l]{7} 
	\Line(20,20)(200,20) \Line(140,20)(140,50)
	\BCirc(20,20){2} \PText(20,10)(0)[b]{1}  
	\BCirc(50,20){2} \PText(50,10)(0)[b]{2}  
	\BCirc(80,20){2} \PText(80,10)(0)[b]{3}  
	\BCirc(110,20){2}\PText(110,10)(0)[b]{4} 
	\BCirc(140,20){2}\PText(141,10)(0)[b]{5} 
	\BCirc(170,20){2}\PText(171,10)(0)[b]{6} 
	\BCirc(200,20){2}\PText(201,10)(0)[b]{7} 
	\BCirc(140,50){2}\PText(145,51)(0)[l]{8} 
\end{picture}
\caption{\label{fig:dynkin}\sl Dynkin diagrams and labelling conventions}
\end{figure}
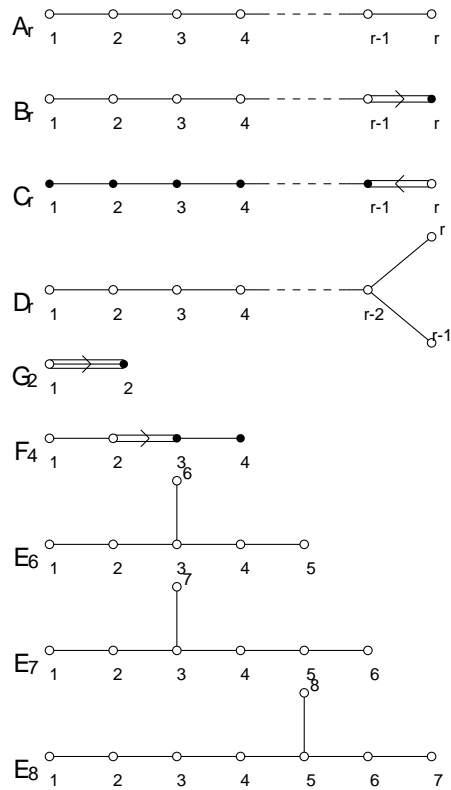
 
Representations are denoted either by their dimensions (if no confusion is 
possible), or by their Dynkin labels $(a_1,\ldots,a_r)$, where $r$ is the 
rank. Our labelling convention is indicated in fig \ref{fig:dynkin}.
A bar denotes the complex conjugate representation; ``$A$" denotes the 
adjoint representation, ``$R$" a generic representation and ``$r$" the 
reference representation to be defined later. The dimensions of these
representations are denoted by $N_A$, $N_R$ or $N_r$; the quadratic
Casimir eigenvalues (defined more precisely below) by $C_A$, $C_R$ and
$C_r$ respectively.

The generators of the adjoint representation are related to the
structure constants
\begin{equation}
\label{eq:tamatrix}
	(T_A)^a_{bc} = - i\ f^{abc}\ .
\end{equation}

\subsection{Invariant Tensors}

We will encounter traces
\begin{equation}
   \Tr T_R^{a_1}\ldots T_R^{a_n}
\end{equation}
in any representation $R$ of any simple Lie algebra.
We wish to express the result in the 
minimal number of quantities.

Every trace defines an invariant tensor $M$:
\begin{equation}
   \Tr T_R^{a_1}\ldots T_R^{a_n} = M_R^{a_1\dots a_n}\ .
\end{equation}
This tensor is invariant because the trace is invariant under the 
replacement
\begin{equation}
    T_R^a \to U_R T_R^a U_R^{-1} = U_A^{ab} T_R^b\ ,
\end{equation}
where $U_R$ is an element of the group in the representation R; $U_A$ is 
the same group element in the adjoint representation. Hence we have
\begin{equation}
   M_R^{a_1\ldots a_n}=U^{a_1b_1}_A\ldots U^{a_nb_n}_A
   M_R^{b_1\ldots b_n}\ ,
\end{equation}
which implies, in infinitesimal form
\begin{equation}
\label{eq:invdef}
   \sum_i f^{cba_i} M_R^{a_1\ldots b\ldots a_n} = 0\ ,
\end{equation} with $b$ inserted at position $i$. This ``generalized 
Jacobi-identity" may be taken as the definition of an invariant tensor.

\subsection{Casimir Operators}

Every invariant tensor $M$ defines a Casimir operator $C(M)$
\begin{equation}
   C_R(M) = \sum_{a_1,\ldots, a_n} T_R^{a_1}\ldots T_R^{a_n} 
      M^{a_1\ldots a_n}\ . 
\end{equation}
It follows from equation (\ref{eq:invdef}) that $C_R(M)$ commutes with all generators 
$T_R$ in the representation $R$. If $R$ is irreducible Schur's lemma 
implies that $C_R(M)$ is constant on the representation space $R$. 
Note that this is true independent of the symmetry of $M$, and irrespective 
of any concrete definition of $M$ in terms of traces. All that is used is 
the Jacobi-identity (equation (\ref{eq:invdef})).

Of special interest are the quadratic Casimir operators 
$C_R$, which we define as
\begin{equation}
	(T_R^a T_R^a)_{ij} = C_R\delta_{ij}\ .
\end{equation}
As a special case of this 
identity we can write, using equation (\ref{eq:tamatrix})
\begin{equation}
\label{eq:twof}
	f^{acd}f^{bcd} = C_A\delta^{ab} \ .
\end{equation}

\subsection{Symmetrized traces}

Not all invariant tensors and Casimir operators constructed so far are 
independent. We would like to express all traces in terms of a minimal set of 
invariant tensors. As a first step one may use the commutation relations to 
express the trace in a completely symmetric trace plus terms of lower order 
in the generators, which in their turn can also be expressed in terms of 
symmetrized traces. 
An efficient algorithm for doing this will be discussed in the next section.
After this step we only need to consider symmetrized traces
\begin{equation}
   \Str T^{a_1}\ldots T^{a_n} \equiv {1 \over n!} \sum_{\pi} 
    \Tr T^{a_{\pi(1)}}\ldots  T^{a_{\pi(n)}}\ , 
\end{equation}
where the sum is over all permutations of the indices (the cyclic 
permutation may of course be factored out using the cyclic property of the 
trace).

For each representation one may define a symmetric invariant tensor $d_R$ 
with
\begin{equation}
    d_R^{a_1\ldots a_n}\equiv \Str T_R^{a_1}\ldots  T_R^{a_n} \ , 
\end{equation}
but this still vastly overparametrizes the problem, because a new tensor is 
defined for every order $n$ and for every representation $R$.

\subsection{Basic Casimir invariants}

It is well-known that the number of independent symmetric invariant tensors 
is equal to the rank of the algebra. This can be seen as follows. For each 
invariant symmetric tensor $d$ of order $n$ define a polynomial
\begin{equation}
\label{eq:polydef}
   P_d(F)=F^{a_1} \ldots F^{a_n} d^{a_1\dots a_n} \ ,
\end{equation}
where $F^a$ is a real vector of dimension equal to the dimension of the 
algebra, $N_A$. The tensor $d$ can be derived from $P_d(F)$ by 
differentiating with respect to $F$:
\begin{equation}
\label{eq:poldef}
   d^{a_1\ldots a_n} = {1\over n!} {\partial \over \partial F^{a_1}}\ldots
   {\partial \over \partial F^{a_n}} P_d(F) \ .
\end{equation}
Although {\it a priori} $P_d(F)$ is a polynomial in $N_A$ variables, 
the fact that $d$ is an invariant tensor implies that $P_d$ depends in fact 
only on $r$ variables, where $r$ is the rank of the algebra. This is true 
because the polynomial is invariant under
\begin{equation}
   F^a \to U_A^{ac} F^c \ ,
\end{equation}
and it is well-known that for 
every $F^a$ one can find a transformation $U_A$ that rotates $F^a$ into the 
Cartan subalgebra. Hence $P_d(F)$ depends only on as many parameters as the 
dimension of the Cartan subalgebra, \ie\ $r$. Therefore it is not 
surprising that any such polynomial can be expressed in terms of $r$ basic 
ones, although the precise details (\eg\ the orders of the basic 
polynomials) don't follow from this simple argument.

The orders of the basic polynomials for each group are known 
\cite{Racah,Gelfand}, and are given 
in the following table (for future purposes this table also gives the 
``dual Coxeter number" $g$).
\begin{table}[htb]
\centering
\begin{tabular}{|c|c|l|}
\hline
Algebra & $g$ & Invariant tensor ranks \\ \hline
$A_r$   & $r+1$  &$2, 3, 4,\ldots,r,r+1$ \\
$B_r$   & $2r-1$   & $2, 4, 6 ,\ldots ,2r$ \\
$C_r$   & $r+1$  & $2, 4, 6 ,\ldots ,2r$  \\
$D_r$  & $2r-2$   & $2,4,6,\ldots,2r-2; r$ \\
$G_2$  & 4  & $2,6$ \\
$F_4$  & 9  & $2,6,8,12$ \\
$E_6$  & 12  & $2,5,6,8,9,12$\\
$E_7$   & 18  & $2,6,8,10,12,14,18$\\
$E_8$   & 30 & $2,8,12,14,18,20,24,30$  \\ \hline
\end{tabular}
\caption{\label{tab:itr} \sl Ranks of basic invariant tensors}
\end{table}
As explained above, each basic polynomial corresponds to an invariant 
tensor, which in its turn corresponds to a Casimir invariant. The 
implication of table \ref{tab:itr} is that for any given algebra the 
polynomials
\begin{equation}
   \Tr F^n_R \equiv \Tr ( \sum_a  F^a T^a_R )^n 
\end{equation}
can be expressed in terms of $r$ basic polynomials of degrees as indicated 
above.  
In section 5 
we will show how to obtain such expressions for any irreducible 
representation (irrep) of any (semi)-simple Lie algebra. The ranks of the 
invariant tensors -- or more accurately those\footnote{We are not aware
of others, but also not of a proof that they
do not exist.}
 that can be written as traces 
over some representation $R$ -- are in fact an outcome of these 
calculations.

If we can express a polynomial corresponding to some invariant tensor $d$ 
in terms of basic polynomials, we can also express the invariant tensors 
into basic ones. Namely, suppose
\begin{equation}
   P_d(F)=\sum \prod_i P_{d_i}(F) 
\end{equation}
where the sum is over various terms of this type, with coefficients. Then 
the differentiation (eq.\ref{eq:poldef}) of a term on the right hand side yields 
precisely the fully symmetrized combination of the tensors $d_i$, with 
weight 1; this means that the overall combinatorial factor equals the 
number of terms. For example
\begin{equation}
\label{eq:polex}
 {1 \over 4!} {\partial \over \partial F^a}{\partial \over \partial F^b}
{\partial \over \partial F^c}{\partial \over \partial F^d} (\sum_e (F^e)^2)^2 =
\frac{1}{3} (\delta^{ab}\delta^{cd} +\delta^{ac}\delta^{bd}+ \delta^{ad}\delta^{bc}) \ . 
\end{equation}

In practice we will express all higher traces in terms of $r$ basic ones, 
but we do not obtain the full dependence on $F$ of the basic traces, and 
consequently we cannot say anything about the explicit form of the basic 
invariant tensors. Given an explicit basis for the Lie-algebra one may 
compute the full $F$-dependence, but that is the same as computing the 
invariant tensor directly by computing a trace.

\subsection{Indices}

Since a Casimir operator is constant on a irrep, its value can be computed 
by taking the symmetrized trace over this irrep. We will see explicitly how 
to expand a trace in terms of fundamental symmetric invariant tensors. In 
general one has
\begin{equation}
\label{eq:symtr}
   \Str T_R^{a_1} \ldots T_R^{a_n} = I_n(R) d^{a_1\ldots a_n} 
   + \hbox{products of lower orders} \ .
\end{equation}
The invariant tensors can be chosen in some representation-independent  
way, for example by computing it for one given reference representation. 
Then all symmetrized traces can be expressed in terms of this basis of 
tensors. The leading term necessarily has the indicated form, with a 
computable representation dependent coefficient $I_n(R)$. This coefficient 
is called the $n^{\rm th}$ index of the representation. If there is no 
fundamental invariant tensor of order $n$ the indices $I_n(R)$ are 
obviously zero for any representation\rlap.\footnote{For $D_r$, $r$ even, there 
are {\it two} indices of order $r$. The additional one will be denoted as 
$\tilde I_r$. We will deal with this case in more detail below.}

The extra terms in equation (\ref{eq:symtr}) are symmetrized products of lower order 
tensors such that the total order is $n$, without contracted indices. 
The coefficients of these terms will be called {\it sub-indices}.

There is a lot of freedom in defining $d^{a_1\ldots a_n}$ since 
we could have modified it by any combination of lower order terms in 
equation (\ref{eq:symtr}). Note that modifying the tensors by lower
order terms does not affect 
the indices, but does change the sub-indices.
This freedom can be used to impose the 
conditions \cite{patera,Azcarr}
\begin{equation}
\label{eq:ortho}
   d_{\ortho}^{a_1\ldots a_l\ldots a_n}
d_{\ortho}^{a_1\ldots a_l} = 0  \quad l < n
\end{equation}
This then defines the symmetrized tensors up to an overall normalization. 
The 
normalization can be fixed by fixing a normalization for the indices. This
basis will be referred to as the {\it orthogonal basis}. It is the
most elegant one from a mathematical point of view, but, as we will see,
not the most convenient one for our purposes. In the following we will
use the notation $d_{\ortho}^{a_1\ldots a_l\ldots a_n}$ for tensors in 
the orthogonal basis.

As mentioned before, tensors defined in any basis can be used to
define
Casimir invariants, but using 
the orthogonal basis has a clear advantage because it leads to a 
simple relation with the 
indices ($N_R$ is the dimension of $R$) :  
\begin{equation}
\label{eq:indexcas}
   I_n(R)= { N_R \over {\cal N}_n} C_p(R) \ , 
\end{equation}
where 
\begin{equation}
\label{eq:normN}
   {\cal N}_n = d_{\ortho}^{a_1\ldots a_n}d_{\ortho}^{a_1\ldots a_n}\ . 
\end{equation}
This relation holds because when contracting with an orthogonal tensor
only the leading terms survive. Note that 
this is true even
if we do not expand equation (\ref{eq:symtr}) in terms of the 
orthogonal basis, but in terms of any other basis.
Hence the Casimir eigenvalues are determined up to a representation 
independent factor once the indices are known. 

The indices are of interest in their own right, 
as was in particular emphasized in \cite{patera}.
In some cases they have a 
topological interpretation via index theorems. Furthermore they satisfy a 
useful tensor product sum rule. If
\begin{equation}
    R_1 \otimes R_1 = \sum_i \oplus R_i 
\end{equation}
then 
\begin{equation}
    N_{R_1} I_p (R_2)+ N_{R_2} I_p (R_1) = \sum_i  I_p(R_i) \ .
\end{equation}
In subgroup embeddings $H \subset G$ there is also such a sum rule for 
branching rules: if $R \rightarrow \sum_i \oplus r_i $ then
\begin{equation}
   I^G_p(R) = I_p(G/H) \sum_i I^H_p(r_i)\ , 
\end{equation}
where $I_p(G/H)$ is the embedding index (here we assume that both $G$
and $H$ have precisely one index of order $p$; other cases require just
slightly more discussion).  

\subsection{Normalization}

To arrive at a universal normalization we make use of the following
general formula for the quadratic Casimir invariant
\begin{equation}
\label{eq:CasNorm}
 C_R = {\eta\over 2} \sum_{i=1}^r \sum_{j=1}^r (a_i+2) G_{ij} a_j   
\end{equation}
Here $a_i$ are the Dynkin labels of the representation $R$,
and $G_{ij}$ is the inverse Cartan matrix.
 The advantage of this
formula is that the Dynkin labels as well as the Cartan matrix have
a fixed normalization that is not subject to conventions. 
Only the overall normalization is convention dependent.  
The factor
$\eta$ is introduced to allow the reader to fix the normalization
according to taste. The dependence on $\eta$ will be shown
explicitly in all formulas.
Given the universality of
(\ref{eq:CasNorm}) it is natural to choose $\eta$ in a group-independent
way. Using (\ref{eq:CasNorm})
get for $C_A$:
$$ C_A=\eta g\ , $$
where $g$ is the dual Coxeter number. 

This convention defines the normalization of the generators once we have
fixed the rank 2 symmetric tensor. The natural definition is
$$ d^{ab}=d^{ab}_{\ortho} = \delta^{ab}. $$
Then ${\cal N}_2 \equiv d^{ab}_{\ortho}d^{ab}_{\ortho}=N_A$. 
Now the second index $I_2(R)$ is also fixed via (\ref{eq:indexcas}):
$$ I_2(R)={N_R\over N_A} C_R\ . $$
For the vector representations $V$ of the classical Lie algebras
we find then $I_2(V)=\half \eta$ for $SU(N)$ and $Sp(N)$, and
$I_2(V)=\eta$ for $SO(N)$. 

There are (at least) two considerations that might lead to a 
choice for $\eta$. First of all it is possible to fix
the conventions in such a way that $I_2(R)$ is always an integer. 
This leads to the choice $\eta=2$. On the other hand there 
are standard choices for the generators of $SU(2)$
namely $T^a=\half \sigma^a$ (where $\sigma^a$ are the Pauli-matrices),
and for $SO(N)$, namely
$T^{\mu\nu}_{ij}=i(\delta^{\mu}_i \delta^{\nu}_j - 
\delta^{\mu}_j\delta^{\nu}_i)$, where the pair
$\mu\nu$ with $\mu < \nu$ represents an adjoint index. Unfortunately these
two choices correspond to different values of $\eta$, namely 
$\eta=1$ for $SU(2)$ and $\eta=2$ for $SO(N)$.

\subsection{Indices versus Casimir invariants}

We conclude this section with a few historical remarks. 

A vast amount of literature exists on the computation and properties 
of Casimir invariants. Most of these papers, [7-19],
 give more 
or less explicit expressions for the Casimir eigenvalues of the
classical Lie Algebras $A_n, B_n, C_n$ and $D_n$ and in one case,
\cite{Okubo}, also for
$G_2$.  
In 
\cite{Bincer,BinRie} formulas for $G_2$ and $F_4$ are obtained, whereas
$E_8$ was considered, up to order 14, in \cite{KarGun}. The issue of 
completeness of a 
set of Casimir operators was studied in \cite{Berdjis,BerBes}.

In applications to Feynman diagrams indices are more important than
Casimir invariants, because traces over matter loops yield indices 
and sub-indices, and not Casimir invariants. Indices have been
discussed most frequently in relation to chiral anomalies.
The second index was
introduced by Dynkin, \cite{Dynkin}, and generalized 
to higher order in \cite{PaShWi}.
Shortly afterwards \cite{patera}
it was realized that the definition of the indices
could be improved by imposing the orthogonality constraint (\ref{eq:ortho}).
In \cite{patera} formulas are given for the indices of classical Lie
algebras. Indices of exceptional algebras have
been studied up to sixth order, mainly for the purpose of 
anomaly cancellation in ten dimensions, relevant for string theory. 
Sub-indices, when defined in the orthogonal basis, can be expressed
in terms of indices. Unfortunately these relations are difficult to obtain,
and become very complicated at higher orders unless some lower indices
vanish. In \cite{patera} formulas for $SU(n)$
have been given up to fifth order. We have computed the sixth order
formula, but the result is rather awkward and does not encourage
extension to higher orders.

Although, as explained above,
indices are closely related to Casimir invariants, the available
formulas for the latter are of little use to us since they do not use
orthogonal tensors for the definition of the Casimirs. 
Even if they did, one would still need the normalization factor ${\cal N}_n$,
(\ref{eq:normN}).
The computation of this factor for all Lie-groups and all
values of $n$ is
a difficult problem, related to the even more difficult problem
of determining the tensors $d_{\ortho}^{a_1\ldots a_n}$ 
explicitly. For recent progress on the latter problem for the classical
Lie algebras see \cite{Azcarr}. 
We do not present 
explicit expressions for the symmetric tensors here.
Since they always appear in contracted form, we never need them
explicitly.

Furthermore the Casimir
eigenvalues give no information on sub-indices.

\section{Reduction to symmetrized traces}

In this section we will discuss the reduction of traces as they occur in 
Feynman diagrams into the invariants of the previous section. This is by no 
means a trivial affair because the necessary symmetrizations make that the 
algorithms typically involve ${\cal O}(n!)$ terms when there are $n$ 
generators in the trace. It is therefore important to choose the method 
carefully. We will have to distinguish two cases. In this
section we will make the reduction of traces of the type $\Str 
T_R^{a_1}\ldots T_R^{a_n}$ in which $R$ can be any representation with the 
exception of 
the adjoint representation. In the next section we will consider such 
traces for the adjoint representation. The special role of the adjoint 
representation lies in the fact that, because of the equation
(\ref{eq:tamatrix}), the commutation relation
\begin{equation}
      [T^a_A, T^b_A] = i\ f^{abc}T^c_A
\end{equation}
does not really diminish the number of generators of the adjoint 
representation. It is actually just a different way of writing the Jacobi 
identity. A related reason for considering the adjoint representation
separately is that the reduction of the other traces generates new 
structure constants.

We will continuously keep in mind that the algorithms we derive are for 
implementation in a symbolic computer program. This means that in many 
cases a recursion type algorithm may suffice, even though it may not be 
very practical for hand calculations.


\subsection{First stage elimination}   
 
The first part of the reduction is dedicated to the replacement of the 
traces over the generators $T_R$ by the invariants $d_R$. For all 
representations except for the adjoint this can be done in a general 
algorithm. One should realize however that for very complicated traces the 
results may not be very short.

In general a trace is not symmetrized. Therefore the introduction of the 
tensors $d_R$ needs some work with commutation relations to make it 
symmetric. On the other hand, computer algebra needs algorithms that work 
from a formula, rather than towards one. Hence one can use the substitution
\begin{eqnarray}
\label{eq:canon}
    {\rm Tr}[T_R^{a_1}\cdots T_R^{a_n}] & = &
    {\rm Tr}[T_R^{a_1}\cdots T_R^{a_n}]
    - \Str T_R^{a_1}\cdots T_R^{a_n}
    + d_R^{a_1\cdots a_n}
\end{eqnarray}
Writing out the symmetrized trace will of course give $n!$ terms, each with 
a factor $1/n!$. Then we can commute the various $T_R^{a_i}T_R^{a_j}$ till 
they are all in the order of the original trace after which the $n!$ terms 
with $n$ generators will cancel the original trace. At this point we are 
left with the symmetric tensor $d_R$ and ${\cal O}(n\ n!)$ terms which all 
have $n-1$ generators. As a recursion it will eventually result in terms 
with only two generators for which we know the trace. This algorithm is 
however rather costly when the number of generators inside the trace is 
large.

The above formula has as its main benefit that it proves that one can 
express a trace of generators T of any representation $R \ne A$
 in terms of symmetrized traces and structure 
constants $f$. For practical purposes we have a better algorithm. It is 
based on the formula:
\begin{eqnarray}
\label{eq:symtrace}
        T_R^{\left\{ a_1\right. }\cdots T_R^{\left. 
         a_n\right\} } T_R^b & = & \sum_{j=0}^{n}\frac{(-1)^j}{j!}
			 B_j E_j^{a_1\cdots a_nb}
\end{eqnarray}
in which $B_j$ is the $j$-th Bernoulli number. 
( $B_0 = 1$, $B_1 = -1/2$, 
$B_2 = 1/6$ and $B_4 = -1/30$) and the function $E_j$ is defined by the 
recursion
\begin{eqnarray}
    E_0^{a_1\cdots a_n} & = &
        T_R^{\left\{ a_1\right. }\cdots T_R^{\left. a_n\right\} } \\
    E_j^{a_1\cdots a_nb} & = & \sum_{i=1}^n
        E_{j-1}^{a_1\cdots a_{i-1}a_{i+1}\cdots a_nk}\ if^{a_ibk}
\end{eqnarray}
Basically in the $E_j$ one extracts a string of $j$ structure constants $f$. 
By writing out the functions $E_j$ one can show with some work that the 
proof of this formula is equivalent to proving the relation
\begin{eqnarray}
R_0 & = & \frac{1}{n+1}B_0\sum_{i=0}^nR_i \nonumber \\ &&
    +\sum_{i=1}^n\frac{n!}{(n+1-i)!}\frac{(-1)^i}{i!}B_i
        \sum_{j=0}^{i-1}(-1)^j\frac{(i-1)!}{j!(i-1-j)!}
            (R_j+(-1)^iR_{n-j})
\end{eqnarray}
This equation should hold for any positive value of $n$ and any choice of 
the $R_i$ (The $R_i$ represent the symmetric combination of $T^{i_1}$ to 
$T^{i_n}$ with a $T^b$ inserted at $i$ places from the right. Hence $R_0$ 
is the left hand side term of equation (\ref{eq:symtrace})). Actually the 
$R_i$ are independent objects and hence we have a set of equations each of 
which is characterized by a value for n and the index $j$ of $R_j$. To 
prove the whole formula we have to prove that all of these equations are 
valid. We use the following approach: One can test their validity for any 
small value of $n$ and all allowed values of $j$ (computer algebra lets one 
check this easily up to $n=100$). Next one takes the special case of only 
$R_0$ not equal to zero and the case of only $R_n$ not equal to zero. These 
are rather easy to prove. Then one takes the case for $R_1$ not equal to 
zero. This is only a little bit more complicated. Finally one can express 
the case for other values of $n$ and $j$ in terms of the equation for $n-1,
j-1$ and $n,j-1$. This then combines into a proper induction proof.
The above formula can give some interesting summations when one selects 
special values for the $R_i$ like $R_i = x^i$ after which the inner sum can 
be done.

An important part in the application of equation (\ref{eq:symtrace}) is how to 
terminate the recursion. We note that due to the cyclic property of traces
\begin{eqnarray}
      {\rm Tr}\left[ T_R^{\left\{ a_1\right. }\cdots T_R^{\left. 
         a_n\right\} } T_R^b \right] & = & d_R^{a_1\cdots a_nb}
\end{eqnarray}
Additionally we can terminate two cases in which there are still two 
generators outside the symmetrization:
\begin{eqnarray}
      {\rm Tr}\left[ T_R^{a_1}T_R^{a_2}T_R^{a_3}\right] & = &
            d_R^{a_1a_2a_3}+\frac{i}{2}f^{a_1a_2a_3}I_2(R) \\
      {\rm Tr}\left[ T_R^{\left\{ a_1\right. }T_R^{\left. a_2\right\} }
             T_R^{a_3}T_R^{a_4} \right] & = &
            d_R^{a_1a_2a_3a_4} +\frac{i}{2}f^{a_3a_4k}d_R^{a_1a_2k}
                \nonumber \\ &&
        +\frac{1}{12}I_2(R)(f^{a_1a_3k}f^{a_2a_4k}+f^{a_1a_4k}f^{a_2a_3k})
\end{eqnarray}
All together this algorithm is far superior. For a trace of 7 
generators $T_R$ all with different indices it is about 50 
times faster than the one based on equation (\ref{eq:canon}).

Just as with traces of $\gamma$-matrices one can make the algorithms much 
faster with a number of supporting tricks for contracted indices. 
This is however not quite as simple as with $\gamma$-matrices. When the 
contracted indices are close to each other, we use
\begin{eqnarray}
    T_R^bT_R^aT_R^aT_R^c & = & C_R T_R^bT_R^c \\
    T_R^aT_R^bT_R^a & = & (C_R-\frac{1}{2}C_A)T_R^b\ . 
\end{eqnarray}
The relation between $I_2(R)$ and $C_R$ is given by 
$N_RC_R = I_2(R)N_A$. 
Additionally we have
\begin{eqnarray}
    T_R^{a_1}T_R^{a_2}T_R^b\cdots T_R^cT_R^{a_1}T_R^{a_2}
    & = &
    T_R^{a_1}T_R^{a_2}T_R^b\cdots T_R^cT_R^{a_2}T_R^{a_1}
    - \frac{1}{2}C_A T_R^{a_1}T_R^b\cdots T_R^cT_R^{a_1}
\end{eqnarray}
And then for contracted indices that are not very close to each other we 
can use:
\begin{eqnarray}
    T^j_R\cdots T^a_RT^b_RT^j_R & = & T^j_R\cdots T^a_RT^j_RT^b_R
            +if^{bjc}T^j_R\cdots T^a_RT^c_R \nonumber \\
        & = & T^j_R\cdots T^a_RT^j_RT^b_R
            +if^{bjc}T^j_R\cdots T^c_RT^a_R
             - f^{bjc}f^{acd}T^j_R\cdots T^d_R
\end{eqnarray}
We commute the $T^c_R$ matrix also towards the $T^j_R$ matrix because
\begin{eqnarray}
\label{eq:TTff}
    if^{bjc}T^j_R T^c_R & = & \frac{i}{2}f^{bjc}(T^j_RT^c_R-T^c_RT^j_R) 
            \nonumber \\
        & = & -\frac{1}{2}f^{bjc}f^{jcd}T^d_R \nonumber \\
        & = & -\frac{1}{2}C_AT^b_R
\end{eqnarray}
Hence we can always eliminate two generators $T_R$ when there is a pair of 
contracted indices. Some terms may however obtain two $f$-matrices in 
exchange. The above equation shows also that the trace of a string of 
generators of which two generators are contracted with the same structure 
constant $f$, will lead to simplifications by commuting the two 
generators towards each other. In this case many of the terms that come 
from a commutator have just one generator fewer than the original term.

By now it should be clear why the adjoint representation cannot be treated 
in exactly the same way. Eliminating two generators at the cost of 
introducing two structure constants $f$ leaves us with exactly the same 
number of generators of the adjoint representation.

We can use a few extra shortcuts for simple cases to avoid the use of 
equation (\ref{eq:symtrace}) in those cases:
\begin{eqnarray}
{\rm Tr}[T_R^{a_1}T_R^{a_2}T_R^{a_3}] & = & d_R^{a_1a_2a_3}
    +\frac{i}{2}I_2(R)f^{a_1a_2a_3} \nonumber \\
{\rm Tr}[T_R^{a_1}T_R^{a_2}T_R^{a_3}T_R^{a_4}] & = & d_R^{a_1a_2a_3a_4}
    +\frac{i}{2}(d_R^{a_1a_4n}f^{a_2a_3n}-d_R^{a_2a_3n}f^{a_1a_4n})
    \nonumber \\ &&
    +\frac{1}{6}I_2(R)(f^{a_1a_4n}f^{a_2a_3n}-f^{a_1a_2n}f^{a_3a_4n})
\end{eqnarray}
To get the last equation into its minimal form we have used the Jacobi 
identities:
\begin{eqnarray}
\label{eq:jacobi}
0 & = & f^{i_1i_2j}f^{i_3i_4j} + f^{i_2i_3j}f^{i_1i_4j}
       +f^{i_3i_1j}f^{i_2i_4j} \nonumber \\
0 & = & d^{i_1i_2j}f^{i_3i_4j} + d^{i_2i_3j}f^{i_1i_4j}
       +d^{i_3i_1j}f^{i_2i_4j}
\end{eqnarray}
It is possible to create similar shortcuts for the higher traces. This 
serves however not much purpose. The majority of cases involves short 
traces and these expressions are rather lengthy.


\section{Reduction of adjoint traces}

At this stage we have only tensors of the type $d_R$ and structure 
constants $f$ left as objects with indices. All these indices are indices in 
the adjoint space and hence all have $N_A$ dimensions. Additionally there can 
be various constants like the second order Casimir's and the second order 
indices $I_2(R)$,
but these do not play a role in the following.

For the other representations the loops that define the trace were rather 
easy to find. For the adjoint representation this is more complicated: 
all three indices of the structure constant $f$ can play a role and 
hence there are more possibilities. The advantage is that very often 
one can find `smaller' loops. If for instance we have a diagram that 
consists of only vertices in the adjoint representation and there are no 
loose ends we have the results of table \ref{tab:girth}
\begin{table}[htb]
\centering
\begin{tabular}{|r|c|c|c|c|c|c|c|c|}
\hline
girth         & 2 & 3 & 4 &  5 &  6 &  7 &  8 &  9 \\ \hline
$n_{\rm min}$ & 2 & 4 & 6 & 10 & 14 & 24 & 30 & 54 \\ \hline
\end{tabular}
\caption{\label{tab:girth}\sl Minimum number of vertices needed for a 
diagram with a given smallest loop.}
\end{table}
in which `girth' is the size of the smallest loop in the diagram and 
$n_{\rm min}$ is the minimal number of vertices needed to construct such a 
diagram. Of course in mixed diagrams in which other 
representations are involved one can get loops in the adjoint 
representation that have up to $n/2$ vertices if $n$ is the total number of 
vertices in the diagram. Such would be the case of there is one loop of 
$n/2$ vertices in a representation R and a parallel loop of adjoint 
vertices. But in that case the loop is easy to find, and the symmetry of 
the invariant $d_R$ that is present already makes the introduction of the 
$d_A$ into a triviality: due to its contraction with $d_R$ the trace over 
the adjoint generators has already been symmetrized.

Let us first have a look at the canonical reduction algorithm of equation 
(\ref{eq:canon}). At first one might be worried that it will not terminate 
for the adjoint representation. After all it does not diminish the number 
of structure constants $f$. One can see quickly however that the commutator 
terms have a simpler 
loop structure. Hence each term will end up having a number of 
$f$'s grouped into an invariant or have a loop with fewer $f$'s even though 
the total number of $f$'s is still the same. Once we have a loop with at 
most three structure constants it can be reduced with the equation
\begin{equation}
\label{eq:threef}
    f^{i_1i_2a_1}f^{i_2i_3a_2}f^{i_3i_1a_3} = \frac{1}{2}C_Af^{a_1a_2a_3}
\end{equation}
which can be derived from the Jacobi identity. 
Hence in all cases the number of structure constants $f$ will become 
less. Similarly the algorithm of equation (\ref{eq:symtrace}) will reduce a 
number of generators of the adjoint representation to a symmetrized trace 
and a smaller number of structure constants $f$. 
Hence also this algorithm will terminate.

It is however possible to be more efficient about the reductions inside the 
adjoint representation. Our first observation is that for loops with an odd 
number of vertices in the adjoint representation reverting the order of the 
vertices gives a minus sign. Hence the fully symmetric object with an odd 
number of indices must be zero. And because of this we do not need a full 
symmetrization to express loops with an odd number of indices. For 
example for five indices we write
\begin{eqnarray}
    F^{i_1i_2i_3i_4i_5} & = & -F^{i_5i_4i_3i_2i_1}
         \nonumber \\ & = &
        -F^{i_1i_5i_4i_3i_2}
         \nonumber \\ & = &
        -F^{i_1i_2i_3i_4i_5} + F^{ki_4i_3i_2}f^{i_1i_5k}
         \nonumber \\ & &
        + F^{i_1i_4ki_5}f^{i_3i_2k}
        + F^{i_1i_3ki_5}f^{i_4i_2k}
        + F^{i_1i_2ki_5}f^{i_4i_3k}
\end{eqnarray}
in which $F$ represents the trace over a number of $f$'s. Its difference 
with a trace over the $T_A$-generators (\ref{eq:tamatrix}) is just powers of $i$.
Similarly one can use the reversal symmetry for traces of an even number of 
$f$'s to reduce the amount of work by a factor two.

Because of the above simplifications the efficiency of the two algorithms 
((\ref{eq:canon}) and (\ref{eq:symtrace})) is not very different for the 
adjoint representation. 

The reduction algorithm should be clear now. One looks for the smallest 
loop among the various $f$'s. Such a loop will not have contracted indices, 
because otherwise there would be a smaller loop. Hence we do not have to 
worry about contracted indices as we had to do for the other 
representations. If the loop has only two or three $f$'s, we can eliminate 
it with either equation (\ref{eq:twof}) or equation (\ref{eq:threef}). Otherwise 
we can use a simplified version of the canonical reduction 
algorithm of equation (\ref{eq:canon}) to obtain an invariant and terms with 
a simpler loop. 
Actually the fastest way here is to tabulate this reduction all the way up 
to loops with 7 $f$'s. For loops of 8 or more vertices in the adjoint 
representation we use an adapted version of equation (\ref{eq:symtrace}).


\section{Computation of symmetrized traces}

At this point our group theory factors consist of combinations
of structure constants, symmetrized traces $d_A^{a_1,\ldots,a_n}$ over
the adjoint representation, and symmetrized traces over one or more
other irreducible representations. We will now show how such traces
can be expressed in terms of $r$ traces over a single representation,
where $r$ is the rank of the algebra.

As explained in section 2, 
in principle there are three quantities one might be interested
in: Casimir invariants, indices and symmetric tensors. 

The results presented here amount to a computation of the coefficients
of combinations of
fundamental traces appearing in the expansion of a trace in
an arbitrary representation. In other words, we compute indices
and sub-indices (but, as explained
in section 2, the latter are basis-choice dependent).  

With our method these quantities can rather straightforwardly
be computed to any desired order,
and for any representation of any Lie algebra.
To demonstrate this we will 
compute all the indices for the lowest-dimensional representations
of the exceptional algebras, including the $30^{\rm th}$ index
of $E_8$. 

The method we follow here is an extension of results of \cite{ScWd} (where
it was used to obtain the "elliptic genus" in string theory), which
in its turn was an extension of results presented in   
\cite{patera} (where it was used for computing the
indices of the classical algebras).  

\subsection{Characters}

An extremely useful tool for computing traces are the characters 
\begin{equation}
   \Ch_R(F)=\Tr e^{F_R} 
\end{equation}
where $F_R=F^a T_R^a$. 
Hence the expansion of the exponential gives us all symmetrized traces
in terms of the polynomials defined in equation (\ref{eq:polydef}).
What makes the characters especially useful is their 
tensor property
\begin{equation}
\label{eq:tensor}
   \Ch_{{R_1}\otimes {R_1}}(F) = \Ch_{R_1}(F) \Ch_{R_2}(F)\ ,
\end{equation}
which follows directly from its definition. In addition characters
are combinations of traces and therefore also have nice properties on
direct sums
\begin{equation}
\label{eq:decomp}
 \Ch_{{R_1}\oplus {R_1}}(F) = \Ch_{R_1}(F) + \Ch_{R_2}(F)\ .
\end{equation}
With a little more effort one can also derive a formula for 
characters of symmetrized and anti-symmetrized tensor products \cite{Weyl}. 
These formulas can be derived from the following generating functions
\begin{equation}
\label{eq:asym}
 \sum_{k=0}^{\infty} x^k \Ch_{{[k]*R}}(F)=\det(1+x e^F_R) = \prod_{l=1}^{\infty}
\exp(-(-x)^{l}\Ch_R(lF))
\end{equation}
\begin{equation}
\label{eq:sym}
 \sum_{k=0}^{\infty} x^k \Ch_{{(k)*R}}(F)=\det(1-x e^F_R)^{-1} = \prod_{l=1}^{\infty}
\exp((x)^{l}\Ch_R(lF))\ .
\end{equation}
Here $[k]$ denotes the order $k$ anti-symmetric tensor product of some
representation $R$, and $(k)$ the order $k$ symmetric product. 
We use the notation $[k]*R$ or $(k)*R$ to denote the anti-symmetrized
or symmetrized tensor product of the representation $R$. 
Note that the sum in equation (\ref{eq:asym}) is in fact always finite.

The generating functions can be expanded explicitly to obtain
\begin{equation}
\label{eq:asymB}
 \Ch_{[k]*R}(F)= -\sum_{{\{n_i,m_i\} \atop k=n_im_i}} \prod_i {1\over m_i !} 
\left( - {\Ch_R(n_i F)\over n_i}\right)^{m_i}
\end{equation} 
\begin{equation}
\label{eq:symB}
 \Ch_{(k)*R}=\phantom{-}  \sum_{{n_i,m_i \atop k=n_im_i}}
 \prod_i {1\over m_i !} 
\left(   {\Ch_R(n_i F)\over n_i}\right)^{m_i}\ ,
\end{equation} 
where the sum is over all partitions of the integer $k$ into
different integers $n_i$, each appearing with multiplicity $m_i$.  

\subsection{Character computation method}

Our method for computing the characters is as follows. We begin 
by choosing a reference representation, which in all cases is the
one of smallest dimension. The reference representations 
we choose for the simple Lie algebras are shown in
table \ref{tab:refrep}.
(the last column of this table is explained later)
\begin{table}[htb]
\centering
\begin{tabular}{|c|c|c|c|}
\hline
Algebra & Reference representation & Dimension & Indices \\ \hline
$A_r$   & $(1,0,\ldots,0)$           & $r+1$   & $1,\ldots,1$    \\
$B_1$   & (1)                      & 2  &  1       \\
$B_2$   & (0,1)                    & 4  &  1,1    \\
$B_3$   & (1,0,0)                    & 7  &  2,2,1    \\
$B_4$   & (1,0,0,0)                    & 9  &  2,1,1,2    \\
$B_r, \ \ r \geq 5$ & $(1,0,\ldots,0)$ & $2r+1$ & $2,1,\ldots,1$ \\
$C_r$   & $(1,0,\ldots,0)$         & $2r$  & $1,\ldots,1$ \\
$D_3$   & $(0,0,1)$                & 4  & 1,1,1     \\
$D_4$   & $(1,0,0,0)$                & 8  & 2,2,1,0   \\
$D_5$   & $(1,0,0,0,0)$                & 10  & 2,1,1,2,0    \\
$D_r,  \ \ r \geq 3$ & $(1,0,\ldots,0)$ & $2r$ & $2,1,\ldots,1,0$ \\
$G_2$   & $(0,1)$                   & 7 & 2,1 \\
$F_4$   & $(0,0,0,1)$               & 26 & 6,1,1,1 \\
$E_6$   & $(1,0,0,0,0,0)$               & 27 & 6,1,1,1,1,1 \\
$E_7$   & $(0,0,0,0,0,1,0)$               & 56 & 12,1,1,1,1,29,1229 \\
$E_8$   & $(1,0,0,0,0,0,0,0)$               & 248 & 60,1,1,1,1,41,199,61
		\\ \hline
\end{tabular}
\caption{\label{tab:refrep}\sl Reference representations, dimensions and 
indices.}
\end{table}
Note that for $SO(N)$ the reference representation is the
vector representation for $N \geq 7$, but for lower values of $N$
it is a spinor representation. Another way of saying this is that
we treat $SO(N), N \leq 6$ according to 
the Lie-algebra 
isomorphisms $D_3 \sim A_3$, $B_2 \sim C_2$ and $B_1 \sim A_1$.
The last column of table \ref{tab:refrep} is discussed below. The algebras 
$B_3,B_4,D_4$ and $D_5$ are listed separately because, although they
have the ``standard" reference representation, they have non-standard
index normalizations.

For the reference representation the character is left in the form
\begin{equation}
\label{eq:FundTrace}
 \Ch(F)=\Tr e^F \ ,
\end{equation}
All traces whose order does
not appear in table \ref{tab:itr} can be expressed in terms of lower 
traces. Hence equation (\ref{eq:FundTrace}) must be supplemented by trace 
identities for those traces. These trace identities will be derived below. 
The remaining traces will be called ``fundamental" and equation 
(\ref{eq:FundTrace}) is taken to be the definition of the corresponding 
polynomials and symmetric tensors. This then defines a set of reference 
tensors:
\begin{equation}
 d^{a_1\ldots,a_n}_r = \Str T^{a_1}_r \ldots T^{a_n}_r 
\end{equation}
where $n$ is the order of a fundamental Casimir operator. The precise
form of this tensor, or equivalently the precise form of the fundamental
polynomials $\Tr F^n$ depends on the details of the Lie algebra basis
choice, but will never be needed. 

Any other character is now written in terms of traces of $F^n$ over
the reference representation, using all available trace identities. 
By differentiating with respect 
to $F$ ({\it c.f.} eq. (\ref{eq:polex})) one can then read off the
expression of any $d_R$ in terms of reference tensors. 

This fails if the reference representation has an index
that is zero. This happens only for the $n^{\rm th}$ index of the
algebra $D_{n}$, and we will deal with that case separately. 

The next step is to express the characters of all ``basic"
representations in terms of the reference character. The $i^{\rm th}$
basic representation is defined by Dynkin labels $a_j=\delta_{ij}$,
$j=1,\ldots,r$. The most important tool for obtaining
these characters is equation (\ref{eq:asymB}). This yields all fundamental 
representations of the algebras $A_r$ and $C_r$, whereas for the orthogonal 
groups only the spinor representations are still missing. The spinors, as
well as the basic representations of the exceptional algebras,
require some extra work, and are discussed below. 

Finally one can compute the characters of all other representations
by using in a systematic way the sum rule for tensor products. 
It can be proved that for any simple Lie-algebra
this allows one to relate
the characters of all other irreps linearly to those
of the basic ones. In principle this still allows for the
possibility that complicated linear 
equations need to be solved. We find however, that
one can organize the tensor products in such a way that only one unknown
character appears at every step. This can be proved for the classical
Lie algebras (see below), and we have checked it empirically for the 
exceptional ones.

Let us contrast this procedure with the computation via Weyl's
character formula
\begin{equation}
\label{eq:WeylForm}
 \Ch_{\Lambda}(h)={\sum_{w \in W} \epsilon_w \exp(w(\Lambda+\rho),h)  
 \over \sum_{w \in W} \epsilon_w \exp(w(\rho),h)}
\end{equation}
where $\Lambda$ is the highest weight of a representation, the summation
is over all elements $w$ in the Weyl group $W$, $\epsilon_w$
is the determinant of w, $\rho$ is the 
Weyl vector  (with Dynkin labels all equal to 1), and $h$ is a 
vector in weight space, which plays the r\^ole of $F$ in the 
foregoing discussion.
One obvious disadvantage of this formula
is the summation over all elements of the Weyl group, although this is
still manageable in most cases of interest. A less obvious disadvantage
is that numerator and denominator both have a zero of order $N_+$, the
number of positive roots, in $h$. For example, to obtain the highest
non-trivial Casimir eigenvalue of $E_8$, which is of order 30, one
needs to expand numerator and denominator to order $N_++30=150$. This
is an impossible task. The method sketched above, and worked out below,
{\it does} allow an expansion of the character to order 30, even for $E_8$.

An important ingredient in our procedure is obviously the computation
of tensor products. Conceptually this is certainly not easier than the 
computation of characters, but nowadays computer programs exist that
can do this very efficiently\rlap.\footnote{We have used the programs
{\tt LiE}~\cite{lie}\ that computes tensor products 
of Lie-algebra representations
directly and {\tt Kac}~\cite{Kac} that
uses the Verlinde formula to compute fusion rules of Kac-Moody
algebras. A subset of these fusion rules coincides with tensor
product rules, and it turns out that this precisely includes the
tensor product rules we need.}
Rather than using characters (and in
particular index sum rules) to compute tensor products, it is then
more efficient to use tensor products to compute characters. The 
procedure described here requires just a small effort to compute
the characters of the basic representations up to a certain desired
order. The computation of the character of any other representation
is then just a matter of simply polynomial operations (multiplications,
additions and subtractions
which can be
efficiently performed by any symbolic manipulation program, such as
{\tt FORM}) guided by the output of a program that computes tensor
products\rlap.\footnote{We have implemented this idea in the program
{\tt Kac}. The results in the appendix were produced in that way.
At a given Kac-Moody level, 
the fusion rules that coincide with tensor products are found to be
sufficient to obtain the
characters of all representations at that level.
For examples and software see 
http://norma.nikhef.nl/$\sim$t58 } 

We will now discuss the various types of algebras in more detail.

\subsection{$A_r$ characters}

Let us now apply these tools first of all to Lie algebras of type $A_r$ 
($SU(r+1)$).  For the reference representation we choose the vector 
representation $(r+1)$. Using equation (\ref{eq:asym}) we can then 
immediately write down the characters for all the anti-symmetric tensor 
product representations $[k]$. In terms of Dynkin labels these are all the 
representations with labels $(0,\ldots,0,1,0,\ldots,0)$, \ie\ a single 
entry 1. These are precisely the basic representations.

Now we can systematically use the tensor product rule (\ref{eq:tensor}) and
the sum rule (\ref{eq:decomp}) to obtain character formulas  for all other 
irreducible representations. If
\begin{equation}
 R_1 \otimes R_2 = \sum_i \oplus n_i R_i  
\end{equation} 
then 
\begin{equation}
\label{eq:sumru}
 \Ch_{R_1}(F)\Ch_{R_2}(F) = \sum_i n_i \Ch_{R_i}(F)
\end{equation} 
By computing the product of two known
characters and subtracting the known characters on the right hand side
one is left with the character of some (in general reducible)
representation, which is thereby determined. 

To show how this works we label
the $SU(r+1)$ irreps by Young tableaux and assign a partial ordering to them.
We use Young tableaux because for $A_r$ they provide a convenient
description of the tensor product rule. 
A Young tableau is ordered above another one if it has more columns; if
the number of columns is the same the one with the largest last column
is ordered above the other one. Suppose now that we know the characters
of all representations ordered below a representation $R$ with
Young tableau $[k_1,\ldots,k_l]$. Consider then the tensor product
$[k_1,\ldots,k_{l-1}] \otimes [k_l]$. Both are ordered below $R$ and hence
their characters are known according to our assumption.  
The tensor product
 yields $[k_1,\ldots,k_l]$ plus representations ordered below 
$[k_1,\ldots,k_l]$, and hence we can now determine the character of $R$.
Proceeding like this we can systematically compute all characters. 

Not only the characters, but
also the trace identities for $A_r$ were obtained in \cite{ScWd}
\begin{equation}
\label{eq:TraceID}
 \sum_{{\{n_i,m_i\} \atop k=n_im_i}} \prod_i {1\over m_i !} 
\left( - {\Tr F^{n_i}\over n_i}\right)^{m_i}=0 \ \ \   (k > r+1) \ ,
\end{equation} 
where the summation is as in equation (\ref{eq:asymB}).
This result was obtained from equation (\ref{eq:asymB}) using the fact that 
for $A_r$ anti-symmetric tensors of rank larger than the rank of the algebra 
are trivial. 

To illustrate this let us return to the Weyl formula, 
equation(\ref{eq:WeylForm}).
For $A_1$ this yields a very simple result for a representation
of spin $j$:
\begin{equation}
\label{eq:Weyl}
 \Ch_{j}(h)={\sinh((2j+1)h) \over \sinh(h)}
\end{equation}
Expanding this for the spin-$\half$ representation $(j=\half)$ we get
\begin{equation}
 \Ch_{\half}(h)=2 + h^2 + \frac{1}{12} h^4 + \frac{1}{360}h^6 + \ldots 
\end{equation}
The spin-$\half$ representation serves as the reference representation
in our method. Hence its character is 
\begin{equation}
\label{eq:CharAOone}
\Tr e^F=2+\frac{1}{2} (TrF^2)
+ \frac{1}{24} (TrF^4)+\frac{1}{720} (TrF^6)+ \ldots 
\end{equation} 
Using the $SU(2)$ trace identities 
(\ref{eq:TraceID}) $\Tr F^4 = \half (Tr F^2)^2$ and 
\begin{eqnarray}
	\Tr F^6 & = & 6(- \frac{1}{48}   ( (Tr F^2)^3 ) 
             + \frac{1}{8} (Tr F^2)  (Tr F^4) \nonumber \\
            & = & \frac{1}{4} (Tr F^2)^3
\end{eqnarray}
we arrive at the answer  
\begin{equation}
\label{eq:CharAO}
\Tr e^F=2+\frac{1}{2} (2 h^2)
+ \frac{1}{24} (2 h^4)+\frac{1}{720} (2 h^6)+ \ldots 
\end{equation} 
were we substituted
$F=h \sigma^3$, so that $\Tr F^2= 2 h^2$ (Obviously we could have
substituted the diagonal form of $F$ directly in equation 
(\ref{eq:CharAOone}), but the
use of trace identities is far more convenient for larger algebras). 

Clearly the equations (\ref{eq:Weyl}) and (\ref{eq:CharAO}) agree, as 
expected. However, the way the agreement comes out is not entirely trivial 
(although 
it can easily be derived).  Note in particular that
the Weyl formula is {\it a priori} expressed in terms of only
$r$ variables, so that all trace identities are already built in. 
On the other hand, in writing down the formal expression $\Tr e^F$ there
is no need to specify the number of variables, and indeed the 
formula is the same for any algebra. The non-trivial group
structure is thus encapsulated in the trace identities. It is instructive 
to compare the two formulations also for other representations.

In this case the Weyl formula is superior in elegance and simplicity,
although it is somewhat more difficult to expand
to higher orders due to its
denominator. For higher rank groups the Weyl formula becomes extremely
cumbersome, as explained earlier, while our method does not grow
in complexity. 

\subsection{$B_r$ characters}

The basic representations are the anti-symmetric tensors
of rank $1\ldots,r-1$ plus the spinor representation. The characters
of the anti-symmetric tensors are related to the vector character as in
the case of $A_r$. The spinor character can be expressed in terms of
traces of the vector representation by explicit computation. The
result is \cite{ScWd}
\begin{equation}
\label{eq:BspinChar}
 \Ch_{(0,\ldots,0,1)}(F) = 2^r \exp\left[\sum_{n=1}^{\infty}
{(2^{2n}-1)B_{2n} \over 4 n (2n)! } \Tr F^{2n} \right] \ ,
\end{equation}
where $B_{2n}$ are the Bernoulli numbers.

For algebras
of type $B_r$ the same trace identity as for $A_r$ holds, but with 
order $k > 2r+1$. This is true because of the embedding $B_r \subset A_{2r}$.
All traces of odd order vanish trivially. 

The demonstration that the other characters can be obtained
recursively from the tensor product rule is similar as for $A_r$,
with some complications due to the spinors. We will omit the
details (and the same holds for $C_r$ and $D_r$).

\subsection{$C_r$ characters}

The fundamental representations are the anti-symmetric tensors
of rank $l=1\ldots,r$ with a symplectic trace removed. The character 
of the fundamental representation $l$ is equal to the $l^{\rm th}$
anti-symmetric tensor power of the vector character minus
the $(l-2)^{\rm th}$ anti-symmetric power of the vector character 
(if $l \geq 2$). 

The $C_r$ trace identities can be derived using the embedding
$C_r \subset A_{2r-1}$, which leads to trace identities for traces
of order $k > 2r$. Just as for $B_r$, the odd traces vanish. 

\subsection{$D_r$ characters}

The fundamental representations are the anti-symmetric tensors
of rank $l=1\ldots,r-2$ plus the two conjugate spinor representations.
The anti-symmetric tensor characters are computed as for $B_r$, but
the spinor characters cannot be expressed completely in terms of traces
over the vector representation. This is because there exists a
symmetric tensor of rank $r$ which never appears in traces over the
vector representation, namely the Levi-Civita tensor. This tensor is
an anti-symmetric tensor of rank $2r$ with vector indices. Combining
the $2r$ vector indices in pairs, with each pair labelling an element
of the adjoint representation, we can view the Levi-Civita tensor also
as a symmetric tensor of rank $r$ with adjoint indices. 

Using this
new invariant, we can write down the spinor character:
\begin{eqnarray}
\label{eq:DspinCharPlus}
\Ch_{(0,\ldots,1,0)}(F) & = & 2^{r-1} \exp\left[\sum_{n=1}^{\infty}
{(2^{2n}-1)B_{2n} \over 4 n (2n)! } \Tr F^{2n} \right]\nn 
+ {1\over r!}\chi_r(F) \exp\left[\sum_{n=1}^{\infty}
{B_{2n} \over 4 n (2n)! } \Tr F^{2n} \right]
\end{eqnarray}
\begin{eqnarray}
\label{eq:DspinCharMin}
\Ch_{(0,\ldots,0,1)}(F) & = & 2^{r-1} \exp\left[\sum_{n=1}^{\infty}
{(2^{2n}-1)B_{2n} \over 4 n (2n)! } \Tr F^{2n} \right] \nn
  - {1\over r!} \chi_r(F) \exp\left[\sum_{n=1}^{\infty}
{  B_{2n} \over 4 n (2n)! } \Tr F^{2n} \right]\ .
\end{eqnarray}
where ${1\over r!}\chi_r(F)$ is a polynomial of order $r$ in $F$
defined by
the leading term in the difference of these expressions. It is
proportional to the Levi-Civita tensor with indices pairwise
contracted with $F^a$. The precise definition of the tensor is
given in appendix E. 

The trace identities for $D_r$ are as those for $B_r$ for $k > 2r$.
However, due to the extra fundamental trace of order $r$, there must
be an additional trace identity to reduce the number of independent 
ones back to $r$. Indeed, it turns out the the trace of order $2r$
can be eliminated using the identity
\begin{equation}
 \sum_{{n_i,m_i \atop n_im_i=2r}} \prod_i {1\over m_i !} 
\left( - {\Tr F^{n_i})\over n_i}\right)^{m_i}= 4 (-1)^r 
[{1\over r!}\chi_r(F)]^2   
\end{equation}
This identity was also obtained in \cite{ScWd} (the coefficient on
the right hand side is incorrect in \cite{ScWd}).

\subsection{$G_2$ characters}

Exceptional group characters can be computed by expressing them in
terms of characters of a regular subalgebra. Since the subalgebra has
the same rank, one gets polynomials in the same number of variables and
hence no information is lost. For $G_2$ the only option is the
subalgebra $A_2$.
We have
\begin{equation}
\Ch_{G_2,7}=\Ch_{A_2,3}+\Ch_{A_2,\bar 3}+\Ch_{A_2,1}\ , 
\end{equation}
denoting representations by their dimension and omitting the 
argument $F$. Since all $G_2$ representations are real, the third
order invariant of $A_2$ is always cancelled out, and 
all other odd invariants vanish as well. The fourth order 
invariant can be expressed in terms of second order ones using the 
$A_2$ trace identity. The sixth order
invariant of $A_2$ can be expressed in terms of lower ones, but this
expression involves the third order invariant which doesn't exist in $G_2$.
Hence in $G_2$ the sixth order invariant is new.
After a little algebra we can write the reference character of $G_2$ as 
\begin{equation}
 \Ch_{G_2,7} =7+
       \frac{1}{2}  \Tr F^2 
       + \frac{1}{4!} \frac{1}{4} ( \Tr F^2 )^2 
       + \frac{1}{6!} \Tr F^6    
       +  \frac{1}{8!}  [ \frac{2}{3} ( \Tr F^2 )( \Tr F^6 )
- \frac{5}{192}( \Tr F^2 )^4 ] + \ldots  
\end{equation}
Here all explicit traces are over the reference representation 
$(0,1)$ of dimension 7. The character of the other fundamental
representation, (1,0) of dimension 14 is easily computed from the
anti-symmetric tensor product $(7 \otimes 7 )_A=(7)+(14)$. 
Explicitly: 
\begin{eqnarray}
   \Ch_{G_2,14} & = & 14+
       \frac{1}{2!} 4 \Tr F^2 
       + \frac{1}{4!} \frac{5}{2} ( \Tr F^2 )^2 
       + \frac{1}{6!} [-26 \Tr F^6+\frac{15}{4}( \Tr F^2 )^3]\nn
       + \frac{1}{8!}  [ -\frac{160}{3} ( \Tr F^2 )( \Tr F^6 )
       - \frac{515}{96}( \Tr F^2 )^4 ] + \ldots
\end{eqnarray}
Note that all traces here are over the reference representation. 
From this expression we read off the second and sixth indices of
the representation $(14)$: they are 4 and $-26$ respectively. 

Since the (7) of $G_2$ can be embedded in the vector representation
of $SO(7)$, $G_2$ inherits all $B_3$ trace identities for traces of
order 8 and higher. There is an additional trace identity for the 
fourth order trace, which can be read off directly from $\Ch_{G_2,7}$:
\begin{equation}
  \Tr F^4 = \frac{1}{4} (\Tr F^2)^2  
\end{equation}
This exhausts the set of trace identities for $G_2$. 

\subsection{$F_4$ characters}
The computation is similar to the previous case, now using the
sub-algebra $B_4$.
We have 
\begin{equation}
\label{eq:chts}
 \Ch_{F_4,26}=\Ch_{B_4,16}+\Ch_{B_4,9}+\Ch_{B_4,1}
\end{equation} 
The vanishing of the fourth order invariant is not obvious
in this case, but follows easily. The sixth and eight order
polynomials are directly related to those of $B_4$. The tenth order
one vanishes again by inspection (\ie\ the tenth order trace can be
expressed in terms of $B_4$ traces of order 2,6 and 8, but not 4), and
the twelfth order trace involves the third power of the fourth order
polynomial of $B_4$, which did not occur before. It is absorbed
in the definition of $\Tr F^{12}$, the $12^{\rm th}$ order term in 
$ \Ch_{F_4,26} $ (up to a factor ${1\over 12!}$).
To obtain the character of $(1,0,0,0)$ (dimension (52)) we use
\begin{equation}
 \Ch_{F_4,52}=\Ch_{B_4,36}+\Ch_{B_4,16}
\end{equation} 
and we express all $B_4$ traces into $F_4$ traces using the
definitions introduced when computing \ref{eq:chts}. 
The other characters of basic representations can be obtained
from the anti-symmetric tensor products
\begin{eqnarray}
    \Ch_{F_4,273} & = &\Ch_{F_4,[2]*26}-\Ch_{F_4,52} \\
\Ch_{F_4,1274} & = &\Ch_{F_4,[3]*26}-\Ch_{F_4,52}\Ch_{F_4,26}+\Ch_{F_4,26}
\end{eqnarray}
The last identity involves a little algebra. In the third order
anti-symmetric tensor power of (26) occurs, in addition to (1274) also
the representations (273) and (1053) (with Dynkin labels (1,0,0,1)). 
The former character is known, the latter can be computed using the
tensor product $(26)\otimes (52)$. 

Of course the characters of (273) and (1274) can also be computed
using the $B_4$ embedding. We have used this as a check.

Just as for $G_2$ one may read off 
the Dynkin indices from the characters. They are shown in the appendix. 
Furthermore there are trace relations for traces of fourth and tenth
order which are read off from $\Ch_{(26)}$. By expanding the characters
to sufficiently high order one obtains trace identities for traces of
order 14 and higher. 
The embedding $F_4 \subset D_{13}$
gives trace identities for all traces of order 26 and higher, namely
precisely those of $D_{13}$. 
We will only present the identities for
orders lower than that of the maximal Casimir operator. 

\subsection{$E_6$ characters}

Here we used the sub-algebra $A1 \oplus A5$, and
the decompositions 
\begin{equation}
 \Ch_{E_6,27}=\Ch_{A_1,2}\Ch_{A_5,\bar 6}+\Ch_{A_5,15}  
\end{equation} 
\begin{equation}
\Ch_{E_6,78}=\Ch_{A_1,2}
          +\Ch_{A_5,35}
          +\Ch_{A_1,2}\Ch_{A_5,20} 
\end{equation}
The computation is very 
similar to the previous cases. We get another basic
representation, the $(\overline{27})$, by conjugation:
\begin{equation}
\Ch_{E_6,\overline{27}}(F)= \Ch_{E_6,{27}}(-F)\ . 
\end{equation} 
Furthermore
the anti-symmetric tensor power of order 2 gives us the representations
(0,1,0,0,0,0) 
$(351)$ and (0,0,0,1,0,0) $(\overline{351})$ 
and the order three anti-symmetric power yields 
precisely the representation  (0,0,1,0,0,0) (2925). 

The indices and trace identities for orders up to 12 are listed in
the appendix.  

\subsection{$E_7$ characters}

Here we used the sub-algebra $A_7$, and
the decompositions or anti-symmetric tensor products
\begin{eqnarray}
\Ch_{E_7,56}&=&\Ch_{A_7,28}+\Ch_{A_7,\overline{28}} \\ 
\Ch_{E_7,133}&=&\Ch_{A_7,70}+\Ch_{A_7,\overline{63}} \\ 
\Ch_{E_7,912}&=&\Ch_{A_7,420}+\Ch_{A_7,\overline{420}}
+\Ch_{A_7,36}+\Ch_{A_7,\overline{36}} \\ 
\Ch_{E_7,1539}&=&\Ch_{E_7,[2]*56}-\Ch_{E_7,1} \\
\Ch_{E_7,8645}&=&\Ch_{E_7,[2]*133}-\Ch_{E_7,133} \\
\Ch_{E_7,27664}&=&\Ch_{E_7,[3]*56}-\Ch_{E_7,56} \\
\Ch_{E_7,365750}&=&\Ch_{E_7,[4]*56}-\Ch_{E_7,1539}-\Ch_{E_7,1}
\end{eqnarray}

\subsection{$E_8$ characters}

Here we used the sub-algebra $D_8$, and
\begin{eqnarray}
\Ch_{E_8,248}&=&\Ch_{D_8,128}+\Ch_{D_8,120} \\ 
\Ch_{E_8,3875}&=&\Ch_{D_8,1920}+\Ch_{D_8,1820}+\Ch_{D_8,135} \\ 
\Ch_{E_8,147250}&=&\Ch_{D_8,60060}+\Ch_{D_8,56320}
+\Ch_{D_8,15360}\nn
+\Ch_{D_8,7020}+\Ch_{D_8,6435}+\Ch_{D_8,1920}+\Ch_{D_8,135} \\
\Ch_{E_8,30380}&=&\Ch_{E_8,[2]*248}-\Ch_{E_8,248} \\
\Ch_{E_8,2450240}&=&\Ch_{E_8,[3]*248}
     -(\Ch_{E_8,248})^2 +\Ch_{E_8,248} \\
\Ch_{E_8,6696000}&=&\Ch_{E_8,[2]*3875}-\Ch_{E_8,3875}(\Ch_{E_8,248}-1)
+\Ch_{E_8,147250}\\
\Ch_{E_8,146325270}&=&\Ch_{E_8,[4]*248}-(\Ch_{E_8,[2]*248}
-\Ch_{E_8,248})(\Ch_{E_8,248}-1)
\\
\Ch_{E_8,6899079264}&=&\Ch_{E_8,[5]*248}
-\Ch_{E_8,248}(\Ch_{E_8,[3]*248}-2\Ch_{E_8,[2]*248}+ \Ch_{E_8,248}-1)
\end{eqnarray}
In the first three lines all $D_8$ spinor representations must
be from the same conjugacy class, which is fixed by the decomposition
one chooses for the (248). 
Since the choice one makes for
the class is irrelevant, there is no need for a label to distinguish
conjugate spinors. The representation (6435) is an (anti-)selfdual
tensor. It belongs to the trivial conjugacy class, but it does carry
a non-trivial chirality. If for the representation
denoted (128)
we choose
the one
with Dynkin labels $(0,0,0,0,0,0,0,1)$, then the 
correct set of Dynkin labels for the 6345 is $(0,0,0,0,0,0,2,0)$.

\subsection{Normalization of indices}

The normalization of the symmetric tensors is fixed
by fixing a normalization for the indices. 
We will do this in such a way that they are always integers, as
the word ``index" suggests.
For
the second index there is a natural normalization in terms of
the Atiyah-Singer index theorem for instantons on $S_3$. For any
representation of any algebra
we can choose the second index equal to the net number
of zero modes of a Weyl fermion in that representation in an instanton
field of minimal non-trivial topological charge (where ``net" means
the difference between the two chiralities). Then the second index
is equal to 1 for the reference representations of $A_r$ and $C_r$,
2 for those of $B_r$ and $C_r$, and 2,6,6,12,60 respectively for
the reference representations of $G_2, F_4, E_6, E_7$ and $E_8$. For
the adjoint representation the second index is always equal to twice the
dual Coxeter number $g$ listed in table \ref{tab:itr}. This choice
corresponds to setting $\eta=2$ in (\ref{eq:CasNorm}). This
value of $\eta$ was used 
in the last column of table \ref{tab:refrep}. 

For the higher indices there is a similar topological interpretation 
in terms of gauge bundles on higher dimensional manifolds, but we will
not explore that here in detail. One may however follow the spirit
of such an interpretation and define all higher indices in such a way
that they are integers. This is automatically true if they are
integers for the basic representations, because the characters of
all representations are polynomials with integer coefficients in terms
of the characters of the basic representations. Furthermore, within the
set of basic representations the ones obtained by means of anti-symmetric
tensor products of a given representation $R$, have indices that are
an integer multiple of those of $R$. Then only the spinor representations
of $SO(N)$, one representation  of $F_4$, $E_6$ and $E_7$
and two of $E_8$ require
special attention.

For the reference representations we choose all higher indices equal 
to 1, except when a larger integer is required to make all indices
integral.    
The $2n^{\rm th}$ index of a spinor representation of $SO(N)$ follows
directly from the character equations (\ref{eq:BspinChar}), 
(\ref{eq:DspinCharPlus}) and (\ref{eq:DspinCharMin}): 
\begin{equation}
 \dim(S) { (2^{2n}-1) B_{2n} \over 4 n }\ ,
\end{equation}
where $\dim(S)$ is the dimension of a spinor representation, \ie\
$\dim(S)=2^r$ for algebras of type $B_r (SO(2r+1)$ and $\dim(S)=2^{r-1}$
for type $D_r  (SO(2r))$.
This assumes that the $(2n)^{\rm th}$ 
index of the vector representation is set to 1.
By 
inspection, this expression is an integer except for the fourth index
for $SO(N), N \leq 8$ and the eighth index for $N \leq 10$. 
We have checked that the spinor index is an integer in all other cases for 
$2n < 100$. Table \ref{tab:SONindices} gives the indices for the $SO(N)$
vector and spinor representations, according to our normalization (the
index $\tilde I_r$ is not listed here; its value is 0 for
the vector representation and chosen $\pm1$ for the two fundamental spinors of
$SO(N)$, $N$ even). Note that for $N=3,\ldots,6$ the spinor
representation, and not the vector is the reference representation, which
automatically leads to the entries in the table.  
\begin{table}[htb]
\centering
\begin{tabular}{|c|c|c|c|c|c|c|c|}
\hline
$N$ & $I_2$ & $I_4$ &  $I_6$ & $I_8$ & $I_{10}$ & $I_{12}$ & $I_{14}$ \\ \hline
3   & 4,1   &  ---  &  ---   &  ---  &  ---     &  ---     &  ---     \\
4   & 4,1   &  ---  &  ---   &  ---  &  ---     &  ---     &  ---     \\
5   & 2,1   &  -4,1 &  ---   &  ---  &  ---     &  ---     &  ---     \\
6   & 2,1   &  -4,1 &  ---   &  ---  &  ---     &  ---     &  ---     \\
7   & 2,2   &  2,--1 &  1,1   &  ---  &  ---     &  ---     &  ---     \\
8   & 2,2   &  2,--1 &  1,1   &  ---  &  ---     &  ---     &  ---     \\
9   & 2,4   &  1,--1 &  1,2   & 2,--17  &  ---     &  ---     &  ---     \\
10  & 2,4   &  1,--1 &  1,2   & 2,--17  &  ---     &  ---     &  ---     \\
11  & 2,8   &  1,--2 &  1,4   & 1,--17  &  1,124   &  ---     &  ---     \\
12  & 2,8   &  1,--2 &  1,4   & 1,--17  &  1,124   &  ---     &  ---     \\
13  & 2,16  &  1,--2 &  1,8  & 1,--34  & 1,248   &  1,--2764   &  ---     \\
14  & 2,16  &  1,--4 &  1,8  & 1,--34  &  1,248  &  1,--2764   &  ---     \\
15  & 2,32  &  1,--4 &  1,16 & 1,--68  &  1,496  &  1,--5528   &  1,87376 \\
16  & 2,32  &  1,--8 &  1,16 & 1,--68  &  1,496 &  1,--5528   &  1,87376 \\ \hline
\end{tabular}
\caption{\label{tab:SONindices} Indices for the $SO(N)$ vector and spinor 
representations. The chiral index $\tilde I_r$ is not listed here;
we choose it equal to $+1$ for the fundamental spinor $(0,\ldots,1,0)$
and equal to $-1$ for $(0,\ldots,0,1)$.}
\end{table}

For the exceptional algebras the two highest indices of $E_7$ and the
three highest ones of $E_8$ come out fractional unless we choose
a different normalization for the reference representation. 

Our preferred index normalization for the reference representations
is summarized in the last column of table \ref{tab:refrep}. 
For $SO(2N)$ the last entry indicates the index $\tilde I_N$ of
the spinor representation $(0,\ldots,1,0)$. Only the second
index is affected by the choice of $\eta$. For all higher order
traces we fix the normalization of the index, and then the
$\eta$-dependence goes into the normalization of the symmetric
tensor. 

This also fixes the normalization of all symmetric tensors as
\begin{equation}
\label{eq:tensordef}
 \Str T^{a_1}_{r }\ldots T^{a_n}_{r } = I_n({r}) 
d^{a_1\ldots a_n}\ . 
\end{equation}
This defines the tensors given a choice of generators in the 
reference representation.  

In Appendix D we present some results for contractions of these 
tensors.


\section{Reduction identities}

At this stage we have terms that contain combinations of the invariants 
$d_r$ and the structure constant $f$. Our task is now to eliminate 
$f$ from the terms as much as possible, and reduce the total
number of invariants in the final answer. 

Unlike the results obtained so far, for these reductions 
we cannot give a general algorithm. In fact, we do not even know
what the desirable outcome is, since we are not aware of a mathematical
theorem that gives us a basic set of invariants in terms of which all
others can be expressed. At any given order we can derive large numbers
of identities among the various invariants, but there will be
new relations at every order. In practice this is not a major problem.
First of all we have obtained results relevant for vacuum bubble Feynman 
diagrams of up to nine loops, and secondly the number of invariants we are 
left with is small, although possibly not minimal.  

The most useful identities for 
doing this are the Jacobi identities. Thus far we have seen two of them in 
equation (\ref{eq:jacobi}). The second equation there can be generalized into
\begin{eqnarray}
\label{eq:jacobin}
0 & = & \sum_{{\rm cyclic\ permutations\ of}\ i_1\cdots i_n}
    d_R^{i_1\cdots i_{n-1}a}f^{i_n b a}
\end{eqnarray}
for all representations $R$, including the adjoint representation.
In fact, there is an advantage to postponing
the replacement of the adjoint symmetric
tensors by reference tensors, since they satisfy additional identities.
Furthermore some identities produce new tensors $d_A$. 
For this reason we present the results in terms of $d_R$ and $d_A$ rather
than $d_r$. 

The first identity that can be derived from this is one for invariants with 
three indices:
\begin{eqnarray}
    d_R^{abi}f^{ajc}f^{blc} & = & \frac{1}{2}C_A\ d_R^{ijk}
\end{eqnarray}
It is actually the simplest identity in a class of triangle reductions that 
involve one or two invariant tensors $d_R$. We have also
\begin{eqnarray}
    d_{R_1}^{i_1j_1\cdots j_nk_1}
            d_{R_2}^{i_2j_1\cdots j_nk_2}f^{k_1k_2i_3} & = &
        \frac{1}{N_A}\frac{1}{n+1}
            d_{R_1}^{j_1\cdots j_{n+2}}d_{R_2}^{j_1\cdots j_{n+2}}
                f^{i_1i_2i_3} \\
    d_{R_1}^{i_1j_1\cdots j_nk_1}d_{R_2}^{i_3\cdots i_mj_1\cdots j_nk_2}
            f^{k_1k_2i_2} & = & \frac{-1}{n+1}
    d_{R_1}^{j_1\cdots j_{n+1}k}d_{R_2}^{i_3\cdots i_mj_1\cdots j_{n+1}}
            f^{ki_1i_2}
\end{eqnarray}
These identities are very powerful when a large number of invariants is 
involved. 

For showing further reduction identities we will use a special notation 
which corresponds closely to a notation that can be used inside a computer 
program. We will represent a (symmetric) invariant by a product of vectors:
\begin{eqnarray}
\label{eq:p-rewrite}
    d_{R_1}^{i_1\cdots i_n} & = & p_1^{i_1}\cdots p_1^{i_n}
\end{eqnarray}
The lower index on the vector refers to the particular invariant. In this 
notation we have no problems with the symmetric property of the invariants. 
Of course we are not implying that each invariant can be mathematically 
written this way. It is just notation.

Additionally we will use Schoonschip notation on contracted indices. That 
is: if the index of a vector is contracted with an index of a tensor we put 
the vector in the place of this index. Hence
\begin{eqnarray}
    f^{p_1 p_2 i} f^{p_1 p_2 i} ( p_1 \cdot p_2 )^n & = &
        d_{R_1}^{i_1\cdots i_nj_1j_2}
        d_{R_2}^{i_1\cdots i_nk_1k_2}f^{j_1k_1i}f^{j_2k_2i}
\end{eqnarray}
Furthermore we can add a weight to each formula. This weight is basically 
the number of vertices (assuming that all vertices are three-point 
vertices) in the diagram before we started the elimination procedures. For 
our current algorithms the weight is the total number of vectors $p_i$ plus 
the number of structure constants $f$. Hence the weight of the above 
formula is $2n+6$. We will present all reduction identities that are 
relevant for weights up to 12. This corresponds to 7-loop vacuum bubbles or 
6-loop propagator diagrams. The derivation of all these identities involves 
the use of the generalized identity of equation (\ref{eq:jacobin}).
\begin{eqnarray}
    f^{p_1 p_2 i} f^{p_1 p_2 i} ( p_1 \dotpr p_2 )^n & = &
            \frac{1}{n+1}\ C_A\ (p_1 \dotpr p_2 )^{n+2}
        \\
    f^{p_1p_2i_1}f^{p_1p_2i_2}f^{p_1i_1i_3}f^{p_1i_2i_3}
         ( p_1 \dotpr p_2 )^n & = &
     \frac{1}{n+1}\ d_A^{p_1p_1p_1p_2}( p_1 \dotpr p_2 )^{n+1}
        \\
    f^{p_1p_2i_1}f^{p_1i_1i_2}f^{p_2i_2i_3}f^{p_1p_2i_3}
         ( p_1 \dotpr p_2 )^n & = &
       \frac{5}{6(n+1)(n+2)}\ C_A^2\ (p_1\dotpr p_2)^{n+3}
        \nonumber \\ & &
       - \frac{1}{n+1}\ d_A^{p_1p_1p_2p_2}(p_1\dotpr p_2)^{n+1}
        \\
    f^{p_1p_2i_1}f^{p_1p_2i_2}f^{p_1p_2i_3}f^{i_1i_2i_3}
         ( p_1 \dotpr p_2 )^n & = & 0 \\
    f^{p_1p_2i_1}f^{p_1p_2i_1}f^{p_1p_2i_2}f^{p_1p_2i_2}
          (p_1\dotpr p_2)^n & = &
       \frac{2\ n!}{(n+2)!}\ d_A^{p_1p_1p_2p_2}(p_1\dotpr p_2)^{n+2}
        \nonumber \\ & &
      +\frac{(3n+1)\ n!}{3\ (n+3)!}\ C_A^2\ (p_1\dotpr p_2)^{n+4}
        \\
    f^{p_1p_2i}f^{p_1p_2i}(p_1\dotpr p_3)(p_2 \dotpr p_3)^n & = &
        \frac{1}{2}\ C_A\ (p_1\dotpr p_2)^2 (p_1\dotpr p_3) (p_2 \dotpr p_3)^n
        \\
    f^{p_1p_2i}f^{p_1p_3i}(p_1\dotpr p_2)(p_2 \dotpr p_3)^n & = &
        \frac{1}{2}\ C_A\ (p_1\dotpr p_2)^2 (p_1\dotpr p_3) (p_2 \dotpr p_3)^n
        \\
    f^{p_1p_2i}f^{p_1p_3i}(p_1\dotpr p_2)(p_2\dotpr p_3)(p_1 \dotpr p_3)^n
         & = & \frac{C_A}{2(n+1)}
            \ (p_1\dotpr p_2)^2(p_2\dotpr p_3)(p_1 \dotpr p_3)^{n+1}
        \\
    f^{p_1p_2p_3}f^{p_1p_2p_3}(p_1\dotpr p_2)(p_1\dotpr p_3)
        & = & \frac{1}{4}\ C_A\ (p_1\dotpr p_2)^2
            (p_1\dotpr p_3)^2(p_2\dotpr p_3)
\end{eqnarray}
The equations with an odd number of $f$'s that are relevant are all zero. 
This has been explicitly shown, but for many of them one can see this 
already on the basis of symmetry principles.

We are not going to present the identities that would be needed for 
diagrams of weight 14 or 16. There would be too many of them and moreover, 
this is not how we have constructed the computer program. In the program we 
have found a way to apply equation (\ref{eq:jacobin}) recursively in such 
a way that it reduces all combinations with the exception of one 
(at weight 14). The 
program has to guess at what combination of invariants and $f$'s to take 
for the application of the formula and we let it guess several times. In 
the end this covers all cases except for the one that we cannot do by these 
methods anyway.

The first object that causes some real problems because the above 
algorithms are not sufficient to handle them, occurs at weight 14. This 
object can either be written as
\begin{equation}
    f^{p_1p_2p_3}f^{p_1p_2p_3}(p_1\dotpr p_2)(p_1\dotpr p_3)(p_2\dotpr p_3)
\end{equation}
or with some rewriting (and omitting trivial terms that are a byproduct of 
the rewriting):
\begin{equation}
\label{eq:aaaff}
    f^{p_1p_2i}f^{p_1p_2i}(p_1\dotpr p_3)^2(p_2\dotpr p_3)^2
\end{equation}
If at least two of the three invariants are in the adjoint representation 
this object can be reduced with the same technique that we use below 
to simplify some combinations of invariants only (see equation 
(\ref{eq:7loops})).


\subsection{Combinations of invariants}

Here we will consider combinations of invariants only. 
The easiest combinations are full contractions between two invariants as in 
\begin{equation}
    d_{R_1}^{i_1\cdots i_n} d_{R_2}^{i_1\cdots i_n}
\end{equation}
Unfortunately, when the weight of the diagrams increases, the complexity of 
the combinations increases correspondingly. In some cases one can make 
reductions. For instance:
\begin{eqnarray}
\label{eq:oneline}
    d_{R_1}^{j\ i_1\cdots i_n} d_{R_2}^{k\ i_1\cdots i_n} & = &
        \frac{1}{N_A}\ \delta^{jk} d_{R_1}^{i_0i_1\cdots i_n} d_{R_2}^{i_0i_1\cdots i_n}
\end{eqnarray}
and hence
\begin{eqnarray}
    d_{R_1}^{a\ i_1\cdots i_n} d_{R_2}^{b\ i_1\cdots i_n}
    d_{R_3}^{a\ j_1\cdots j_m} d_{R_4}^{b\ j_1\cdots j_m} & = &
    \frac{1}{N_A}\ d_{R_1}^{i_0i_1\cdots i_n} d_{R_2}^{i_0i_1\cdots i_n}
    d_{R_3}^{j_0j_1\cdots j_m} d_{R_4}^{j_0j_1\cdots j_m}
\end{eqnarray}
but for objects of the type
\begin{equation}
    d_{R_1}^{i_1i_2i_3i_4}d_{R_2}^{i_1i_2i_5}d_{R_3}^{i_3i_4i_5}
\end{equation}
or
\begin{equation}
    d_{R_1}^{i_1i_2i_3i_4i_5}d_{R_2}^{i_1i_2i_3i_6}d_{R_3}^{i_4i_5i_6}
\end{equation}
there does not seem to be a general simplification of this type. 
In the case of all invariants belonging to the adjoint representation we 
can still do things as we see in the next formula:
\begin{eqnarray}
\label{eq:6loops}
 d_A^{abcdef}d_A^{abcdef}-\frac{5}{8} d_A^{abcd}d_A^{cdef}d_A^{efab}
    +\frac{7}{240}C_A^2d_A^{abcd}d_A^{abcd} +\frac{1}{864}C_A^6N_A & = & 0
\end{eqnarray}
The derivation of this formula is a matter of evaluating a circular ladder 
with 6 rungs in two different ways. In the first way one sees it as two 
loops with 6 vertices, and in the second way as three loops with 4 
vertices. The algorithms of the previous sections are then sufficient to 
obtain this formula. Note however that such derivations usually need the 
use of a computer program: the intermediate stages can contain large 
numbers of terms. 
A similar technique can be used for the object in 
equation (\ref{eq:aaaff}). We look at a circular ladder with 7 rungs. If we 
see this as three loops with 4 vertices (and two $f$'s left) we get the form 
of the equation, and if we see it as two loops with 7 vertices, we get a 
representation involving two $d_A$ invariants with 6 indices (the ones with 
7 indices are zero for the adjoint representation). The result is (after a 
rather lengthy calculation, applying equation (\ref{eq:6loops}) and 
normalizing):
\begin{eqnarray}
\label{eq:7loops}
       d_R^{abcd}d_A^{cdef}d_A^{efgh}f^{agi}f^{bhi}
       + \frac{2}{27} C_A^3 d_R^{abcd}d_A^{abcd} && \nonumber \\
       - \frac{19}{15} C_A d_R^{abcd}d_A^{cdef}d_A^{efab}
       + \frac{8}{9} d_A^{abcdef}d_R^{abcg}d_A^{defg}
       & = & 0
\end{eqnarray}
with R replaced by A. 
Because a similar exchange of the order of evaluation in one of the 
examples below gives the same equation with a slightly 
extended generality we have already presented this more general form with 
the representation R. One can also derive this equation from 
equation (\ref{eq:dA4dA6}), but the derivation of that formula uses basically 
the same technique.

Considering that the algorithms we have presented can reduce all 
combinations of invariants and $f$'s, with the exception of the above 
combination with at least two invariants not in the adjoint representation, 
up to weight 14 we have only a limited number of 
topologies (contractions between invariants) left. We will show them 
graphically, omitting the very trivial ones that can be reduced with 
equation \ref{eq:oneline} and objects of the type $d_R^{aai_1\cdots i_n}$.
The elements of this pictorial language are
\begin{eqnarray}
    \begin{picture}(20,20)(0,0)
    \Line(10,10)(10,0) \Line(10,10)(3,17) \Line(10,10)(17,17) \dReg(10,10)
    \end{picture} & = & d_R^{i_1i_2i_3} \nonumber \\
    \begin{picture}(20,20)(0,0)
    \Line(10,20)(10,0) \Line(0,10)(20,10) \dReg(10,10)
    \end{picture} & = & d_R^{i_1i_2i_3i_4} \nonumber \\
    \begin{picture}(20,20)(0,0)
    \Line(10,20)(10,0) \Line(0,10)(20,10) \dAdj(10,10)
    \end{picture} & = & d_A^{i_1i_2i_3i_4} \nonumber \\
    \begin{picture}(20,20)(0,0)
    \Line(10,20)(10,0) \Line(0,10)(20,10) \dEmp(10,10)
    \end{picture} & = & d_{\ortho}^{i_1i_2i_3i_4} \nonumber \\
    \begin{picture}(20,20)(0,0)
    \Line(10,10)(10,0) \Line(10,10)(3,17) \Line(10,10)(17,17) \freg(10,10)
    \end{picture} & = & f^{i_1i_2i_3} \nonumber \\
    \begin{picture}(20,20)(0,0)
    \Line(10,10)(10,0) \Line(10,10)(3,17) \Line(10,10)(17,17) \fanti(10,10)
    \end{picture} & = & f^{i_3i_2i_1}
\end{eqnarray}
in which we assume the indices of $f$ to run counterclockwise in the 
diagram.
For weight 6 we have
\begin{eqnarray}
    d_{33}(p_1,p_2) & = & d_{R_1}^{ijk}d_{R_2}^{ijk} \nonumber \\
        & = &
    \raisebox{-7pt}{ \begin{picture}(40,20)(0,0)
    \Line(5,13.5)(35,13.5) \Line(5,6.5)(35,6.5)
    \dReg(5,10) \dReg(35,10)
    \Text(20,10.5)[]{\footnotesize 3}
        \Text(5,5)[t]{\footnotesize 1}
        \Text(35,5)[t]{\footnotesize 2}
    \end{picture}}
\end{eqnarray}
For weight 8 there is also only one topology:
\begin{eqnarray}
    d_{44}(p_1,p_2) & = &
    \raisebox{-7pt}{ \begin{picture}(40,20)(0,0)
    \Line(5,13.5)(35,13.5) \Line(5,6.5)(35,6.5)
    \dReg(5,10) \dReg(35,10)
    \Text(20,10.5)[]{\footnotesize 4}
        \Text(5,5)[t]{\footnotesize 1}
        \Text(35,5)[t]{\footnotesize 2}
    \end{picture}}
\end{eqnarray}
For weight 10 there are two topologies:
\begin{eqnarray}
    d_{55}(p_1,p_2) & = &
    \raisebox{-7pt}{ \begin{picture}(40,20)(0,0)
    \Line(5,13.5)(35,13.5) \Line(5,6.5)(35,6.5)
    \dReg(5,10) \dReg(35,10)
    \Text(20,10.5)[]{\footnotesize 5}
        \Text(5,5)[t]{\footnotesize 1}
        \Text(35,5)[t]{\footnotesize 2}
    \end{picture}} \\
    d_{433}(p_1,p_2,p_3) & = &
    \raisebox{-12pt}{ \begin{picture}(40,30)(0,0)
        \Line(20,27)(5,7) \Line(20,23)(5,3)
        \Line(20,27)(35,7) \Line(20,23)(35,3)
        \Line(5,5)(35,5)
        \dReg(5,5) \dReg(35,5) \dReg(20,25)
        \Text(16,26)[rb]{\footnotesize 1}
        \Text(2,3)[rt]{\footnotesize 2}
        \Text(38,3)[lt]{\footnotesize 3}
    \end{picture}}
\end{eqnarray}
For weight 12 we have 5 topologies:
\begin{eqnarray}
    d_{66}(p_1,p_2) & = &
    \raisebox{-7pt}{ \begin{picture}(40,20)(0,0)
    \Line(5,13.5)(35,13.5) \Line(5,6.5)(35,6.5)
    \dReg(5,10) \dReg(35,10)
    \Text(20,10.5)[]{\footnotesize 6}
        \Text(5,5)[t]{\footnotesize 1}
        \Text(35,5)[t]{\footnotesize 2}
    \end{picture}} \\
    d_{633}(p_1,p_2,p_3) & = &
    \raisebox{-7pt}{ \begin{picture}(60,20)(0,0)
    \Line(5,13)(55,13)\Line(5,10)(55,10)\Line(5,7)(55,7)
    \dReg(5,10) \dReg(30,10) \dReg(55,10)
        \Text(5,5)[t]{\footnotesize 2}
        \Text(30,5)[t]{\footnotesize 1}
        \Text(55,5)[t]{\footnotesize 3}
    \end{picture}} \\
    d_{543}(p_1,p_2,p_3) & = &
    \raisebox{-12pt}{ \begin{picture}(52,35)(0,0)
        \Line(25,25)(5,5) \Line(25,28)(5,8) \Line(25,22)(5,2)
        \Line(25,27)(45,7) \Line(25,23)(45,3)
        \Line(5,5)(45,5)
        \dReg(5,5) \dReg(45,5) \dReg(25,25)
        \Text(22,26)[rb]{\footnotesize 1}
        \Text(2,3)[rt]{\footnotesize 2}
        \Text(48,3)[lt]{\footnotesize 3}
    \end{picture}} \\
    d_{444}(p_1,p_2,p_3) & = &
    \raisebox{-12pt}{ \begin{picture}(52,35)(0,0)
        \Line(25,27)(5,7) \Line(25,23)(5,3)
        \Line(25,27)(45,7) \Line(25,23)(45,3)
        \Line(5,6.5)(45,6.5) \Line(5,3.3)(45,3.3)
        \dReg(5,5) \dReg(45,5) \dReg(25,25)
        \Text(22,26)[rb]{\footnotesize 1}
        \Text(2,3)[rt]{\footnotesize 2}
        \Text(48,3)[lt]{\footnotesize 3}
    \end{picture}} \\
    d_{3333}(p_1,p_2,p_3,p_4) & = &
    \raisebox{-16pt}{ \begin{picture}(45,40)(0,0)
        \Line(10,5)(35,5) \Line(10,5)(10,30) \Line(10,5)(35,30)
        \Line(10,30)(35,5) \Line(35,5)(35,30) \Line(10,30)(35,30)
        \dReg(10,5) \dReg(35,5) \dReg(10,30) \dReg(35,30)
        \Text(7,33)[rb]{\footnotesize 1}
        \Text(7,3)[rt]{\footnotesize 2}
        \Text(38,3)[lt]{\footnotesize 3}
        \Text(38,32)[lb]{\footnotesize 4}
    \end{picture}}
\end{eqnarray}
Finally for weight 14 we have 9 topologies:
\begin{eqnarray}
    d_{77}(p_1,p_2) & = &
    \raisebox{-7pt}{ \begin{picture}(40,20)(0,0)
    \Line(5,13.5)(35,13.5) \Line(5,6.5)(35,6.5)
    \dReg(5,10) \dReg(35,10)
    \Text(20,10.5)[]{\footnotesize 7}
        \Text(5,5)[t]{\footnotesize 1}
        \Text(35,5)[t]{\footnotesize 2}
    \end{picture}} \\
    d_{743}(p_1,p_2,p_3) & = &
    \raisebox{-7pt}{ \begin{picture}(60,20)(0,0)
    \Line(5,13.5)(30,13.5) \Line(5,6.5)(30,6.5)
    \Line(30,13)(55,13)\Line(30,10)(55,10)\Line(30,7)(55,7)
    \dReg(5,10) \dReg(30,10) \dReg(55,10)
    \Text(17.5,10.5)[]{\footnotesize 4}
        \Text(5,5)[t]{\footnotesize 2}
        \Text(30,5)[t]{\footnotesize 1}
        \Text(55,5)[t]{\footnotesize 3}
    \end{picture}} \\
    d_{653}(p_1,p_2,p_3) & = &
    \raisebox{-12pt}{ \begin{picture}(52,35)(0,0)
        \Line(24,26)(4,6) \Line(26,24)(6,4)
        \Line(25,25)(45,5)
        \Line(5,8)(45,8) \Line(5,2)(45,2)
        \dReg(5,5) \dReg(45,5) \dReg(25,25)
        \Text(22,26)[rb]{\footnotesize 3}
        \Text(2,3)[rt]{\footnotesize 1}
        \Text(48,3)[lt]{\footnotesize 2}
        \Text(25,5.5)[]{\footnotesize 4}
    \end{picture}} \\
    d_{644}(p_1,p_2,p_3) & = &
    \raisebox{-12pt}{ \begin{picture}(52,35)(0,0)
        \Line(25,25)(5,5) \Line(25,28)(5,8) \Line(25,22)(5,2)
        \Line(25,25)(45,5) \Line(25,28)(45,8) \Line(25,22)(45,2)
        \Line(5,5)(45,5)
        \dReg(5,5) \dReg(45,5) \dReg(25,25)
        \Text(22,26)[rb]{\footnotesize 1}
        \Text(2,3)[rt]{\footnotesize 2}
        \Text(48,3)[lt]{\footnotesize 3}
    \end{picture}} \\
    d_{554}(p_1,p_2,p_3) & = &
    \raisebox{-12pt}{ \begin{picture}(52,35)(0,0)
        \Line(25,27)(5,7) \Line(25,23)(5,3)
        \Line(25,27)(45,7) \Line(25,23)(45,3)
        \Line(5,7.5)(45,7.5) \Line(5,5)(45,5) \Line(5,2.5)(45,2.5)
        \dReg(5,5) \dReg(45,5) \dReg(25,25)
        \Text(22,26)[rb]{\footnotesize 3}
        \Text(2,3)[rt]{\footnotesize 1}
        \Text(48,3)[lt]{\footnotesize 2}
    \end{picture}} \\
    d_{5333}(p_1,p_2,p_3,p_4) & = &
    \raisebox{-12pt}{ \begin{picture}(52,35)(0,0)
        \Line(25,27)(5,7) \Line(25,23)(5,3)
        \Line(25,27)(45,7) \Line(25,23)(45,3)
        \Line(25,25)(25,5)
        \Line(5,5)(45,5)
        \dReg(5,5) \dReg(45,5) \dReg(25,25) \dReg(25,5)
        \Text(22,26)[rb]{\footnotesize 1}
        \Text(2,3)[rt]{\footnotesize 2}
        \Text(48,3)[lt]{\footnotesize 3}
        \Text(22,2)[t]{\footnotesize 4}
    \end{picture}} \\
    d_{4433a}(p_1,p_2,p_3,p_4) & = &
    \raisebox{-16pt}{ \begin{picture}(45,40)(0,0)
        \Line(8.5,5)(8.5,30) \Line(10,5)(35,30)
        \Line(11.5,5)(11.5,30)
        \Line(10,30)(35,5) \Line(36.5,5)(36.5,30) \Line(33.5,5)(33.5,30)
        \Line(10,30)(35,30) \Line(10,30)(35,30) 
        \dReg(10,5) \dReg(35,5) \dReg(10,30) \dReg(35,30)
        \Text(7,33)[rb]{\footnotesize 1}
        \Text(7,3)[rt]{\footnotesize 3}
        \Text(38,3)[lt]{\footnotesize 4}
        \Text(38,32)[lb]{\footnotesize 2}
    \end{picture}} \\
    d_{4433b}(p_1,p_2,p_3,p_4) & = &
    \raisebox{-16pt}{ \begin{picture}(45,40)(0,0)
        \Line(10,5)(35,5) \Line(10,5)(10,30) \Line(10,5)(35,30)
        \Line(10,30)(35,5) \Line(35,5)(35,30)
        \Line(10,28)(35,28) \Line(10,32)(35,32) 
        \dReg(10,5) \dReg(35,5) \dReg(10,30) \dReg(35,30)
        \Text(7,33)[rb]{\footnotesize 1}
        \Text(7,3)[rt]{\footnotesize 3}
        \Text(38,3)[lt]{\footnotesize 4}
        \Text(38,32)[lb]{\footnotesize 2}
    \end{picture}} \\
    d_{4433c}(p_1,p_2,p_3,p_4) & = &
    \raisebox{-16pt}{ \begin{picture}(45,40)(0,0)
        \Line(10,5)(35,5) \Line(8.5,5)(8.5,30) \Line(11.5,5)(11.5,30)
        \Line(33.5,5)(33.5,30) \Line(36.5,5)(36.5,30)
        \Line(10,28)(35,28) \Line(10,32)(35,32) 
        \dReg(10,5) \dReg(35,5) \dReg(10,30) \dReg(35,30)
        \Text(7,33)[rb]{\footnotesize 1}
        \Text(7,3)[rt]{\footnotesize 3}
        \Text(38,3)[lt]{\footnotesize 4}
        \Text(38,32)[lb]{\footnotesize 2}
    \end{picture}}
\end{eqnarray}
For weight 16 there are more than 20 non-trivial topologies. We will not 
show them here.

In terms of these diagrams of invariants we can compose a few extra 
equations that can be very useful:
\begin{eqnarray}
\label{eq:reduA4}
    \raisebox{-7pt}{ \begin{picture}(55,20)(0,0)
    \Line(5,10)(20,10)
    \Line(20,13)(35,13)\Line(20,7)(35,7)
    \Line(35,13)(50,18)\Line(35,7)(50,2)
    \dReg(20,10) \dAdj(35,10)
    \end{picture}}
 & = &
    \frac{1}{6}C_A^2\raisebox{-7pt}{ \begin{picture}(40,20)(0,0)
    \Line(5,10)(20,10)
    \Line(20,13)(35,18)\Line(20,7)(35,2)
    \dReg(20,10)
    \end{picture}} \\
    \raisebox{-7pt}{ \begin{picture}(45,20)(0,0)
    \Line(20,13)(35,18) \Line(20,7)(35,2)
    \Line(20,13)(10,13) \Line(20,7)(10,7) \CArc(10,10)(3,90,270)
    \dReg(20,10)
    \Text(38,19)[lt]{$i_1$}
    \Text(38,1)[lb]{$i_2$}
    \end{picture}}  & = &
            (C_R-\frac{1}{6}C_A)\ I_2(R)\ \delta^{i_1i_2} \\
    \raisebox{-7pt}{ \begin{picture}(45,20)(0,0)
    \Line(20,13)(35,18) \Line(20,7)(35,2) \Line(20,10)(35,10)
    \Line(20,13)(10,13) \Line(20,7)(10,7) \CArc(10,10)(3,90,270)
    \dReg(20,10)
    \end{picture}}  & = &
        (C_R-\frac{1}{4}C_A)
    \raisebox{-7pt}{ \begin{picture}(25,20)(0,0)
    \Line(5,10)(20,10)
    \Line(5,13)(20,18)\Line(5,7)(20,2)
    \dReg(5,10)
    \end{picture}} \\
\label{eq:d62}
    \raisebox{-7pt}{ \begin{picture}(45,20)(0,0)
    \Line(20,13)(35,18) \Line(20,7)(35,2)
    \Line(20,11)(35,12.7) \Line(20,9)(35,7.3)
    \Line(20,13)(10,13) \Line(20,7)(10,7) \CArc(10,10)(3,90,270)
    \dReg(20,10)
    \end{picture}}  & = &
        (C_R-\frac{1}{3}C_A)
    \raisebox{-7pt}{ \begin{picture}(25,20)(0,0)
    \Line(5,13)(20,18) \Line(5,7)(20,2)
    \Line(5,11)(20,12.7) \Line(5,9)(20,7.3)
    \dReg(5,10)
    \end{picture}} + \frac{1}{30} I_2(R)
    \raisebox{-7pt}{ \begin{picture}(25,20)(0,0)
    \Line(5,13)(20,18) \Line(5,7)(20,2)
    \Line(5,11)(20,12.7) \Line(5,9)(20,7.3)
    \dAdj(5,10)
    \end{picture}} \\
    \raisebox{-7pt}{ \begin{picture}(45,20)(0,0)
    \Line(20,13)(35,19) \Line(20,7)(35,1)
    \Line(20,11.5)(35,14.5) \Line(20,8.5)(35,5.5) \Line(20,10)(35,10)
    \Line(20,13)(10,13) \Line(20,7)(10,7) \CArc(10,10)(3,90,270)
    \dReg(20,10)
    \end{picture}}  & = &
        (C_R-\frac{5}{12}C_A)
    \raisebox{-7pt}{ \begin{picture}(25,20)(0,0)
    \Line(5,13)(20,19) \Line(5,7)(20,1)
    \Line(5,11.5)(20,14.5) \Line(5,8.5)(20,5.5) \Line(5,10)(20,10)
    \dReg(5,10)
    \end{picture}} + \frac{1}{12}
    \raisebox{-7pt}{ \begin{picture}(40,20)(0,0)
    \Line(20,13)(35,18) \Line(20,7)(35,2) \Line(5,10)(35,10)
    \Line(5,13)(20,18) \Line(5,7)(20,2)
    \dReg(5,10) \dAdj(20,10)
    \end{picture}} \\
\label{eq:d63}
    \raisebox{-7pt}{ \begin{picture}(40,20)(0,0)
    \Line(5,10)(35,10)
    \Line(5,13)(20,13)\Line(5,7)(20,7)
    \Line(20,13)(35,18)\Line(20,7)(35,2)
    \dReg(5,10) \dReg(20,10)
        \Text(5,4)[t]{\footnotesize 2}
        \Text(20,4)[t]{\footnotesize 1}
    \end{picture}}
 & = &
    -\frac{3}{8}C_A
    \raisebox{-7pt}{ \begin{picture}(55,20)(0,0)
    \Line(5,10)(20,10)
    \Line(20,13)(35,13)\Line(20,7)(35,7)
    \Line(35,13)(50,18)\Line(35,7)(50,2)
    \dReg(20,10) \dReg(35,10)
        \Text(20,4)[t]{\footnotesize 2}
        \Text(35,4)[t]{\footnotesize 1}
    \end{picture}}
    -\frac{1}{40}I_2(R_1)C_A^2
    \raisebox{-7pt}{ \begin{picture}(25,20)(0,0)
    \Line(5,10)(20,10)
    \Line(5,13)(20,18)\Line(5,7)(20,2)
    \dReg(5,10)
    \Text(5,4)[t]{\footnotesize 2}
    \end{picture}}
    +\frac{1}{N_{R_1}}
    \raisebox{-7pt}{ \begin{picture}(55,20)(0,0)
    \Line(5,13)(20,13) \Line(5,10)(20,10) \Line(5,7)(20,7)
    \dReg(5,10) \dReg(20,10)
        \Text(5,4)[t]{\footnotesize 2}
        \Text(20,4)[t]{\footnotesize 1}
    \Line(35,10)(50,10)
    \Line(35,13)(50,18)\Line(35,7)(50,2)
    \dReg(35,10)
        \Text(35,4)[t]{\footnotesize 1}
    \end{picture}} \\
\label{eq:d73}
    \raisebox{-7pt}{ \begin{picture}(40,20)(0,0)
    \Line(5,10)(20,10)
    \Line(5,13)(20,13)\Line(5,7)(20,7)
    \Line(20,13)(35,18)\Line(20,7)(35,2)
    \Line(20,11)(35,12.7)\Line(20,9)(35,7.3)
    \dReg(5,10) \dReg(20,10)
        \Text(5,4)[t]{\footnotesize 2}
        \Text(20,4)[t]{\footnotesize 1}
    \end{picture}}
 & = &
    -\frac{1}{2}C_A
    \raisebox{-7pt}{ \begin{picture}(55,20)(0,0)
    \Line(5,10)(20,10)
    \Line(20,13)(35,13)\Line(20,7)(35,7)
    \Line(35,13)(50,18)\Line(35,7)(50,2)\Line(35,10)(50,10)
    \dReg(20,10) \dReg(35,10)
        \Text(20,4)[t]{\footnotesize 2}
        \Text(35,4)[t]{\footnotesize 1}
    \end{picture}}
    +\frac{1}{N_{R_1}}
    \raisebox{-7pt}{ \begin{picture}(55,20)(0,0)
    \Line(5,13)(20,13) \Line(5,10)(20,10) \Line(5,7)(20,7)
    \dReg(5,10) \dReg(20,10)
        \Text(5,4)[t]{\footnotesize 2}
        \Text(20,4)[t]{\footnotesize 1}
    \Line(35,11)(50,12.7)\Line(35,9)(50,7.3)
    \Line(35,13)(50,18)\Line(35,7)(50,2)
    \dReg(35,10)
        \Text(35,4)[t]{\footnotesize 1}
    \end{picture}}
    -\frac{43}{120}\frac{1}{N_{A}}
    \raisebox{-7pt}{ \begin{picture}(55,20)(0,0)
    \Line(5,13)(20,13) \Line(5,10)(20,10) \Line(5,7)(20,7)
    \dReg(5,10) \dReg(20,10)
        \Text(5,4)[t]{\footnotesize 2}
        \Text(20,4)[t]{\footnotesize 1}
    \Line(35,11)(50,12.7)\Line(35,9)(50,7.3)
    \Line(35,13)(50,18)\Line(35,7)(50,2)
    \dAdj(35,10)
    \end{picture}} \nonumber \\ & &
    -\frac{23}{1440}C_A^2
    \raisebox{-7pt}{ \begin{picture}(55,20)(0,0)
    \Line(5,18)(20,13)\Line(5,2)(20,7)
    \Line(20,10)(35,10)
    \Line(35,13)(50,18)\Line(35,7)(50,2)
    \dReg(20,10) \dReg(35,10)
        \Text(20,4)[t]{\footnotesize 2}
        \Text(35,4)[t]{\footnotesize 1}
    \end{picture}}
    +\frac{49}{60}
    \raisebox{-12pt}{ \begin{picture}(70,30)(0,0)
        \Line(35,25)(20,5) \Line(5,5)(20,5)
        \Line(35,25)(50,5) \Line(50,5)(65,5)
        \Line(20,5)(50,5)  \Line(20,25)(50,25)
        \dReg(20,5) \dReg(50,5) \dAdj(35,25)
        \Text(17,3)[rt]{\footnotesize 1}
        \Text(53,3)[lt]{\footnotesize 2}
    \end{picture}}
\end{eqnarray}
We assume symmetrization over the external legs in the terms of the 
right hand sides.

One of the spin-offs of equation (\ref{eq:d63}) is
\begin{eqnarray}
\label{eq:d363}
    \raisebox{-7pt}{ \begin{picture}(40,20)(0,0)
    \Line(5,10)(35,10) \Line(5,13)(35,13) \Line(5,7)(35,7)
    \dReg(5,10) \dReg(20,10) \dReg(35,10)
        \Text(5,4)[t]{\footnotesize 2}
        \Text(20,4)[t]{\footnotesize 1}
        \Text(35,4)[t]{\footnotesize 3}
    \end{picture}}
 & = &
    -\frac{3}{8}C_A
    \raisebox{-12pt}{ \begin{picture}(40,30)(0,0)
        \Line(20,27)(5,7) \Line(20,23)(5,3)
        \Line(20,27)(35,7) \Line(20,23)(35,3)
        \Line(5,5)(35,5)
        \dReg(5,5) \dReg(35,5) \dReg(20,25)
        \Text(16,26)[rb]{\footnotesize 1}
        \Text(2,3)[rt]{\footnotesize 2}
        \Text(38,3)[lt]{\footnotesize 3}
    \end{picture}}
    -\frac{1}{40}I_2(R_1)C_A^2
    \raisebox{-7pt}{ \begin{picture}(25,20)(0,0)
    \Line(5,10)(20,10)
    \Line(5,13)(20,13)\Line(5,7)(20,7)
    \dReg(5,10)\dReg(20,10)
    \Text(5,4)[t]{\footnotesize 2}
    \Text(20,4)[t]{\footnotesize 3}
    \end{picture}}
    +\frac{1}{N_{R_1}}
    \raisebox{-7pt}{ \begin{picture}(55,20)(0,0)
    \Line(5,13)(20,13) \Line(5,10)(20,10) \Line(5,7)(20,7)
    \dReg(5,10) \dReg(20,10)
        \Text(5,4)[t]{\footnotesize 2}
        \Text(20,4)[t]{\footnotesize 1}
    \Line(35,10)(50,10)
    \Line(35,13)(50,13)\Line(35,7)(50,7)
    \dReg(35,10) \dReg(50,10)
        \Text(35,4)[t]{\footnotesize 1}
        \Text(50,4)[t]{\footnotesize 3}
    \end{picture}}
\end{eqnarray}
This allows us to eliminate the topology $d_{633}$ completely. For groups 
for which $d^{ijk}$ exists and $I_4(A)$ is not zero we can use the 
techniques of the next section to also eliminate the topologies $d_{433}$, 
$d_{4433a}$ and $d_{4433c}$. Similarly equation (\ref{eq:d73}) gives
\begin{eqnarray}
    \raisebox{-7pt}{ \begin{picture}(60,20)(0,0)
    \Line(5,13.5)(30,13.5) \Line(5,6.5)(30,6.5)
    \Line(30,13)(55,13)\Line(30,10)(55,10)\Line(30,7)(55,7)
    \dReg(5,10) \dReg(30,10) \dReg(55,10)
    \Text(17.5,10.5)[]{\footnotesize 4}
        \Text(5,5)[t]{\footnotesize 3}
        \Text(30,5)[t]{\footnotesize 1}
        \Text(55,5)[t]{\footnotesize 2}
    \end{picture}} & = & -\frac{1}{2}C_A
    \raisebox{-12pt}{ \begin{picture}(52,35)(0,0)
        \Line(25,25)(5,5) \Line(25,28)(5,8) \Line(25,22)(5,2)
        \Line(25,27)(45,7) \Line(25,23)(45,3)
        \Line(5,5)(45,5)
        \dReg(5,5) \dReg(45,5) \dReg(25,25)
        \Text(22,26)[rb]{\footnotesize 1}
        \Text(2,3)[rt]{\footnotesize 3}
        \Text(48,3)[lt]{\footnotesize 2}
    \end{picture}} + \frac{1}{N_{R_1}}
    \raisebox{-7pt}{ \begin{picture}(55,20)(0,0)
    \Line(5,13)(20,13) \Line(5,10)(20,10) \Line(5,7)(20,7)
    \dReg(5,10) \dReg(20,10)
        \Text(5,4)[t]{\footnotesize 2}
        \Text(20,4)[t]{\footnotesize 1}
    \Line(35,13.5)(50,13.5)\Line(35,6.5)(50,6.5)
    \dReg(35,10) \dReg(50,10)
    \Text(42.5,10.5)[]{\footnotesize 4}
        \Text(35,4)[t]{\footnotesize 1}
        \Text(50,4)[t]{\footnotesize 3}
    \end{picture}} -\frac{43}{120}\frac{1}{N_A}
    \raisebox{-7pt}{ \begin{picture}(55,20)(0,0)
    \Line(5,13)(20,13) \Line(5,10)(20,10) \Line(5,7)(20,7)
    \dReg(5,10) \dReg(20,10)
        \Text(5,4)[t]{\footnotesize 2}
        \Text(20,4)[t]{\footnotesize 1}
    \Line(35,13.5)(50,13.5)\Line(35,6.5)(50,6.5)
    \dAdj(35,10) \dReg(50,10)
    \Text(42.5,10.5)[]{\footnotesize 4}
        \Text(50,4)[t]{\footnotesize 3}
    \end{picture}} \nonumber \\ &&
    -\frac{23}{1440}C_A^2
    \raisebox{-12pt}{ \begin{picture}(40,30)(0,0)
        \Line(20,27)(5,7) \Line(20,23)(5,3)
        \Line(20,27)(35,7) \Line(20,23)(35,3)
        \Line(5,5)(35,5)
        \dReg(5,5) \dReg(35,5) \dReg(20,25)
        \Text(16,26)[rb]{\footnotesize 3}
        \Text(2,3)[rt]{\footnotesize 1}
        \Text(38,3)[lt]{\footnotesize 2}
    \end{picture}} + \frac{49}{60}
    \raisebox{-16pt}{ \begin{picture}(45,40)(0,0)
        \Line(10,5)(35,5) \Line(10,5)(10,30) \Line(10,5)(35,30)
        \Line(10,30)(35,5) \Line(35,5)(35,30)
        \Line(10,28)(35,28) \Line(10,32)(35,32) 
        \dReg(10,5) \dReg(35,5) \dAdj(10,30) \dReg(35,30)
        \Text(7,3)[rt]{\footnotesize 1}
        \Text(38,3)[lt]{\footnotesize 2}
        \Text(38,32)[lb]{\footnotesize 3}
    \end{picture}}
\end{eqnarray}
and therefore also the topology $d_{743}$ can be eliminated.

An equation that is also rather interesting is
\begin{eqnarray}
\label{eq:dA4dA6}
    \raisebox{-7pt}{ \begin{picture}(55,20)(0,0)
    \Line(5,10)(50,10)
    \Line(20,13)(35,13)\Line(20,7)(35,7)
    \Line(35,13)(50,18)\Line(35,7)(50,2)
    \dAdj(20,10) \dAdj(35,10)
        \Text(5,6)[t]{\footnotesize 1}
    \end{picture}}
 & = &
	-\frac{191}{864}C_A^3
    \raisebox{-7pt}{ \begin{picture}(40,20)(0,0)
    \Line(5,10)(35,10)
    \Line(20,13)(35,18)\Line(20,7)(35,2)
    \dAdj(20,10)
        \Text(5,6)[t]{\footnotesize 1}
    \end{picture}}
	+\frac{9}{5}C_A
    \raisebox{-7pt}{ \begin{picture}(55,20)(0,0)
		\Line(20,8)(5,5)\Line(20,12)(5,15)
		\Line(35,8)(50,5)\Line(35,12)(50,15)
		\Line(20,8)(35,8) \Line(20,12)(35,12)
		\dAdj(20,10) \dAdj(35,10)
        \Text(4,5)[r]{\footnotesize 1}
    \end{picture}}
	-\frac{119}{288}C_A^2
    \raisebox{-7pt}{ \begin{picture}(55,20)(0,0)
		\Line(20,8)(5,5)\Line(20,12)(5,15)
		\Line(20,8)(35,3) \Line(20,12)(35,17) \Line(35,3)(35,17)
		\Line(35,3)(50,3)\Line(35,17)(50,17)
		\dAdj(20,10) \freg(35,17) \freg(35,3)
        \Text(4,5)[r]{\footnotesize 1}
    \end{picture}}  \nonumber \\ &&
	+\frac{3}{2}
    \raisebox{-7pt}{ \begin{picture}(70,20)(0,0)
		\Line(20,8)(5,5)\Line(20,12)(5,15)
		\Line(20,8)(35,3) \Line(20,12)(35,17) \Line(35,3)(35,17)
		\Line(35,3)(50,8)\Line(35,17)(50,12)
		\Line(50,8)(65,5)\Line(50,12)(65,15)
		\dAdj(20,10) \freg(35,17) \freg(35,3) \dAdj(50,10)
        \Text(4,5)[r]{\footnotesize 1}
    \end{picture}}
	+\frac{3}{4}
    \raisebox{-7pt}{ \begin{picture}(85,30)(0,0)
		\Line(5,10)(80,10) \Line(20,7)(65,7)
		\Line(20,13)(35,18)\Line(35,18)(50,18)\Line(50,18)(65,13)
		\Line(35,18)(35,30)\Line(50,18)(50,30)
		\dAdj(20,10) \freg(35,18) \freg(50,18) \dAdj(65,10)
        \Text(5,6)[t]{\footnotesize 1}
    \end{picture}}
\end{eqnarray}
It can be used to derive equation (\ref{eq:7loops}), but it can also be useful 
for bigger diagrams. It is derived by writing a diagram with 10 $f$'s into 
loops in two different ways. After that the application of the reduction 
algorithms and some rewriting leads to this formula. 



\section{Representation independent invariants}

The invariants that we have used thus far were only symmetrized traces. It 
is possible to define a new set of invariants that is not only symmetric, 
but also orthogonal (see equation (\ref{eq:ortho})). Because the invariant 
with two indices is proportional to the Kronecker delta in the adjoint 
space, this means for instance that these invariants have a zero trace (a 
contraction of any two indices gives zero). But also contractions with all 
the indices of invariants with fewer indices than the invariant under study 
should give zero. 
For most algebras we can define (the exceptions are certain
$SO(4N)$ algebras with {\it two} independent tensors of order $2N$):
\begin{eqnarray}
    d_R^{i_1i_2i_3} & = & I_3(R)\ d_{\ortho}^{i_1i_2i_3} \\
\label{eq:d4}
    d_R^{i_1i_2i_3i_4} & = & I_4(R)\ d_{\ortho}^{i_1i_2i_3i_4}
            +I_{2,2}(R)\ (\delta^{i_1i_2}\delta^{i_3i_4}
                +\delta^{i_1i_3}\delta^{i_2i_4}
                +\delta^{i_1i_4}\delta^{i_2i_3} )/3 \\
    d_R^{i_1i_2i_3i_4i_5} & = & I_5(R)\ d_{\ortho}^{i_1i_2i_3i_4i_5}
            +I_{3,2}(R)\ (d_{\ortho}^{i_1i_2i_3}\delta^{i_4i_5}+\cdots)/10 \\
\label{eq:d6}
    d_R^{i_1i_2i_3i_4i_5i_6} & = & I_6(R)\ d_{\ortho}^{i_1i_2i_3i_4i_5i_6}
            +I_{4,2}(R)\ (d_{\ortho}^{i_1i_2i_3i_4}\delta^{i_5i_6}+\cdots)/15
                            \nonumber \\ &&
            +I_{3,3}(R)\ (d_{\ortho}^{i_1i_2i_3}d_{\ortho}^{i_4i_5i_6}+\cdots)/10
            +I_{2,2,2}(R)
\ (\delta^{i_1i_2}\delta^{i_3i_4}\delta^{i_5i_6}
                    +\cdots)/15 \\
\label{eq:d7}
    d_R^{i_1i_2i_3i_4i_5i_6i_7} & = 
& I_7(R)\ d_{\ortho}^{i_1i_2i_3i_4i_5i_6i_7}
            +I_{5,2}(R)\ 
(d_{\ortho}^{i_1i_2i_3i_4i_5}\delta^{i_6i_7}+\cdots)/21
                            \nonumber \\ &&
            +I_{4,3}(R)\ (d_{\ortho}^{i_1i_2i_3i_4}
d_{\ortho}^{i_5i_6i_7}+\cdots)/35
            +I_{3,2,2}(R)\ (d_{\ortho}^{i_1i_2i_3}
\delta^{i_4i_5}\delta^{i_6i_7}
                    +\cdots)/105
\end{eqnarray}
and higher ones are defined analogously. The composite constants like 
$I_{2,2}(R)$ can now be derived by the orthogonality conditions. When we 
multiply equation (\ref{eq:d4}) by $\delta^{i_3i_4}$ we obtain
\begin{eqnarray}
    \frac{N_A+2}{3}I_{2,2}(R)\delta^{i_1i_2} & = & d_R^{i_1i_2i_3i_3} \nonumber \\
    & = & (C_R-C_A/6)d_R^{i_1i_2} \nonumber \\
    & = & (C_R-C_A/6)I_2(R)\delta^{i_1i_2} \nonumber \\
    \frac{N_A+2}{3}I_{2,2}(R) & = & (I_2(R)\frac{N_A}{N_R} - \frac{1}{6}I_2(A))I_2(R)
\end{eqnarray}
Similarly we can derive
\begin{eqnarray}
I_{3,2}(R) & = & \frac{10}{N_A+6}(C_R-\frac{1}{4}C_A) I_3(R)
\end{eqnarray}
For the next level of constants we get coupled equations. The easiest way 
to derive them is with the use of the equations (\ref{eq:d62}) and 
(\ref{eq:d363}). If we contract the first equation once with 
$d_{\ortho}^{i_1i_2i_3i_4}$ and once with $\delta^{i_1i_2}\delta^{i_3i_4}$ and 
substitute equation (\ref{eq:d6}) in all three of the equations that we obtain 
this way we have enough to come to a solution. Unfortunately this solution 
is not very elegant. In many cases one can make simplifications. For 
instance one can write (using equation (\ref{eq:reduA4}))
\begin{eqnarray}
    \raisebox{-12pt}{ \begin{picture}(40,30)(0,0)
        \Line(20,27)(5,7) \Line(20,23)(5,3)
        \Line(20,27)(35,7) \Line(20,23)(35,3)
        \Line(5,5)(35,5)
        \dEmp(5,5) \dEmp(35,5) \dAdj(20,25)
    \end{picture}} & = & \frac{1}{6}C_A^2
    \raisebox{-7pt}{ \begin{picture}(40,20)(0,0)
    \Line(5,13.5)(35,13.5) \Line(5,10.0)(35,10.0) \Line(5,6.5)(35,6.5)
    \dEmp(5,10) \dEmp(35,10)
    \end{picture}} \nonumber \\ & = &
    I_4(A)
    \raisebox{-12pt}{ \begin{picture}(40,30)(0,0)
        \Line(20,27)(5,7) \Line(20,23)(5,3)
        \Line(20,27)(35,7) \Line(20,23)(35,3)
        \Line(5,5)(35,5) \dEmp(5,5) \dEmp(35,5) \dEmp(20,25)
    \end{picture}}
        + \frac{2}{3}I_{2,2}(A)
    \raisebox{-7pt}{ \begin{picture}(40,20)(0,0)
    \Line(5,13.5)(35,13.5) \Line(5,10.0)(35,10.0) \Line(5,6.5)(35,6.5)
    \dEmp(5,10) \dEmp(35,10)
    \end{picture}}
\end{eqnarray}
which then reduces to
\begin{eqnarray}
\label{eq:simsun}
    I_4(A)\raisebox{-12pt}{ \begin{picture}(40,30)(0,0)
        \Line(20,27)(5,7) \Line(20,23)(5,3)
        \Line(20,27)(35,7) \Line(20,23)(35,3)
        \Line(5,5)(35,5)
        \dEmp(5,5) \dEmp(35,5) \dEmp(20,25)
    \end{picture}} & = & \frac{1}{6}C_A^2 \frac{N_A-8}{N_A+2}
    \raisebox{-7pt}{ \begin{picture}(40,20)(0,0)
    \Line(5,13.5)(35,13.5) \Line(5,10.0)(35,10.0) \Line(5,6.5)(35,6.5)
    \dEmp(5,10) \dEmp(35,10)
    \end{picture}}
\end{eqnarray}
We notice that whenever $I_4(A) = 0$ or $d_{\ortho}^{i_1i_2i_3i_4} = 0$ we must have 
that either $ \raisebox{-7pt}{ \begin{picture}(40,20)(0,0)
    \Line(5,13.5)(35,13.5) \Line(5,10.0)(35,10.0) \Line(5,6.5)(35,6.5)
    \dEmp(5,10) \dEmp(35,10)
    \end{picture}} = 0 $ or $N_A = 8$ (assuming that $C_A$ is never zero).
The last is indeed the case for $SU(3)$. Hence a general application of this 
relation is rather dangerous.

The equations for $I_{4,2}(R)$, $I_{3,3}(R)$ and $I_{2,2,2}(R)$ can be 
derived by taking the definition of the decomposition in equation 
(\ref{eq:d6}) and multiplying either by
$\delta^{i_1i_2}\delta^{i_3i_4}\delta^{i_5i_6}$ or
$d_{\ortho}^{i_1i_2i_3i_4}\delta^{i_5i_6}$ or 
$d_{\ortho}^{i_1i_2i_3}d_{\ortho}^{i_4i_5i_6}$.
With the use of the equations (\ref{eq:d62}) and 
(\ref{eq:d63}) we obtain:
\begin{eqnarray}
	0 & = &
    \raisebox{-7pt}{ \begin{picture}(40,20)(0,0)
    \Line(5,13.5)(35,13.5) \Line(5,10.0)(35,10.0) \Line(5,6.5)(35,6.5)
    \dEmp(5,10) \dEmp(35,10)
    \end{picture}} \ ( \frac{2}{5}I_{3,3}(R) )
		\nonumber \\ &&
	+N_A(N_A+2)(
		\frac{N_A+4}{15}I_{2,2,2}(R)
		-\frac{1}{3}(C_R-\frac{1}{3}C_A)I_{2,2}(R)
		-\frac{1}{90}I_{2,2}(A)I_2(R) )
		\\
	0 & = &
    \raisebox{-12pt}{ \begin{picture}(40,30)(0,0)
        \Line(20,27)(5,7) \Line(20,23)(5,3)
        \Line(20,27)(35,7) \Line(20,23)(35,3)
        \Line(5,5)(35,5)
        \dEmp(5,5) \dEmp(35,5) \dEmp(20,25)
    \end{picture}}\ (\frac{3}{5}I_{3,3}(R))
		\nonumber \\ &&
	+ \raisebox{-7pt}{ \begin{picture}(40,20)(0,0)
    \Line(5,13.5)(35,13.5) \Line(5,11.2)(35,11.2)
	\Line(5,8.8)(35,8.8) \Line(5,6.5)(35,6.5)
    \dEmp(5,10) \dEmp(35,10)
    \end{picture}} \ (
		\frac{N_A+8}{15}I_{4,2}(R)
		-(C_R-\frac{1}{3}C_A)I_4(R)
		-\frac{1}{30}I_4(A)I_2(R)
	) \\
	0 & = &
    \raisebox{-12pt}{ \begin{picture}(40,30)(0,0)
        \Line(20,27)(5,7) \Line(20,23)(5,3)
        \Line(20,27)(35,7) \Line(20,23)(35,3)
        \Line(5,5)(35,5)
        \dEmp(5,5) \dEmp(35,5) \dEmp(20,25)
    \end{picture}}\ (\frac{3}{5}I_{4,2}(R)+\frac{3}{8}C_AI_4(R))
		\nonumber \\ &&
	+ ( \raisebox{-7pt}{ \begin{picture}(40,20)(0,0)
    \Line(5,13.5)(35,13.5) \Line(5,10.0)(35,10.0) \Line(5,6.5)(35,6.5)
    \dEmp(5,10) \dEmp(35,10)
    \end{picture}}\ )^2 (
		\frac{N_A+9}{10N_A}I_{3,3}(R)
		-\frac{1}{N_R}(I_3(R))^2
	)  \nonumber \\ &&
    +\raisebox{-7pt}{ \begin{picture}(40,20)(0,0)
    \Line(5,13.5)(35,13.5) \Line(5,10.0)(35,10.0) \Line(5,6.5)(35,6.5)
    \dEmp(5,10) \dEmp(35,10)
    \end{picture}}\ (
		\frac{2}{5}I_{2,2,2}(R)
		+\frac{1}{4}C_AI_{2,2}(R)
		+\frac{1}{4}C_A^2I_2(R)
	)
\end{eqnarray}
Because for each group either some of the objects in these equations are zero,
or there are simplifications, it does not seem wise to solve this system in 
this form. For all groups with the exception of $SU(N)$ we have that 
$d_{\ortho}^{i_1i_2i_3} = 0$, and hence the system reduces to two equations in 
$I_{4,2}(R)$ and $I_{2,2,2}(R)$. For $SU(3)$ we have that $d_{\ortho}^{i_1i_2i_3i_4} 
= 0$ and again we have a simpler system. For the other $SU(N)$ groups we 
can apply equation (\ref{eq:simsun}) improving the solutions somewhat. 
Because of the singularity (zero divided by zero) of the solutions for all 
groups but $SU(N), N > 3$, we do not present the general solution here. It 
serves no purpose because we will not use them.
Along the same lines one can derive the equations for
 $I_{5,2}(R)$, $I_{4,3}(R)$ and $I_{3,2,2}(R)$.

The expressions at rank 6 and 7 become rather complicated due to the fact 
that the tensors in the right hand side of the equations (\ref{eq:d6}) and 
(\ref{eq:d7}) are not orthogonal. Hence the various orthogonality relations 
mix and this gives the complicated result. It should be clear that 
invariants with a higher rank will give even more complicated relations. 
The exception is the adjoint representation. For this representation all 
invariants with an odd number of indices are zero. Hence $I_3(A) = I_5(A) = 
I_{3,2}(A) = \cdots = 0$.

We can use the above relations for the reduction of some contractions of 
invariants. We have seen in the previous section that for the adjoint 
representation more things are possible than for the other representations. 
However we can rewrite the invariants only to invariants of the adjoint 
representation when the corresponding $d$'s can indeed be expressed as 
such. This means that the corresponding $I(A)$ should never be zero. 
Unfortunately this excludes many contractions of invariants. Hence we do 
not see many benefits here.


\subsection{Orthogonal versus Reference tensors}

Here we will compare the computation of a symmetrized trace in
two ways, using the orthogonal basis (satisfying (\ref{eq:ortho}))
and using characters, with
tensors defined for a reference representation. The latter 
will be referred to as ``reference tensors". 
We consider fourth
order traces in $SU(N)$. In this case an explicit expression exists
for any representation
\begin{equation}
\label{eq:fourtr}
 \Str T_R^a T_R^b T_R^c T_R^d = I_4(R) d_{\ortho}^{abcd} 
     + {3\over N_A+2}I_2(R)^2 \left[{N_A\over N_R}
-\frac{1}{6} {I_2(A)\over I_2(R)}\right] d_{2,2}^{abcd} \ ,
\end{equation}
where 
\begin{equation}
 d_{2,2}^{abcd} \equiv \frac{1}{3} ( \delta^{ab}\delta^{cd}
   +\delta^{ac}\delta^{bd}
    + \delta^{ad}\delta^{bc} ) \ . 
\end{equation}
Here ``$d$" without subscript denotes the orthogonal tensor. 
The reference representation is the vector, and the reference
fourth rank tensor is by definition equal to (\ref{eq:fourtr}) with ``$R$"
equal to the reference representation:
\begin{equation}
\label{eq:refdef}
 d_r^{abcd} = d_{\ortho}^{abcd} + {3\over N_A+2} \left(
	{\frac{2}{3}N^2-1 \over N}\right) d_{2,2}^{abcd}
\end{equation}
(Note that $d_{2,2}$ is the same in both bases). This allows
us to express (\ref{eq:fourtr}) in terms of reference rather than orthogonal
tensors:
\begin{eqnarray}
\label{eq:fourtrref}
   \Str T_R^a T_R^b T_R^c T_R^d &=& I_4(R) d_r^{abcd} \nn
     + {3\over N_A+2}\left( I_2(R)^2 \left[{N_A\over N_R}
     -\frac{1}{6} {I_2(A)\over I_2(R)}\right] - {\frac{2}{3} N^2-1\over N}
   \right)  d_{2,2}^{abcd}
\end{eqnarray}

Consider now for example 
the anti-symmetric tensor representation. Its character,
expanded to fourth order is
\begin{eqnarray}
\Ch_{[2]}(F)&=&\half N(N-1)+\half (N-2) \Tr F^2 + \frac{1}{6} (N-4) \Tr F^3 
\nn
    + \frac{1}{24} (N-8) \Tr F^4 + \frac{1}{8} \Tr F^2 \Tr F^2 + \ldots \ ,
\end{eqnarray}
where all traces are over the vector representation. From
the fourth order terms we deduce, by differentiating with respect to
$F^a,\ldots,F^b$:
\begin{equation}
\label{eq:charres}
 \Tr T^a_{[2]} T^b_{[2]}T^c_{[2]}T^d_{[2]} = ( N-8 ) d_r^{abcd}
+ 3 d_{2,2}^{abcd}
\end{equation}
We may now verify this using (\ref{eq:fourtrref}). Although the coefficient of
the second term does not look very encouraging, substituting $I_2(R)=N-2$,
$N_A=N^2-1$ and $N_R=\half N(N-1)$, it does indeed produce the
coefficient 3 in (\ref{eq:charres}). 

This illustrates several points. Trace formulas in terms of orthogonal
tensors such as (\ref{eq:fourtr}) have a simpler form than those in terms of
reference tensors, if one tries to write down expressions for
arbitrary representations $R$. However, expressions such as 
(\ref{eq:charres})  
can be written down fairly easily for any representation although
not (easily) in closed form. Furthermore they can be extended to 
arbitrary order in a straightforward way while this rapidly
becomes extremely difficult for (\ref{eq:fourtr}) or (\ref{eq:fourtrref}).  

Note that the indices (the coefficients of fundamental tensors)
are basis independent (apart from normalizations), whereas
the sub-indices (coefficients of combinations of fundamental
tensors) are not. In the orthogonal basis it is not
hard to see that all sub-indices can in fact be expressed in terms
of indices, so that they do not constitute an additional set
of variables. In any reference basis the same is then true, since
it can be related to an orthogonal basis, but the expression are
(even) more complicated, as in (\ref{eq:fourtrref}). For orders larger
than six, expression of sub-indices in terms of indices are
not available and hard to obtain.  

In our method the sub-indices are essentially treated as 
additional variables, which can be computed for any 
representation as easily as the indices themselves. This allows
the computation of a trace for any representation, which was our
goal.
The result
is a combination of symmetric fundamental tensors with 
explicit numerical
coefficients, as in (\ref{eq:charres}), or an expression involving both
indices and sub-indices.  
Unfortunately it is much harder to present the result
in minimal form, with all representation dependence encapsulated in 
the indices. 

\appendix
\section{Indices and Trace identities for Exceptional Algebras}

In this appendix we summarize our results on traces for
exceptional algebras. All the indices for the lowest dimensional
representations are given, including all basic representations.
Trace identities are given for all traces of order less than
the dual Coxeter number $g$.

For the classical algebras $A\ldots D$ the trace identities were
already given in chapter 6, and index formulas
for some representations
are given in \cite{patera}. 

The indices provide only a small part of the information
contained in the full characters, but it is impractical to
present the latter in printed form. We do have an efficient procedure
to generate the characters of any representation of any simple
Lie-algebra to any desired order. This procedure uses a combination
of {\tt Kac} \cite{Kac} (to compute tensor products) and {\tt FORM}
\cite{form} (to multiply,
add and subtract characters according to these tensor products), and 
is available via http://norma.nikhef.nl/$\sim$t58.
Obviously this then also provides all indices for algebras and 
representations not listed in this appendix.   

\subsection{Indices and trace identities for $G_2$}
Trace identity in the representation (7):
\begin{equation}
   \Tr F^4=\frac{1}{4} (\Tr F^2)^2
\end{equation}
The indices of the lowest-dimensional representations are shown in 
table 5.
\begin{table}[htb]
\centering
\begin{tabular}{|c|c|c|c|}
\hline
   Rep. & Dimension     & ${I_2 \over 2}$ & ${I_6}$          \\ \hline
  (0,1)
\intab(7) 
\intab(1) 
\intab(1) 
\\
  (1,0)
\intab(14) 
\intab(4) 
\intab(-26) 
\\
  (0,2)
\intab(27) 
\intab(9) 
\intab(39) 
\\
  (1,1)
\intab(64) 
\intab(32) 
\intab(-208) 
\\
  (0,3)
\intab(77) 
\intab(44) 
\intab(494) 
\\
  (2,0)
\intab(77) 
\intab(55) 
\intab(-1235) 
\\
  (0,4)
\intab(182) 
\intab(156) 
\intab(3666) 
\\
  (1,2)
\intab(189) 
\intab(144) 
\intab(-456) 
\\
  (3,0)
\intab(273) 
\intab(351) 
\intab(-20709) 
\\
  (2,1)
\intab(286) 
\intab(286) 
\intab(-7904) 
\\
  (0,5)
\intab(378) 
\intab(450) 
\intab(19500) 
\\
  (1,3)
\intab(448) 
\intab(480) 
\intab(2640) 
\\
  (0,6)
\intab(714) 
\intab(1122) 
\intab(82212) 
\\
  (2,2)
\intab(729) 
\intab(972) 
\intab(-27378) 
\\
  (4,0)
\intab(748) 
\intab(1496) 
\intab(-193324) 
\\
  (3,1)
\intab(896) 
\intab(1472) 
\intab(-109408) 
\\
\hline
\end{tabular}
\caption{\sl Indices for $G_2$.}
\end{table}

\subsection{Indices and trace identities for $F_4$}
Trace identities in the representation (26):
\begin{equation}
 \Tr F^4=3 (\frac{1}{6} \Tr F^2)^2
\end{equation}
\begin{equation}
 \Tr F^{10}= \frac{9}{4} (\frac{1}{6} \Tr F^2) (\Tr F^8) 
- \frac{7}{4} (\frac{1}{6} \Tr F^2)^2 (\Tr F^6) 
+ \frac{21}{16}
         (\frac{1}{6} \Tr F^2)^5
\end{equation}
The indices are listed in 
table 6.
\begin{table}[htb]
\centering
\begin{tabular}{|c|c|c|c|c|c|}
\hline
Rep. & Dimension & ${I_2 \over 6}$ & ${I_6}$ & ${I_8}$ & $I_{12}$ \\ \hline
  (0,0,0,1)
\intab(26) 
\intab(1) 
\intab(1) 
\intab(1) 
\intab(1) 
\\
  (1,0,0,0)
\intab(52) 
\intab(3) 
\intab(-7) 
\intab(17) 
\intab(-63) 
\\
  (0,0,1,0)
\intab(273) 
\intab(21) 
\intab(1) 
\intab(-119) 
\intab(-1959) 
\\
  (0,0,0,2)
\intab(324) 
\intab(27) 
\intab(57) 
\intab(153) 
\intab(2073) 
\\
  (1,0,0,1)
\intab(1053) 
\intab(108) 
\intab(-132) 
\intab(612) 
\intab(372) 
\\
  (2,0,0,0)
\intab(1053) 
\intab(135) 
\intab(-645) 
\intab(2907) 
\intab(-134373) 
\\
  (0,1,0,0)
\intab(1274) 
\intab(147) 
\intab(-133) 
\intab(-1309) 
\intab(125811) 
\\
 \hline
\end{tabular}
\caption{\sl Indices for $F_4$.}
\end{table}

\subsection{Indices and trace identities for $E_6$}
Trace identities in the representation (56):
\begin{equation}
 \Tr F^4={12} (\frac{1}{12} \Tr F^2)^2
\end{equation}
\begin{equation}
 \Tr F^7=\frac{7}{2} (\Tr F^5)(\frac{1}{12} \Tr F^2)  
\end{equation}
\begin{equation}
  \Tr F^{10}=
  \frac{9}{2} (\Tr F^8) (\frac{1}{12} \Tr F^2)
 -7 (\Tr F^6) (\frac{1}{12} \Tr F^2)^2
 +\frac{7}{40} (\Tr F^5)^2+42 (\frac{1}{12} \Tr F^2)^5
\end{equation}
\begin{equation}
 \Tr F^{11}=\frac{11}{36}(\Tr F^6)(\Tr F^5)+
\frac{605}{126} (\Tr F^9) (\frac{1}{12} \Tr F^2)-\frac{55}{2} (\Tr F^5)
(\frac{1}{12} \Tr F^2)^3
\end{equation}
The indices are in table 7.
\begin{table}[htb]
\centering
\begin{tabular}{|c|c|c|c|c|c|c|c|}
\hline
Rep. & Dimension & ${I_2\over 6}$ & ${I_5}$ & $I_6$ & $I_{8}$ 
         & $I_{9}$ & $I_{12}$  \\ \hline
  (1,0,0,0,0,0)
\intab(27) 
\intab(1) 
\intab(1) 
\intab(1) 
\intab(1) 
\intab(1) 
\intab(1) 
\\
  (0,0,0,0,1,0)
\intab(27) 
\intab(1) 
\intab(-1) 
\intab(1) 
\intab(1) 
\intab(-1) 
\intab(1) 
\\
  (0,0,0,0,0,1)
\intab(78) 
\intab(4) 
\intab(0) 
\intab(-6) 
\intab(18) 
\intab(0) 
\intab(-62) 
\\
  (0,1,0,0,0,0)
\intab(351) 
\intab(25) 
\intab(11) 
\intab(-5) 
\intab(-101) 
\intab(-229) 
\intab(-2021) 
\\
  (0,0,0,1,0,0)
\intab(351) 
\intab(25) 
\intab(-11) 
\intab(-5) 
\intab(-101) 
\intab(229) 
\intab(-2021) 
\\
  (0,0,0,0,2,0)
\intab(351) 
\intab(28) 
\intab(-44) 
\intab(58) 
\intab(154) 
\intab(-284) 
\intab(2074) 
\\
  (2,0,0,0,0,0)
\intab(351) 
\intab(28) 
\intab(44) 
\intab(58) 
\intab(154) 
\intab(284) 
\intab(2074) 
\\
  (1,0,0,0,1,0)
\intab(650) 
\intab(50) 
\intab(0) 
\intab(60) 
\intab(36) 
\intab(0) 
\intab(116) 
\\
  (0,0,0,0,1,1)
\intab(1728) 
\intab(160) 
\intab(-88) 
\intab(-80) 
\intab(664) 
\intab(152) 
\intab(424) 
\\
  (1,0,0,0,0,1)
\intab(1728) 
\intab(160) 
\intab(88) 
\intab(-80) 
\intab(664) 
\intab(-152) 
\intab(424) 
\\
  (0,0,0,0,0,2)
\intab(2430) 
\intab(270) 
\intab(0) 
\intab(-720) 
\intab(3672) 
\intab(0) 
\intab(-131928) 
\\
  (0,0,1,0,0,0)
\intab(2925) 
\intab(300) 
\intab(0) 
\intab(-270) 
\intab(-918) 
\intab(0) 
\intab(122202) 
\\
\hline
\end{tabular}
\caption{\sl Indices for $E_6$.}
\end{table}

\subsection{Indices and trace identities for $E_7$}
Trace identities in the representation (56):
\begin{equation}
 \Tr F^4={24} (\frac{1}{24} \Tr F^2)^2
\end{equation}
\begin{eqnarray}
 \Tr  F^{16} & = &
           -\frac{8567}{5220} (\frac{1}{24} \Tr F^2)  (\Tr F^6)  (\Tr F^8) 
          \nn +\frac{2360}{319} (\frac{1}{24} \Tr F^2)  (\Tr F^{14}) 
          \nn +\frac{61607}{23490} (\frac{1}{24} \Tr F^2) ^2 (\Tr F^6) ^2
          \nn -\frac{63700}{2871} (\frac{1}{24} \Tr F^2) ^2 (\Tr F^{12}) 
          \nn +\frac{21164}{783} (\frac{1}{24} \Tr F^2) ^3 (\Tr F^{10}) 
          \nn +\frac{7397}{522} (\frac{1}{24} \Tr F^2) ^4 (\Tr F^8) 
          \nn -\frac{72254}{783} (\frac{1}{24} \Tr F^2) ^5 (\Tr F^6) 
          \nn +\frac{222898}{261} (\frac{1}{24} \Tr F^2) ^8
          \nn +\frac{13}{54} (\Tr F^6)  (\Tr F^{10}) 
          \nn +\frac{13}{160} (\Tr F^8) ^2
\end{eqnarray}
The indices are in table 8. 
{\small
\begin{table}[htb]
\centering
\begin{tabular}{|c|c|c|c|c|c|c|c|c|}
\hline
Rep. & Dim. & ${I_2 \over 12}$ & ${I_6}$ & $I_8$ & $I_{10}$ 
         & $I_{12}$ & $I_{14}$ & $I_{18}$ \\ \hline
  (0,0,0,0,0,1,0)
\intab(56) 
\intab(1) 
\intab(1) 
\intab(1) 
\intab(1) 
\intab(1) 
\intab(29) 
\intab(1229) 
\\
  (1,0,0,0,0,0,0)
\intab(133) 
\intab(3) 
\intab(-2) 
\intab(10) 
\intab(-2) 
\intab(-30) 
\intab(542) 
\intab(-111658) 
\\
  (0,0,0,0,0,0,1)
\intab(912) 
\intab(30) 
\intab(-10) 
\intab(-82) 
\intab(230) 
\intab(-2082) 
\intab(-39170) 
\intab(96018190) 
\\
  (0,0,0,0,0,2,0)
\intab(1463) 
\intab(55) 
\intab(90) 
\intab(174) 
\intab(570) 
\intab(2134) 
\intab(238650) 
\intab(161267970) 
\\
  (0,0,0,0,1,0,0)
\intab(1539) 
\intab(54) 
\intab(24) 
\intab(-72) 
\intab(-456) 
\intab(-1992) 
\intab(-235944) 
\intab(-161018664) 
\\
  (1,0,0,0,0,1,0)
\intab(6480) 
\intab(270) 
\intab(30) 
\intab(774) 
\intab(-210) 
\intab(534) 
\intab(73350) 
\intab(-102108810) 
\\
  (2,0,0,0,0,0,0)
\intab(7371) 
\intab(351) 
\intab(-354) 
\intab(2682) 
\intab(-834) 
\intab(-63438) 
\intab(4748094) 
\intab(-14489069226) 
\\
  (0,1,0,0,0,0,0)
\intab(8645) 
\intab(390) 
\intab(-200) 
\intab(40) 
\intab(760) 
\intab(57480) 
\intab(-4368520) 
\intab(14620498520) 
\\
  (0,0,0,0,0,3,0)
\intab(24320) 
\intab(1440) 
\intab(3600) 
\intab(10176) 
\intab(50160) 
\intab(292896) 
\intab(59512080) 
\intab(167838228720) 
\\
  (0,0,0,1,0,0,0)
\intab(27664) 
\intab(1430) 
\intab(-10) 
\intab(-3442) 
\intab(-7450) 
\intab(63998) 
\intab(32976190) 
\intab(149694252430) 
\\
  (0,0,0,0,0,1,1)
\intab(40755) 
\intab(2145) 
\intab(530) 
\intab(-3658) 
\intab(13490) 
\intab(-171138) 
\intab(2436850) 
\intab(-9081228710) 
\\
  (0,0,0,0,1,1,0)
\intab(51072) 
\intab(2832) 
\intab(2872) 
\intab(256) 
\intab(-16568) 
\intab(-172464) 
\intab(-46178632) 
\intab(-158703316792) 
\\
  (1,0,0,0,0,0,1)
\intab(86184) 
\intab(4995) 
\intab(-3165) 
\intab(963) 
\intab(36195) 
\intab(-366717) 
\intab(-37725705) 
\intab(-137019575865) 
\\
  (1,0,0,0,0,2,0)
\intab(150822) 
\intab(9450) 
\intab(8400) 
\intab(41328) 
\intab(59280) 
\intab(410928) 
\intab(30093840) 
\intab(30366263760) 
\\
  (1,0,0,0,1,0,0)
\intab(152152) 
\intab(9152) 
\intab(-328) 
\intab(9320) 
\intab(-78088) 
\intab(-197560) 
\intab(-28617992) 
\intab(-27126731432) 
\\
  (3,0,0,0,0,0,0)
\intab(238602) 
\intab(17940) 
\intab(-26380) 
\intab(271676) 
\intab(-116620) 
\intab(-13615284) 
\intab(1492228660) 
\intab(-16354668799100) 
\\
  (0,0,0,0,0,0,2)
\intab(253935) 
\intab(17820) 
\intab(-9000) 
\intab(-94824) 
\intab(404280) 
\intab(-6024744) 
\intab(-323856360) 
\intab(12685209865560) 
\\
  (0,0,0,0,0,4,0)
\intab(293930) 
\intab(24310) 
\intab(88400) 
\intab(329888) 
\intab(2153840) 
\intab(17066368) 
\intab(4880546320) 
\intab(30243257914480) 
\\
  (2,0,0,0,0,1,0)
\intab(320112) 
\intab(21762) 
\intab(-9318) 
\intab(155826) 
\intab(-75318) 
\intab(-3178974) 
\intab(303759378) 
\intab(-674257133022) 
\\
  (0,1,0,0,0,1,0)
\intab(362880) 
\intab(23760) 
\intab(600) 
\intab(12672) 
\intab(22440) 
\intab(3531792) 
\intab(-239671080) 
\intab(806089955880) 
\\
  (0,0,1,0,0,0,0)
\intab(365750) 
\intab(24750) 
\intab(-9000) 
\intab(-63240) 
\intab(79800) 
\intab(2601720) 
\intab(278208600) 
\intab(-12473996293800) 
\\
\hline
\end{tabular}
\caption{\sl Indices for $E_7$.}
\end{table}
}

\subsection{Indices and trace identities for $E_8$}
Trace identities in the representation (248):
\begin{equation}
\Tr F^4=  36 (\frac{1}{60} \Tr F^2)^2
\end{equation}
\begin{equation}
 \Tr F^6= 30 (\frac{1}{60} \Tr F^2)^3
\end{equation}
\begin{equation}
  \Tr F^{10} =
\frac{15}{4} (\Tr F^8) (\frac{1}{60} \Tr F^2)
 -\frac{315}{4} (\frac{1}{60} \Tr F^2)^5
\end{equation}
\begin{eqnarray}
 \Tr F^{16} & = &
           \frac{3003}{64} (\Tr F^8) (\frac{1}{60} \Tr F^2)^4
          \nn +\frac{143}{1920} (\Tr F^8)^2
          \nn -\frac{273}{20} (\Tr F^{12}) (\frac{1}{60} \Tr F^2)^2
          \nn +\frac{29}{5} (\Tr F^{14}) (\frac{1}{60} \Tr F^2)
          \nn -\frac{147147}{128} (\frac{1}{60} \Tr F^2)^8
\end{eqnarray}
\begin{eqnarray}
 \Tr F^{22} & = &
            -\frac{29393}{39360} (\Tr F^8) (\Tr F^{12}) (\frac{1}{60} \Tr F^2)
          \nn +\frac{323}{1920} (\Tr F^8) (\Tr F^{14})
          \nn +\frac{3193461271}{1679360} (\Tr F^8) (\frac{1}{60} \Tr F^2)^7
          \nn +\frac{361151791}{50380800} (\Tr F^8)^2 (\frac{1}{60} \Tr F^2)^3
          \nn -\frac{298544701}{524800} (\Tr F^{12}) (\frac{1}{60} \Tr F^2)^5
          \nn +\frac{3438981}{16400} (\Tr F^{14}) (\frac{1}{60} \Tr F^2)^4
          \nn -\frac{163457}{6560} (\Tr F^{18}) (\frac{1}{60} \Tr F^2)^2
          \nn +\frac{623}{82} (\Tr F^{20}) (\frac{1}{60} \Tr F^2)
          \nn -\frac{164475086139}{3358720} (\frac{1}{60} \Tr F^2)^{11}
\end{eqnarray}
\begin{eqnarray}
  \Tr F^{26} & = &
           \frac{1888509563603}{44213686272} (\Tr F^8) (\Tr F^{12})
 (\frac{1}{60} \Tr F^2)^3
          \nn -\frac{13299145169}{1347978240} (\Tr F^8) (\Tr F^{14})
 (\frac{1}{60} \Tr F^2)^2
          \nn +\frac{22517}{145152} (\Tr F^8) (\Tr F^{18})
          \nn -\frac{400077914058102709}{7074189803520} (\Tr F^8) 
(\frac{1}{60} \Tr F^2)^9
          \nn -\frac{10415984073858649}{35370949017600} (\Tr F^8)^2 
(\frac{1}{60} \Tr F^2)^5
          \nn -\frac{15835512121}{221838704640} (\Tr F^8)^3 
(\frac{1}{60} \Tr F^2)
          \nn +\frac{5083}{42336} (\Tr F^{12}) (\Tr F^{14})
          \nn +\frac{96960058189033}{5554483200} (\Tr F^{12}) (\frac{1}{60} \Tr F^2)^7
          \nn -\frac{3481855}{6318648} (\Tr F^{12})^2 (\frac{1}{60} \Tr F^2)
          \nn -\frac{28373592046607}{4386278400} (\Tr F^{14}) (\frac{1}{60} \Tr F^2)^6
          \nn +\frac{1446721465417}{1973825280} (\Tr F^{18}) (\frac{1}{60} \Tr F^2)^4
          \nn -\frac{1354577341}{7441008} (\Tr F^{20}) (\frac{1}{60} \Tr F^2)^3
          \nn +\frac{929825}{105868} (\Tr F^{24}) (\frac{1}{60} \Tr F^2)
          \nn +\frac{506251846536678653}{336866181120} (\frac{1}{60} \Tr F^2)^{13}
\end{eqnarray}
\begin{eqnarray}
 \Tr F^{28} & = &
           \frac{666245292706591}{2412453120000} (\Tr F^8) (\Tr F^{12}) 
(\frac{1}{60} \Tr F^2)^4
          \nn -\frac{575488458230809}{9649812480000} (\Tr F^8) (\Tr F^{14}) 
(\frac{1}{60} \Tr F^2)^3
          \nn +\frac{572769851}{1766476800} (\Tr F^8) (\Tr F^{18})
 (\frac{1}{60} \Tr F^2)
          \nn +\frac{1090453}{6133600} (\Tr F^8) (\Tr F^{20})
          \nn -\frac{20281446070347775183}{56144363520000} (\Tr F^8)
 (\frac{1}{60} \Tr F^2)^{10}
          \nn -\frac{19127927}{2770944000} (\Tr F^8)^2 (\Tr F^{12})
          \nn -\frac{1076029152338172071}{561443635200000} (\Tr F^8)^2
 (\frac{1}{60} \Tr F^2)^6
          \nn -\frac{13886644775887}{36092805120000} (\Tr F^8)^3 
(\frac{1}{60} \Tr F^2)^2
          \nn +\frac{3432221}{8131200} (\Tr F^{12}) (\Tr F^{14}) 
(\frac{1}{60} \Tr F^2)
          \nn +\frac{108447101840059177}{969830400000} (\Tr F^{12})
 (\frac{1}{60} \Tr F^2)^8
          \nn -\frac{7079594327}{1941730560} (\Tr F^{12})^2 
(\frac{1}{60} \Tr F^2)^2
          \nn -\frac{23843919029848581}{574393600000} (\Tr F^{14}) 
(\frac{1}{60} \Tr F^2)^7
          \nn +\frac{157757}{1936000} (\Tr F^{14})^2
          \nn +\frac{1607603265138373}{344636160000} (\Tr F^{18})
 (\frac{1}{60} \Tr F^2)^5
          \nn -\frac{61788727534443}{54567392000} (\Tr F^{20})
 (\frac{1}{60} \Tr F^2)^4
          \nn +\frac{286680771}{6654560} (\Tr F^{24})
 (\frac{1}{60} \Tr F^2)^2
          \nn +\frac{565720659186764144207}{58817904640000}
 (\frac{1}{60} \Tr F^2)^{14}
\end{eqnarray}
{ \small
\begin{table}[htb]
\centering
\begin{tabular}{|c|c|c|c|c|c|}
\hline
Rep. & Dimension & ${I_2\over 30}$ & ${I_8}$ & $I_{12}$ & $I_{14}$ 
         \\ \hline
  (1,0,0,0,0,0,0,0) & \hfill $248$        & \hfill $1$        & \hfill $1$         & \hfill $1$          & \hfill $1$           \\
  (0,0,0,0,0,0,1,0) & \hfill $3875$       & \hfill $25$       & \hfill $-17$       & \hfill $223$        & \hfill $-521$        \\
  (2,0,0,0,0,0,0,0) & \hfill $27000$      & \hfill $225$      & \hfill $393$       & \hfill $2073$       & \hfill $8961$        \\
  (0,1,0,0,0,0,0,0) & \hfill $30380$      & \hfill $245$      & \hfill $119$       & \hfill $-1801$      & \hfill $-7945$       \\
  (0,0,0,0,0,0,0,1) & \hfill $147250$     & \hfill $1425$     & \hfill $-801$      & \hfill $-3921$      & \hfill $90423$       \\
  (1,0,0,0,0,0,1,0) & \hfill $779247$     & \hfill $8379$     & \hfill $357$       & \hfill $64677$      & \hfill $-207291$     \\
  (3,0,0,0,0,0,0,0) & \hfill $1763125$    & \hfill $22750$    & \hfill $64330$     & \hfill $653050$     & \hfill $3872050$     \\
  (0,0,1,0,0,0,0,0) & \hfill $2450240$    & \hfill $29640$    & \hfill $576$       & \hfill $-300624$    & \hfill $-407160$     \\
  (1,1,0,0,0,0,0,0) & \hfill $4096000$    & \hfill $51200$    & \hfill $59264$     & \hfill $-176896$    & \hfill $-1416448$    \\
  (0,0,0,0,0,0,2,0) & \hfill $4881384$    & \hfill $65610$    & \hfill $-68202$    & \hfill $1623078$    & \hfill $-5978610$    \\
  (0,0,0,0,0,1,0,0) & \hfill $6696000$    & \hfill $88200$    & \hfill $-64176$    & \hfill $344544$     & \hfill $2464392$     \\
  (1,0,0,0,0,0,0,1) & \hfill $26411008$   & \hfill $372736$   & \hfill $12544$     & \hfill $-928256$    & \hfill $20640256$    \\
  (2,0,0,0,0,0,1,0) & \hfill $70680000$   & \hfill $1083000$  & \hfill $991440$    & \hfill $15398400$   & \hfill $1956600$     \\
  (0,1,0,0,0,0,1,0) & \hfill $76271625$   & \hfill $1148175$  & \hfill $-64071$    & \hfill $732969$     & \hfill $-67564287$   \\
  (4,0,0,0,0,0,0,0) & \hfill $79143000$   & \hfill $1404150$  & \hfill $6100842$   & \hfill $97389402$   & \hfill $723747954$   \\
  (0,0,0,1,0,0,0,0) & \hfill $146325270$  & \hfill $2360085$  & \hfill $-942669$   & \hfill $-20062029$  & \hfill $81822195$    \\
  (0,2,0,0,0,0,0,0) & \hfill $203205000$  & \hfill $3441375$  & \hfill $3576615$   & \hfill $-53721225$  & \hfill $-402564225$  \\
  (2,1,0,0,0,0,0,0) & \hfill $281545875$  & \hfill $4843800$  & \hfill $10500696$  & \hfill $50453496$   & \hfill $233862408$   \\
  (0,0,0,0,0,0,1,1) & \hfill $301694976$  & \hfill $5068800$  & \hfill $-4540800$  & \hfill $36284160$   & \hfill $244435200$   \\
  (1,0,1,0,0,0,0,0) & \hfill $344452500$  & \hfill $5740875$  & \hfill $3591945$   & \hfill $-51773175$  & \hfill $-133939575$  \\
  (1,0,0,0,0,0,2,0) & \hfill $820260000$  & \hfill $14773500$ & \hfill $-7295820$  & \hfill $368355300$  & \hfill $-1650963300$ \\
  (1,0,0,0,0,1,0,0) & \hfill $1094951000$ & \hfill $19426550$ & \hfill $-3552406$  & \hfill $74388314$   & \hfill $342570098$   \\
  (2,0,0,0,0,0,0,1) & \hfill $2172667860$ & \hfill $40883535$ & \hfill $36197469$  & \hfill $182867949$  & \hfill $3494811285$  \\
  (0,1,0,0,0,0,0,1) & \hfill $2275896000$ & \hfill $42214200$ & \hfill $-2179296$  & \hfill $-439704336$ & \hfill $1139298552$  \\
  (0,0,0,0,0,0,3,0) & \hfill $2903770000$ & \hfill $60885500$ & \hfill $-95237740$ & \hfill $3610174100$ & \hfill $-17484769900$\\
  (3,0,0,0,0,0,1,0) & \hfill $3929713760$ & \hfill $79228100$ & \hfill $167887580$ & \hfill $2727186380$ & \hfill $10466026340$ \\
  (0,0,0,0,0,0,0,2) & \hfill $4076399250$ & \hfill $83281275$ & \hfill $-77203203$ & \hfill $-459950403$ & \hfill $17875089309$ \\
  (0,0,1,0,0,0,1,0) & \hfill $4825673125$ & \hfill $93400125$ & \hfill $-36251565$ & \hfill $-180072525$ & \hfill $-4026565725$ \\
  (0,0,0,0,1,0,0,0) & \hfill $6899079264$ & \hfill $139094340$& \hfill $-107301348$& \hfill $-484327668$ & \hfill $13082745060$ \\
\hline
\end{tabular}
\caption{\sl Indices for $E_8$, part one.}
\end{table} }

{\small
\begin{table}[htb]
\centering
\begin{tabular}{|c|c|c|c|}
\hline
 $I_{18}$ & $I_{20}$ & $I_{24}$ & $I_{30}$ \\ \hline
 \hfill $1$              & \hfill $41$               & \hfill $199$                  & \hfill $61$              \\
 \hfill $-281$           & \hfill $720023$           & \hfill $-8538743$             & \hfill $107370139$       \\
 \hfill $131601$         & \hfill $20785953$         & \hfill $1677921087$           & \hfill $32641770621$     \\
 \hfill $-130825$        & \hfill $-21485681$        & \hfill $-1669283839$          & \hfill $-32749110565$    \\
 \hfill $-3057657$       & \hfill $1091333799$       & \hfill $69614416281$          & \hfill $-13332825829797$ \\
 \hfill $3122949$        & \hfill $-891843603$       & \hfill $-70053431037$         & \hfill $13392095601009$ \\
 \hfill $158684770$      & \hfill $53898275690$      & \hfill $19220216027590$       & \hfill $4181215457807530$ \\
 \hfill $96664440$       & \hfill $42322995216$      & \hfill $18320504001024$       & \hfill $4178331237939240$ \\
 \hfill $-129144448$     & \hfill $-47849709056$     & \hfill $-18734041431424$      & \hfill $-4186512285307648$ \\
 \hfill $-131658210$     & \hfill $336853672758$     & \hfill $-90053043268518$      & \hfill $53479289344535190$ \\
 \hfill $32750232$       & \hfill $-373796000256$    & \hfill $71666802925296$       & \hfill $-57656847733698648$ \\
 \hfill $-887500544$     & \hfill $601951099904$     & \hfill $-72720785548544$      & \hfill $50171949070921216$ \\
 \hfill $1422281640$     & \hfill $-496637707680$    & \hfill $79406160046320$       & \hfill $-50034441438444840$ \\
 \hfill $368688753$      & \hfill $-283297282191$    & \hfill $-5188200203889$       & \hfill $7356020195237373$ \\
 \hfill $52000832994$    & \hfill $23915855965002$   & \hfill $18712064549246118$    & \hfill $18617881056766529514$ \\
 \hfill $10867212915$    & \hfill $-105084506709$    & \hfill $-9407941453913691$    & \hfill $-16544807547516665625$ \\
 \hfill $-32456806065$   & \hfill $-12291541050705$  & \hfill $-4580281352771055$    & \hfill $-1096019298618319485$ \\
 \hfill $-14097200472$   & \hfill $-10058442950664$  & \hfill $-14025344636105496$   & \hfill $-17530899917498537112$ \\
 \hfill $-22110000000$   & \hfill $4164533921280$    & \hfill $9773219242968960$     & \hfill $16439802976484140800$ \\
 \hfill $13626625545$    & \hfill $10661778671985$   & \hfill $13958084762321535$    & \hfill $17581170727669309605$ \\
 \hfill $-10809258420$   & \hfill $79696508512740$   & \hfill $-32082728508187860$   & \hfill $-3180131072643324180$ \\
 \hfill $19892530658$    & \hfill $-97084816917686$  & \hfill $17486065830031526$    & \hfill $-14251422467900989462$ \\
 \hfill $-107677764315$  & \hfill $118967521198869$  & \hfill $-19927893621180789$   & \hfill $12855700625814459585$ \\
 \hfill $-134688633528$  & \hfill $112636179273744$  & \hfill $-29916313535701344$   & \hfill $-3637801450738060968$ \\
 \hfill $-675793438060$  & \hfill $2261894204588980$ & \hfill $-1168369134581957620$ & \hfill $7589921677844796014660$ \\
 \hfill $721988940500$   & \hfill $90443981815180$   & \hfill $94451334234961220$    & \hfill $3296282167094693780$ \\
 \hfill $-1346442094371$ & \hfill $689963459361477$  & \hfill $507354515738178603$   & \hfill $-7162275330089795125911$ \\
 \hfill $518531279235$   & \hfill $147692804386155$  & \hfill $73753893160393605$    & \hfill $17639532662269149015$    \\
 \hfill $104563589460$   & \hfill $-617925299730228$ & \hfill $-535510167643211772$  & \hfill $7128890471280203011860$  \\ \hline
\end{tabular}
\caption{\sl Indices for $E_8$, part two.}
\end{table}
}

.

\section{The computer program for the reduction into invariants}

We have implemented the reduction algorithms into a computer 
program\footnote{This program can be obtained from 
"http://norma.nikhef.nl/$\sim$t68/FORMapplications/Color"} in the 
language of FORM~\cite{form}. This language is particularly suited for 
these types of problems. Because of all the problems with reduction 
identities when the number of vertices becomes large, we have restricted 
the program to the case of no more than 16 vertices. If the user needs to 
run the program with more vertices, it can be extended by analogy, but many 
new reduction identities would have to be derived. Alternatively one could 
decide to not reduce a number of contractions with $f$'s in them and leave 
them for later evaluation. The user should be warned however that some 
diagrams with 16 vertices may need quite some computer time and 
resources for their evaluation. 

The program consists of three parts. The first part reduces all traces of 
matrices which do not belong to the adjoint representation. Much attention 
is given to a potential contraction of indices. The special cases have been 
written out in one highly nested loop to take the maximum benefit of these 
contractions. This saves much work when we have to use the algorithm of 
equation (\ref{eq:symtrace}) for the remaining trace. It is quite useful to 
rewrite each invariant immediately with the notation of equation 
(\ref{eq:p-rewrite}). This removes invariants which have more than one line 
contracted with the same $f$ in a very natural way without any extra 
pattern matching. For this type of rewriting {\tt FORM} even has a special 
statement (ToVector).

The second part eliminates the loops of $f$'s. Here we do not have to worry 
  about contracted indices inside the trace, because {\tt FORM}
 looks each time 
for the smallest loop to make its next reduction. Hence this part of the 
program is much simpler.
Because this reduction is much faster than the general reduction the first 
routine calls this reduction routine each time after it has removed 
contracted indices and generated more $f$'s. Very often such a removal 
generates a loop of $f$'s and if this is removed it may introduce new 
contracted indices. The net result can be a significant increase in speed. 
In some cases this is not faster however. Hence there is the option not to 
hunt for $f$-loops until all other representations have been rewritten.

The third part of the program contains the reduction identities. Here the 
program tries to eliminate contractions with $f$'s that are obviously zero 
and to rewrite the contractions for which it can construct meaningful 
identities. This is a rather peculiar piece of programming. The derivation 
of the reduction identities is by no means a fixed algorithm. The equation 
(\ref{eq:jacobin}) can be applied to one of the $d_R$ and one of the $f$'s, 
but it is not always clear which pair will give good results and which will 
make things worse. In general it seems to be a good strategy to try to 
increase the number of contractions between the invariants (number of 
common indices). In that case there will be more contractions between the 
$f$'s and hence more chance of loops that contain only $f$'s. This is not 
always possible in a direct way, and sometimes we have to just try 
equation (\ref{eq:jacobin}) in the hope that in the next pass the improvement 
will follow. The selection of the invariant and the $f$ that take part in 
this game has to be done carefully and the code consists of two pieces that 
make a slightly different choice. By running a loop that contains the first 
choice twice and the second choice once, we could reduce nearly everything 
up to 16 vertices. Of course we had to define the object of formula 
(\ref{eq:aaaff}) as a separate entity. Similarly we had to define three 
such objects at level 16. The program also defines the topologies at level 
16 that we did not present in the text.


\section{Some examples}

We have run a number of color traces with the program. Here we present the 
results with some timing information. All runs were done on a PentiumPro 200 
processor running NextStep. First we look at traces of the type
\[ Tr [ T_R^{i_1}\cdots T_R^{i_n}T_R^{i_1}\cdots T_R^{i_n} ] \]
as such traces represent some type of maximal complexity. 
Here we show the results to $n = 7$ in table \ref{tab:qloop}.
\begin{table}[htb]
\centering
\begin{tabular}{|c|c|l|}
\hline
n & time & result \\ \hline
2 & 0.23 s & \begin{minipage}[t]{12.5cm}
$ N_AI_2(R)(C_R-C_A/2) $
 \end{minipage} \\ \hline
3 & 0.23 s & \begin{minipage}[t]{12.5cm}
$ N_AI_2(R)(C_R-C_A)(C_R-C_A/2) $
 \end{minipage} \\ \hline
4 & 0.25 s & \begin{minipage}[t]{12.5cm}
$d_R^{abcd}d_A^{abcd} + N_AI_2(R)(C_R^3 - 3\ C_R^2C_A +11/4\ C_RC_A^2
		-19/24\  C_A^3)$
 \end{minipage} \\ \hline
5 & 0.95 s & \begin{minipage}[t]{12.5cm}
$d_R^{abcd}d_A^{abcd}(5C_R-6C_A) +1/3I_2(R)\  d_A^{abcd}d_A^{abcd}$ \\
$+N_AI_2(R)(C_R^4-5\ C_R^3C_A+35/4\ C_R^2C_A^2-155/24\ C_RC_A^3 +125/72\ C_A^4)$
 \end{minipage} \\ \hline
6 & 2.59 s & \begin{minipage}[t]{12.5cm}
$ -8d_R^{abcdef}d_A^{abcdef}+6 d_R^{abcd}d_A^{abef}d_A^{cdef}
+I_2(R)d_A^{abef}d_A^{cdef}(2\ C_R-199/60\ C_A) $ \\
$ +d_R^{abef}d_A^{cdef}(15\ C_R^2-87/2\ C_RC_A+179/6\ C_A^2)
  +N_AI_2(R)(C_R^5 $\\ $-15/2\ C_R^4C_A+85/4\ C_R^3C_A^2
	-115/4\ C_R^2C_A^3 +905/48\ C_RC_A^4\ -1405/288\ C_A^5)$
 \end{minipage} \\ \hline
7 & 34.9 s & \begin{minipage}[t]{12.5cm}
$	+112/3\ d_R^{abcdef}d_A^{abcg}d_A^{defg}
	-328/9\ d_A^{abcdef}d_R^{abcg}d_A^{defg} $ \\
$  +d_R^{abcdef}d_A^{abcdef}(-56\ C_R+296/3\ C_A) $ \\
$ +d_R^{abcd}d_A^{abef}d_A^{cdef}(42\ C_R-749/10\ C_A)
    +67/15\ I_2(R)d_A^{abcd}d_A^{abef}d_A^{cdef} $\\
$ +d_R^{abcd}d_A^{abcd} (35\ C_R^3-357/2 C_R^2C_A+868/3\ C_RC_A^2
		-2695/18\ C_A^3) $ \\
$ +I_2(R)d_A^{abcd}d_A^{abcd} (7\ C_R^2-1603/60 C_RC_A+497/20\ C_A^2 )
		 $ \\
$ +N_AI_2(R)(
       + C_R^6
       - 21/2        \ C_R^5C_A  
       + 175/4       \ C_R^4C_A^2
       - 280/3       \ C_R^3C_A^3 $ \\ $
       + 5215/48     \ C_R^2C_A^4
       - 19075/288   \ C_R  C_A^5
       + 43357/2592  \ C_A^6 )$
 \end{minipage} \\
\hline
\end{tabular}
\caption{\sl \label{tab:qloop} Results for traces of the type
$ Tr [ T_R^{i_1}\cdots T_R^{i_n}T_R^{i_1}\cdots T_R^{i_n} ] $.}
\end{table}

\noindent Actually the program can go to $n=8$. For this it took about 
$1520$ sec. We do not give the answer here.
Similarly we can calculate this in the adjoint representation only. 
This is of course much faster because the program selects 
automatically the smallest loops. These results can be found in table 
\ref{tab:gloop}.
\begin{table}[htb]
\centering
\begin{tabular}{|c|c|l|}
\hline
n & time & result \\ \hline
2 & 0.15 s & \begin{minipage}[t]{7.5cm}
$\frac{1}{2}N_AC_A^2$
 \end{minipage} \\ \hline
3 & 0.20 s & \begin{minipage}[t]{7.5cm}
$0$
 \end{minipage} \\ \hline
4 & 0.23 s & \begin{minipage}[t]{7.5cm}
$d_A^{abcd}d_A^{abcd} - \frac{1}{24}N_AC_A^4$
 \end{minipage} \\ \hline
5 & 0.78 s & \begin{minipage}[t]{7.5cm}
$\frac{2}{3}C_Ad_A^{abcd}d_A^{abcd} - \frac{1}{36}N_AC_A^5$
 \end{minipage} \\ \hline
6 & 0.81 s & \begin{minipage}[t]{7.5cm}
$	d_A^{abcd}d_A^{abef}d_A^{cdef}
	+\frac{1}{4}C_A^2d_A^{abcd}d_A^{abcd}
	-\frac{13}{864}N_AC_A^6$
 \end{minipage} \\ \hline
7 & 0.89 s & \begin{minipage}[t]{7.5cm}
$		-\frac{8}{9}d_A^{abcdef}d_A^{abcg}d_A^{defg}
		+\frac{53}{30}C_A\ d_A^{abcd}d_A^{abef}d_A^{cdef} $ \hfill \\
$		-\frac{5}{648}N_AC_A^7 $
 \end{minipage} \\ \hline
\end{tabular}
\caption{\label{tab:gloop} Like the previous table but now in the adjoint 
representation.}
\end{table}
For $n=8$ the program took $1.5$ sec
We notice that here the computer time does not increase very much with the 
number of crossing lines (the number of vertices and hence the weight is 
$2n$). There is actually more `compilation time' than `execution' time. 
The jump in time going from $n=4$ to $n=5$ represents the use of the
reduction algorithms to eliminate $f$'s. In that case the program needs
considerably more compilation time.

The fact that diagrams with only vertices in the adjoint representation are 
easier to evaluate than the diagrams with vertices in the other 
representations is exactly the opposite of what happens with the Cvitanovic 
algorithms~\cite{cvitanovic}. For them each $f$ is converted to one or more 
terms with one or more matrices in the fundamental representation. This can 
lead to an avalanche of terms at the intermediate stages, because no 
advantage is taken from the potentially simpler structures. In the case of 
the traces in the fundamental representation the Cvitanovic algorithms are 
much faster. These algorithms do not worry about symmetrizations and are 
directly applicable to such traces.

As an example of high complexity for purely adjoint diagrams we take 
the one topology of girth 6 with 14 vertices. It is also called the Coxeter 
graph. In this the smallest loop has 6 vertices. 
The result is rather short:
\begin{eqnarray}
	G_6(n=14) & = &
		\frac{16}{9}d_A^{abcdef}d_A^{abcg}d_A^{defg}
		-\frac{8}{15}C_A\ d_A^{abcd}d_A^{abef}d_A^{cdef}
		+\frac{1}{648}N_AC_A^7
\end{eqnarray}
This took 1.6 sec.

\section{Explicit expressions}

Here we present some expressions for a number of invariants. These are 
mostly invariants for representations that can be used as reference 
representations. The expressions are given in terms of the
normalization factor $\eta$ defined in (\ref{eq:CasNorm}).

In all cases the tensors (referred to as $d_n(R)$) are defined as  
\begin{equation}
 d_R^{a_1\ldots a_n}=\Str T^{a_1}_{R}\ldots T^{a_n}_{R} \ . 
\end{equation}
In particular no traces are subtracted and no overall
factors are included. For $SO(N)$ we deviate from the preferred index
normalization of table (\ref{tab:SONindices}), since
otherwise we would have to deal with a few low-$N$ $SO(N)$
cases separately. 

Results for the fundamental (vector) representation $V$ of SU(N):
\begin{eqnarray}
C_V & = & \frac{a}{N}(N^2-1) \\
C_A & = & 2aN \\
d_{33}(VV) & = & \frac{a^3}{2N}(N^2-1)(N^2-4) \\
d_{44}(VV) & = & \frac{a^4}{6N^2}(N^2-1)(N^4-6N^2+18) \\
d_{55}(VV) & = & \frac{a^5}{24N^3}(N^2-1)(N^2-2)(N^4+24) \\
d_{433}(VVV) & = & \frac{a^5}{6N^2}(N^2-1)(N^2-4)(N^2-6) \\
d_{66}(VV) & = & \frac{a^6}{120N^4}(N^2-1)(N^8+6N^6-60N^4+600) \\
d_{633}(VVV) & = & \frac{a^6}{480N^3}(N^2-1)^2(N^2-4)^2 \\
d_{543}(VVV) & = & \frac{a^6}{288N^3}(N^2-1)(N^2-4)(N^4-6N^2+18) \\
d_{444}(VVV) & = & \frac{a^6}{27N^3}(N^2-1)(N^6-9N^4+81N^2-189) \\
d_{3333}(VVVV) & = & \frac{a^6}{8N^2}(N^2-1)(N^2-4)(N^2-12)
\end{eqnarray}
with $a=\frac{1}{2}\eta$ (see equation(\ref{eq:CasNorm})).
The choice $\eta=1; a=\half$ corresponds to the most commonly
used normalization. Then $\Tr T_V^a T_V^b = \half \delta^{ab}$. 
In SU(N) the vector representation is always equal to the
reference representation. 

For the vector representation of SO(N) we have:
\begin{eqnarray}
C_V & = & \frac{a}{2}(N-1) \\
C_A & = & a(N-2) \\
N_A & = & \frac{1}{2}N(N-1) \\
d_{44}(VV) & = & \frac{a^4}{24}N_A(N^2-N+4) \\ & = &
	\frac{a^4}{12}N_A(N_A+2)  \\
d_{66}(VV) & = & \frac{a^6}{1920}N_A(N^4-2N^3+33N^2-32N+52) \\ & = &
	\frac{a^6}{480}N_A(N_A^2+16N_A+13)\\
d_{444}(VVV) & = & \frac{a^6}{432}N_A(2N^3-3N^2+33N-16)
\\
\end{eqnarray}
with $a=\eta$ (see equation(\ref{eq:CasNorm})). In this case
$a=\eta=2$ is the most frequently used convention. Then
$\Tr T^a_V T^b_V= 2 \delta^{ab}$.

Note that for SO(N), $N \leq 6$ the vector representation is
{\it not} the
reference representation. 
The formulas for the reference representation
for those groups can be read off from the appropriate SU(N) or
Sp(N) results. The tensors used in the foregoing formulas
are normalized so that $I_4(V)=I_6(V)=1$ in (\ref{eq:tensordef}). 
As explained above, this 
differs from the index normalization chosen in 
table (\ref{tab:SONindices}) for $SO(7)$ and $SO(8)$. For these 
groups our convention is to make $I_4$ twice as large, and hence
$d_4$ twice as small.

For the vector representation  of Sp(N) we have:
\begin{eqnarray}
C_V & = & \frac{a}{2}(N+1) \\
C_A & = & a(N+2) \\
N_A & = & \frac{1}{2}N(N+1) \\
d_{44}(VV) & = & \frac{a^4}{24}N_A(N^2+N+4) \\ & = &
	\frac{a^4}{12}N_A(N_A+2)  \\
d_{66}(VV) & = & \frac{a^6}{1920}N_A(N^4+2N^3+33N^2+32N+52) \\ & = &
	\frac{a^6}{480}N_A(N_A^2+16N_A+13)\\
d_{444}(VVV) & = & \frac{a^6}{432}N_A(2N^3+3N^2+33N+16) \\
\end{eqnarray}
with $a=\frac{1}{2}\eta$ (see equation(\ref{eq:CasNorm})).
In this case the vector coincides with the reference representation.

For all groups for which $I_4(A)=0$ we can derive a number of invariants with 
relatively simple methods. 
This is of particular interest for the exceptional algebras, which have
$I_4(R)=0$ for {\it  any} representation. We will present the following
formulas for tensors defined in the adjoint representation, which is not
the reference representation (except for $E_8$). The reason for doing
this is that it allows us to write a single set of relations for all
algebras. It is straightforward to re-express these results in terms of
the reference representation. To do so one needs the relation between
adjoint tensors and reference tensors, which follows directly from
the characters of both representations; the latter can be computed
by means of the methods used in Appendix A.
The adjoint representation
is unsuitable for the odd traces of $E_6$, which are discussed
separately below.

To do the computations,
first we notice that
\begin{equation}
	d_A^{i_1i_2i_3i_4} = I_{22}(A)(\delta^{i_1i_2}\delta^{i_3i_4}
	+\delta^{i_1i_3}\delta^{i_2i_4}+\delta^{i_1i_4}\delta^{i_2i_3})/3
\end{equation}
with $I_{22}(A) = \frac{5}{2}C_A^2/(N_A+2)$. Using this and 
equation (\ref{eq:6loops}) we can determine $d_{66}(AA)$. Next we can go even 
further by using a technique similar to the one used to derive 
equation(\ref{eq:6loops}): We run the program for the product of two traces 
with 8 vertices. First we run it for a representation R in one trace and 
the adjoint representation in the other. After the run we put R equal to A. 
This gives an expression that includes $d_{88}(AA)$ and $d_{844}(AAA)$ and 
objects that contain combinations of $d_4$ and $d_6$. We can run the same 
traces with the program, but starting with both of them in the adjoint 
representation. In that case we obtain an expression that does not contain 
$d_{88}(AA)$. This gives us the required equation. Now we substitute $d_4$ 
and we need an equation for $d_8^{jji_1\cdots i_6}$ which is also easy to 
obtain with the program:
\begin{equation}
	d_A^{jji_1\cdots i_6} = \frac{10}{21}d_A^{i_1\cdots \i_6}
		+\frac{1}{6}(d_A^{ji_1i_2i_3}d_A^{ji_4i_5i_6}+\cdots)/10
\end{equation}
in which we have to take the 10 symmetric combinations over the indices in 
the last term. For $d_6$ we have a similar equation which is given by 
equation (\ref{eq:d62}). In total we obtain:
\begin{eqnarray}
d_{44}(AA) & = & \frac{25\ C_A^4}{12(N_A+2)}
	\\
d_{66}(AA) & = & \frac{C_A^6N_A}{(N_A+2)^2}(\frac{797}{288} + \frac{8}{27}N_A
       - \frac{1}{864}N_A^2)
	\\
d_{444}(AAA) & = & \frac{C_A^6 N_A}{(N_A+2)^2} (\frac{125}{27} + \frac{125}{216}N_A )
	\\
d_{644}(AAA) & = & \frac{175}{48}\frac{C_A^7 N_A}{(N_A+2)^2}
	\\
d_{88}(AA) & = & \frac{C_A^8 N_A}{(N_A+2)^3} (
       \frac{3425}{1008}
       + \frac{111025}{145152}N_A
       + \frac{125}{6804}N_A^2
       + \frac{25}{435456} N_A^3 )
	\\
d_{844}(AAA) & = & \frac{C_A^8 N_A}{(N_A+2)^3}(
		\frac{125}{24} + \frac{625}{288}N_A )
	\\
d_{664}(AAA) & = & \frac{C_A^8 N_A}{(N_A+2)^3}(
       \frac{5455}{864}
       + \frac{3485}{2592}N_A
       - \frac{5}{2592}N_A^2)
	\\
d_{4444a}(AAAA) & = & \frac{C_A^8 N_A}{(N_A+2)^3}(
       \frac{3125}{324}
       + \frac{625}{216}N_A
       + \frac{625}{1296}N_A^2 )
	\\
d_{4444b}(AAAA) & = & \frac{C_A^8 N_A}{(N_A+2)^3}(
       \frac{6875}{648}
       + \frac{3125}{1296}N_A )
\end{eqnarray}
The last two topologies are defined as
\begin{eqnarray}
    d_{4444a}(p_1,p_2,p_3,p_4) & = &
    \raisebox{-16pt}{ \begin{picture}(45,40)(0,0)
        \Line(10,6.5)(35,6.5) \Line(10,3.5)(35,3.5) 
		\Line(8.5,5)(8.5,30) \Line(11.5,5)(11.5,30)
        \Line(33.5,5)(33.5,30) \Line(36.5,5)(36.5,30)
        \Line(10,28)(35,28) \Line(10,32)(35,32) 
        \dReg(10,5) \dReg(35,5) \dReg(10,30) \dReg(35,30)
        \Text(7,33)[rb]{\footnotesize 1}
        \Text(7,3)[rt]{\footnotesize 3}
        \Text(38,3)[lt]{\footnotesize 4}
        \Text(38,32)[lb]{\footnotesize 2}
    \end{picture}} \\
    d_{4444b}(p_1,p_2,p_3,p_4) & = &
    \raisebox{-16pt}{ \begin{picture}(45,40)(0,0)
        \Line(10,6.5)(35,6.5) \Line(10,3.5)(35,3.5)
		\Line(10,5)(10,30) \Line(10,5)(35,30)
        \Line(10,30)(35,5) \Line(35,5)(35,30)
        \Line(10,28)(35,28) \Line(10,32)(35,32) 
        \dReg(10,5) \dReg(35,5) \dReg(10,30) \dReg(35,30)
        \Text(7,33)[rb]{\footnotesize 1}
        \Text(7,3)[rt]{\footnotesize 3}
        \Text(38,3)[lt]{\footnotesize 4}
        \Text(38,32)[lb]{\footnotesize 2}
    \end{picture}}
\end{eqnarray}

The appropriate values to be substituted for the various groups are given 
in table \ref{tab:cavalues}.
\begin{table}[htb]
\centering
\begin{tabular}{|r|c|c|c|c|c|}
\hline
group   & $G_2$     & $F_4$     & $E_6$      & $E_7$      & $E_8$      \\ \hline
$N_A$   & $14$      & $52$      & $78$       & $133$      & $248$      \\ \hline
$C_A$   & $4\ \eta$ & $9\ \eta$ & $24\ \eta$ & $18\ \eta$ & $30\ \eta$ \\ \hline
\end{tabular}
\caption{\label{tab:cavalues}\sl Values of $N_A$ and $C_A$ for the 
exceptional groups.}
\end{table}

For $E_6$ we also have to consider the invariants with 5 indices. These we 
can obtain with the reference representation $r$. For 
invariants that involve this representation we have:
\begin{eqnarray}
C_r & = & \frac{52}{3}\ \eta \\
N_r & = & 27 \\
d_{44}(rr) & = & 18720\ \eta^4 \\
d_{55}(rr) & = & 291200\ \eta^5 \\
d_{66}(rr) & = & 1023152\ \eta^6 \\
d_{444}(rrr) & = & 536640\ \eta^6 \\
d_{77}(rr) & = & \frac{11211200}{3}\ \eta^7 \\
d_{644}(rrr) & = & \frac{20616544}{5}\ \eta^7\\
d_{554}(rrr) & = & 582400\ \eta^7 \\
d_{88}(rr) & = & \frac{466346192}{27}\ \eta^8
\end{eqnarray}
A number of these quantities can be obtained in various ways and serve as a 
check of our programs.

\section{Chiral representations of $SO(2m)$}

The algebra $D_m$ has an index of order $m$ that vanishes for
the vector representation, and is non-zero only for chiral
representations. The latter are characterized by having 
unequal values for the last two Dynkin
labels, $a_{m-1}\not= a_m$. The simplest representations of this
type are the spinors, $S=(0,\ldots,0,1,0)$ and $S'=(0,\ldots,0,1)$.
If $m$ is odd $S'$ is the complex conjugate of $S$. Note that for odd
$m$ the extra index (henceforth referred to as the ``chiral index") has
odd order, unlike all other $SO(N)$ indices, whereas for $m$ even there
are two distinct indices of even order, namely the chiral index and
one of the regular indices.

Since the chiral index vanishes for the vector representation, whereas
all indices are non-zero for the spinor, it might be argued that the
latter is perhaps a more suitable choice for the reference representation.
However, the vector representation has other advantages, perhaps 
most importantly that the trace-identities for traces of order 
larger than $2m$ are simpler. The main drawback of this choice is
that it
requires a separate discussion of chiral traces, which we give here.   

Our conventions for $SO(2m)$ are as follows.
The generators in the vector representation are
\begin{equation}
\label{eq:Ogenerator} M^{\mu\nu}_{ij}=i\sqrt{\eta/2}(\delta^{\mu}_i\delta^{\nu}_j-
\delta^{\nu}_i\delta^{\mu}_j) 
\end{equation}
This is a complete set of generators for $\mu < \nu$. The normalization
factor $\sqrt{\eta/2}$ is introduced in (\ref{eq:CasNorm}). Clearly
for $SO(N)$ the most attractive choice is $\eta=2$.
 
The generators of the (chiral plus anti-chiral) 
spinor representation are then
\begin{equation}
 \Sigma^{\mu\nu}={i\over 4}\sqrt{\eta/2}
 \left[\gamma^{\mu},\gamma^{\nu}\right] \ ,
\end{equation}
with $\{\gamma^{\mu},\gamma^{\nu}\}=2g^{\mu\nu}$. The dimension of 
this Clifford algebra is $2^m$.

For the chiral $\gamma$-matrix 
(often referred to as $\gamma_5$)
 we choose
\begin{equation} 
\gamma_c=(i)^m\gamma^1 \ldots \gamma^{2m}\ \ \ \ 
\end{equation}
The phase is chosen so that $\gamma_c$ is hermitean (assuming the
other $\gamma$-matrices are chosen hermitean as well).
Hermiticity fixes the overall factor
up to an $m$-dependent sign, for which we have made a conventional choice.  

The chiral spinor generators are
\begin{equation}
\Sigma^{\mu\nu}_{\pm} = \half \Sigma^{\mu\nu}(1\pm \gamma_c) 
\end{equation}
The symmetrized trace of $m$ such generators is
\begin{equation}
{\rm Str} \Sigma^{\mu_1\nu_1}_{\pm} \ldots  \Sigma^{\mu_m\nu_m}_{\pm} 
= \half {\rm Str} \Sigma^{\mu_1\nu_1} \ldots  \Sigma^{\mu_m\nu_m} 
\pm\half {\rm Str} \Sigma^{\mu_1\nu_1} \ldots  \Sigma^{\mu_m\nu_m}
\gamma_c 
\end{equation}
The first term yields (for $m$ even)
an ordinary $d$-tensor, the second one yields
the $\tilde d$ tensor. 
By general arguments,
\begin{eqnarray} 
\nonumber
 {\rm Str} \Sigma^{\mu_1\nu_1}_{\pm} \ldots  \Sigma^{\mu_n\nu_n}_{\pm}
&=I_n(S_{\pm}) d^{[\mu_1\nu_1],\ldots,[\mu_n\nu_n]}+
{\rm lower~order~}\nonumber
\\
&\pm\tilde I_n(S_{\pm}) 
\tilde d^{[\mu_1\nu_1],\ldots,[\mu_n\nu_n]}\ ,\hfill 
\end{eqnarray}
where $[\mu_1\nu_1]$ denotes an adjoint index.
By convention 
\begin{equation}
 \tilde I_n(S_{\pm}) = \pm 1 
\end{equation}
then
\begin{equation}
\tilde d^{[\mu_1\nu_1],\ldots,[\mu_m\nu_m]}=
\half {\rm Str} \Sigma^{\mu_1\nu_1} \ldots  \Sigma^{\mu_m\nu_m}
\gamma_c =
\half (\eta/2)^{m/2} \epsilon^{\mu_1,\ldots,\nu_m}\ ,
\end{equation}
with the definition $\epsilon^{1,\ldots,2m}=1$. Note that
the interpretation requires {\it pairs} of indices to be identified
with an adjoint index, and that the $\epsilon$ tensor is indeed fully
symmetric under permutation of pairs of indices. 

This gives us an explicit expression for the extra tensor, and
computing invariants that involve this tensor is now straightforward.
One can either do that by computing all other tensors also in the
spinor representation. Then computing the invariants amounts to 
simple $\gamma$-algebra. Since expressions
for the spinor characters in terms of
vector traces are explicitly known
(see (\ref{eq:DspinCharPlus}) and (\ref{eq:DspinCharMin})), all 
tensors in the spinor
representation can be re-expanded in terms of reference tensors.
Alternatively it is also very easy to work directly with the
reference tensors, which, using (\ref{eq:Ogenerator}) can be 
expressed completely in terms of kronecker $\delta$'s with vector indices.
These indices are to be taken in pairs and identified with adjoint indices,
and can then be pairwise contracted with the $\epsilon$ tensor.  

For the normalization of the tensor $\tilde d$ of $SO(2m)$ we get 
$$ d_{mm}(CC) = {1\over4}  2^{-m} (2m)! $$
The factor $2^{-m}$ compensates the double counting of index
pairs $\mu\nu$ as compared to adjoint indices. The argument ``CC"
indicates that two chiral tensors $\tilde d$ are used. Furthermore
we have 
$$ d_{mm}(CV) = d_{mm}(CS)=0 \ . $$ 
Here ``V" is the fundamental representation, as before, and ``S" refers
to the non-chiral tensor computed in the spinor representation,
$$ d_S^{[\mu_1\nu_1],\ldots,[\mu_m\nu_m]}=
\half {\rm Str} \Sigma^{\mu_1\nu_1} \ldots  \Sigma^{\mu_m\nu_m} $$
This tensor is related to the one in the reference representation $V$.

To 
order 12, the maximal order we used for the other $SO(N)$ tensors,
the only other quantity of interest is $d_{444}$, which has
chiral tensor contributions only for $SO(8)$.  Here is a complete
set of results for that group.
\begin{eqnarray}
 d_{44}(VV)&=70 \eta^4 \\
 d_{44}(SS)&={245\over 8}  \eta^4\\
 d_{44}(CC)&={315\over 8 } \eta^4\\
 d_{66}(VV)&={581\over 8 } \eta^6\\
 d_{444}(VVV)&={70 \eta^6} \\
 d_{444}(SSS)&={665\over 32} \eta^6 \\
 d_{444}(SSC)=d_{444}(VVC)&=0 \\
 d_{444}(SCC)&={525\over 32} \eta^6 \\
 d_{444}(VCC)&={-105\over 8 }\eta^6 \\
 d_{444}(CCC)&=0 
\end{eqnarray}
The fundamental quantities here are the ones involving $d_4(V), d_6(V)$
and $d_4(C)$. The tensor $d_4(S)$ is related to $d_4(V)$:
$$ d_S^{abcd}=-\half d_V^{abcd}+{3\over2} d_{2,2}^{abcd}$$
The results involving $d_4(S)$ are given here because
one may use them to check the triality relations
$$ d_{44}(VV)=d_{44}(SS)+2 d_{444}(SC)+d_{44}(CC) $$
and 
$$ d_{444}(VVV)=d_{444}(SSS)+3 d_{444}(SSC)+3 d_{444}(SCC)+d_{444}(CCC) $$
which are due to the fact that the tensors $d_4(V)$ and the
combinations $d_4(S)\pm d_4(C)$ are given by symmetrized traces
of the triality-related representations $(8_v)$, $(8_s)$ and $(8_c)$.


\end{document}